\documentclass{article}
\usepackage[left=1in,right=1in,top=1in,bottom=1in]{geometry}
\usepackage{amsmath}

\usepackage{stmaryrd}
\usepackage{mdframed}
\usepackage{mathpartir}
\usepackage{irule}

\usepackage{hhline}
\usepackage{amssymb,makecell}
\usepackage{url}
\usepackage[bookmarks=true,colorlinks=true, linkcolor=red, citecolor=cyan]{hyperref}
\usepackage[color,all,poly,matrix,arrow]{xy}
\usepackage{tikz}
\usepackage{pgfplots}
\pgfplotsset{width=8cm,compat=1.9}
\usetikzlibrary{pgfplots.dateplot}
\usepackage{pgfplotstable}

\newcommand{\LTO}[1]{\overset{\sf{#1}}{\bullet}}
\newcommand{\DTO}[1]{\underset{\sf{#1}}{\bullet}}
\newcommand{\LTOO}[1]{\stackrel{\sf{#1}}{\circ}}

\newcommand{\Too}[1]{\xrightarrow{\ \ #1\ \ }}

\newcommand{\Fromm}[1]{\xleftarrow{\ \ #1\ \ }}

\def\Inst{{\bf Inst}}
\def\iinst{{\text{\textendash}\bf \Inst}}
\def\taking{:}

\title
      {\vspace{-.7in}  Algebraic Data Integration\footnote{Expanded and corrected version of the 2017 Journal of Functional Programming Article \url{http://doi.org/10.1017/S0956796817000168}}}

\author
       {  Patrick Schultz \\
          Department of Mathematics \\ Massachusetts Institute of Technology
        \\ \and 
        Ryan Wisnesky\\
          Conexus AI \\ {\sf ryan@wisnesky.net}
        }

\begin{document}

\label{firstpage}
\maketitle

\begin{abstract}
In this paper we develop an algebraic approach to data integration by combining techniques from functional programming, category theory, and database theory. In our formalism, database schemas and instances are algebraic (multi-sorted equational) theories of a certain form.  Schemas denote categories, and instances denote their initial (term) algebras.  The instances on a schema $S$ form a category, $S\iinst$, and a morphism of schemas $F : S \to T$ induces three adjoint data migration functors:  $\Sigma_F : S\iinst \to T\iinst$, defined by substitution along $F$, which has a right adjoint $\Delta_F : T\iinst \to S\iinst$, which in turn has a right adjoint $\Pi_F : S\iinst \to T\iinst$.  We present a query language based on for/where/return syntax where each query denotes a sequence of data migration functors; a pushout-based design pattern for performing data integration using our formalism; and describe the implementation of our formalism in a tool we call CQL (Categorical Query Language).
\end{abstract}


\section{Introduction}

In this paper we develop an algebraic approach to data integration by combining techniques from functional programming, category theory, and database theory.  By {\it data integration} we mean combining separate but related database schemas and instances to form a coherent, unified schema and instance, and we consider query and data migration to be special cases of data integration.  By {\it algebraic} we mean that our schemas are algebraic (purely equational) theories and our instances denote algebras (models) of our schemas.  We use category theory to define the semantics of our approach: schemas and instances form categories, and data integration operations are characterized with categorical constructions such as adjoint functors and pushouts.  We use techniques from functional programming to implement our approach by constructing syntactic objects and reasoning about them, both on paper and using automated techniques.  We use database theory as a baseline with which to compare our approach.

The mathematics of the semantics of our approach are worked out in detail in Schultz {\it et al.}~\cite{patrick} using sophisticated category theory.  In this paper, our goal is to implement a syntax for this semantics, and crucially, to do so in a {\it computable} way.  How to do this is not obvious, because the mathematical objects defined in Schultz {\it et al.}~\cite{patrick} are almost always infinite and not computable.  This paper is a comprehensive description of the implementation of the CQL tool (\url{http://categoricaldata.net/}), an open-source data integration tool capable of solving problems similar to those solved by relational data integration tools such as Clio~\cite{Haas:2005:CGU:1066157.1066252} and Rondo~\cite{Melnik:2003:RPP:872757.872782}, as well as query languages such as SQL and LINQ~\cite{monad}.  

Because our approach draws on functional programming, category theory, and database theory, the more knowledge a reader has about each of these fields the more the reader will get out of the paper.  These three theories are used in a deep, rather than wide, way: we use mostly basic categorical concepts such as category, functor, natural transformation, and adjunction; we use mostly basic functional programming concepts such as equational logic and algebraic data types; and we use mostly basic database theory concepts such as conjunctive queries and labeled nulls.  For this reason, we believe that a reader well-versed in only one of these areas can still get something out of this paper, and will be rewarded with a deeper insight into the other areas, at least from a data integration perspective.  We include primers on category theory (section~\ref{cat}) and equational logic (section~\ref{alg}), and connections to database theory are made as remarks in the text.  

\subsection{Background}

Our data model extends a particular category-theoretic data model that we call the {\it functorial data model} (FDM)~\cite{Spivak:2012:FDM:2324905.2325108}.  Originating in the late 1990s~\cite{Fleming02adatabase}, the FDM defines a schema $S$ to be a finite presentation of a  category~\cite{BW}: a directed, labeled multi-graph and a set of path equality constraints such as shown in Figure~\ref{fdm1}.  In this figure, the nodes of the graph indicate ``entities'' of employees, departments, and strings, and the arrows represent functions between entities, such as the assignment of a name to each employee.  The equations in the graph serve as data integrity constraints, indicating, for example, that secretaries work in the same department they are the secretary for.  An instance $I$ on $S$ (a.k.a, an $S$-instance) is a set of tables, one per node in S, and a set of columns, one per edge in $S$, that obey the equations in $S$.  Such an instance determines a functor $S \to {\bf Set}$, where {\bf Set} is the category of sets and functions.  In this paper, we will write both ``.'' and ``;'' to refer to left to right composition in a category, and ``$\circ$'' to refer to right to left composition.   

\begin{figure}[hb]
\begin{mdframed}
$$
\xymatrix@=9pt{&\LTO{{\sf Emp}}\ar@<.5ex>[rrrrr]^{\sf wrk}\ar@(l,u)[]+<0pt,13pt>^{\sf mgr}\ar[dddr]^{\sf ename}&&&&&\LTO{{\sf Dept}}\ar@<.5ex>[lllll]^{\sf secr}\ar[dddllll]^{\sf dname}\\\\\\&&\DTO{\sf String}&~&~&~&}
$$
$$
{\sf mgr} ; {\sf wrk} = {\sf wrk}  \ \ \ \ \  
id_{\sf Dept} = {\sf secr} ; {\sf wrk} \ \ \ \ \  
{\sf mgr} ; {\sf mgr} = {\sf mgr}
$$ 
\begin{centering}
 
\begin{footnotesize}
\begin{tabular}{|c|c|c|}  
\multicolumn{3}{c}{{\sf Dept}}   \\  \hline 
\hspace{.1in} {\sf  ID} \hspace{.1in} & \hspace{.1in}{\sf dname} \hspace{.1in}& \hspace{.1in} {\sf secr} \hspace{.1in} \\  \hhline{|=|=|=|}
 m & Math &  b  \\ \hline
 s & $\text{null}_1$ &  c \\ \hline
\end{tabular}
\hspace*{.4in} 
\begin{tabular}{|c|c|c|c|}
\multicolumn{4}{c}{{\sf Emp}} \\ \hline
\hspace{.2in} {\sf  ID} \hspace{.2in} & \hspace{.2in} {\sf ename} \hspace{.2in} & \hspace{.2in} {\sf mgr} \hspace{.2in} & \hspace{.02in} {\sf wrk} \hspace{.02in} \\ \hhline{|=|=|=|=|}
a & Al &  b &  m  \\ \hline
b & $\text{null}_2$ &  b &  m \\  \hline
c & Carl &c  & s \\ \hline
\end{tabular}
\end{footnotesize}
   
\end{centering}

\caption{A Schema and Instance in the Original Functorial Data Model}
\label{fdm1}
\end{mdframed}
\end{figure}

In the FDM, the database instances on a schema $S$ (i.e., functors $S \to {\bf Set}$) constitute a category, denoted $S\iinst$, and a functor (a.k.a schema mapping~\cite{FKMP05}) $F\taking S \to T$ between schemas $S$ and $T$ induces three adjoint data migration functors: $\Delta_F\taking T\iinst \to S\iinst$, defined as $\Delta_F(I) := I \circ F$ (note that $F : S \to T$ and $I : T \to {\bf Set}$), and the left and right adjoints to $\Delta_F$, respectively: $\Sigma_F\taking S\iinst \to T\iinst$ and $\Pi_F\taking S\iinst \to T\iinst$.  The $\Sigma, \Delta, \Pi$ data migration functors provide a category-theoretic alternative to the traditional relational operations for both querying data (relational algebra) and migrating / integrating data (``chasing'' embedded dependencies~\cite{FKMP05}).  Their relative advantages and disadvantages over the relational operations are still being studied, but see Spivak~\cite{Spivak:2012:FDM:2324905.2325108} for a preliminary discussion.  At a high-level, $\Delta$ can be thought of as a projection, $\Pi$ as join, and $\Sigma$ as union/chase.

An example schema mapping $F : S \to T$ is shown in Figure~\ref{fmd}.  A full description of this figure is given in Section~\ref{fdmsec}; here we sketch an overview of the figure to give intuition about $\Delta_F, \Pi_F$, and $\Sigma_F$.  The functor $F : S \to T$ is the unique edge-preserving map from $S$ to $T$.  The $\Delta_F$ operation takes the table {\sf N} and projects it to two tables.  The $\Pi_F$ operation performs a cartesian product of {\sf N1} and {\sf N2}, and the $\Sigma_F$ operation performs an ``outer join''~\cite{Garcia-Molina:2008:DSC:1450931} of {\sf N1} and {\sf N2}; i.e., it unions two tables that have different columns by adding null values as necessary.  When there is an edge ${\sf f} : {\sf N1} \to {\sf N2}$ (Figure~\ref{fkmex}), $\Delta_F$ performs a ``lossless join decomposition''~\cite{Garcia-Molina:2008:DSC:1450931} along ${\sf f}$ of {\sf N}, $\Pi_F$ performs a join along ${\sf f}$, and $\Sigma_F$ performs a union-then-merge~\cite{tuplemerge}, resulting in a join for the particular instance in the figure.

The FDM's basic idea of schemas-as-categories and three adjoint data migration functors $\Delta,\Sigma,\Pi$ recurs in our data model, but we base our formalism entirely on algebraic (equational) logic and therefore diverge from the original FDM.  We define database schemas and instances to be equational theories of a certain kind.  A schema mapping $F \taking C \to D$ is defined as a morphism (provability-respecting translation) of equational theories $C$ and $D$, and we define the $\Sigma_F$ data migration as substitution along $F$.  The conditions we impose on our equational theories guarantee that $\Sigma_F$ has a right adjoint, $\Delta_F$, which in turn has a right adjoint, $\Pi_F$.  In practice, programming with $\Delta$, $\Sigma$, and $\Pi$ is verbose, and so we define a terse query language where each query is a collection of generalized for/where/return expressions~\cite{DBLP:books/aw/AbiteboulHV95}).  Each query can be evaluated into a data migration of the form $\Delta \circ \Pi$ (and vice-versa) and each query can be ``co-evaluated'' into a data migration of the form $\Delta \circ \Sigma$ (and vice-versa).  To integrate schema we must go beyond $\Delta,\Sigma,\Pi$, and we use pushouts~\cite{BW} of schemas and instances as the basis for a schema and data integration design pattern suitable for building data warehouses.  

\begin{figure}[b]
\begin{mdframed}
\hspace{1.2in}
\parbox{1.2in}{\fbox{\xymatrix@=8pt{
& \LTO{String}\\
\LTO{N1} \ar[ur]^{\sf name} \ar[dr]_{\sf salary} & & \LTO{N2}\ar[dl]^{\sf age}\\
&  \DTO{Nat}}} }
$\hspace{.4in} \Too{\ \ F \ \ } \hspace{.8in}$
\parbox{1.2in}{\fbox{\xymatrix@=8pt{
\LTO{String}\\
\LTO{N}\ar[u]^{\sf name} \ar@/^/[d]^{\sf age} \ar@/_/[d]_{\sf salary}\\
\DTO{Nat}}}}
\vspace*{.1in}

\hrulefill
\vspace*{.05in}
\begin{footnotesize}

\begin{tabular}{|c|c|c|}
\multicolumn{3}{c}{{\sf N1}} \vspace{.01in} \\\hline
\hspace{.01in} {\sf ID} \hspace{.01in} & \hspace{.1in} {\sf Name} \hspace{.1in} &  {\sf Salary}  \\\hhline{|=|=|=|}
1&Alice&\$100\\\hline 
2&Bob&\$250\\\hline 
3&Sue&\$300\\\hline
\end{tabular}
\hspace{.25in}
\begin{tabular}{|c|c|}
\multicolumn{2}{c}{{\sf N2}} \vspace{.01in} \\\hline
\hspace{.01in} {\sf ID} \hspace{.01in} &   {\sf Age} \\\hhline{|=|=|}
1&20\\\hline 
2&20\\\hline 
3&30\\\hline
\end{tabular}
\hspace{.21in}
$\Fromm{ \llbracket \Delta_F \rrbracket }$
\hspace{.21in}
\begin{tabular}{|c|c|c|c|}
\multicolumn{4}{c}{{\sf N}} \vspace{.01in} \\\hline
\hspace{.01in} {\sf ID} \hspace{.01in} &  \hspace{.1in} {\sf Name}  \hspace{.1in} &  \hspace{.1in} {\sf Salary}  \hspace{.1in} &  {\sf Age} \\\hhline{|=|=|=|=|} 
1&Alice&\$100&20\\\hline 
2&Bob&\$250&20\\\hline 
3&Sue&\$300&30\\\hline
\end{tabular}

\vspace{.05in}
\hrulefill
\vspace{.05in}

\begin{tabular}{|c|c|c|}
\multicolumn{3}{c}{{\sf N1}} \vspace{.01in} \\\hline
\hspace{.01in} {\sf ID} \hspace{.01in} & \hspace{.1in} {\sf Name} \hspace{.1in} &  {\sf Salary}\\\hhline{|=|=|=|} 
1&Alice&\$100\\\hline 
2&Bob&\$250\\\hline 
3&Sue&\$300\\\hline
\end{tabular}
\hspace{.25in}
\begin{tabular}{|c|c|}
\multicolumn{2}{c}{{\sf N2}} \vspace{.01in} \\\hline
\hspace{.01in} {\sf ID} \hspace{.01in} &  {\sf Age} \\\hhline{|=|=|} 
1&20\\\hline 
2&20\\\hline 
3&30\\\hline
\end{tabular}
\hspace{.21in}
$\Too{ \llbracket \Sigma_F \rrbracket }$
\hspace{.21in}
\begin{tabular}{|c|c|c|c|}
\multicolumn{4}{c}{{\sf N}} \vspace{.01in} \\\hline
\hspace{.01in} {\sf ID} \hspace{.01in} &  \hspace{.1in} {\sf Name}  \hspace{.1in} &  \hspace{.1in} {\sf Salary}  \hspace{.1in} &   {\sf Age}   \\\hhline{|=|=|=|=|} 
a&Alice&\$100& 1.age\\\hline 
b&Bob&\$250& 2.age \\\hline 
c&Sue&\$300&  3.age \\\hline
d& 4.name & 4.salary & 20\\\hline 
e& 5.name & 5.salary & 20\\\hline 
f& 6.name & 6.salary & 30\\\hline
\end{tabular}

\vspace{.05in}
\hrulefill
\vspace{.05in}

\begin{tabular}{|c|c|c|}
\multicolumn{3}{c}{{\sf N1}} \vspace{.01in} \\\hline
\hspace{.01in} {\sf ID} \hspace{.01in} & \hspace{.1in} {\sf Name} \hspace{.1in} &  {\sf Salary} \\\hhline{|=|=|=|}
1&Alice&\$100\\\hline 
2&Bob&\$250\\\hline 
3&Sue&\$300\\\hline
\end{tabular}
\hspace{.25in}
\begin{tabular}{|c|c|}
\multicolumn{2}{c}{{\sf N2}} \vspace{.01in} \\\hline
\hspace{.01in} {\sf ID} \hspace{.01in} &   {\sf Age}  \\\hhline{|=|=|}
1&20\\\hline 
2&20\\\hline 
3&30\\\hline
\end{tabular}
\hspace{.21in}
$\Too{ \llbracket \Pi_F \rrbracket }$
\hspace{.21in}
\begin{tabular}{|c|c|c|c|}
\multicolumn{4}{c}{{\sf N}} \vspace{.01in} \\\hline
\hspace{.01in} {\sf ID} \hspace{.01in} &  \hspace{.1in} {\sf Name}  \hspace{.1in} &  \hspace{.1in} {\sf Salary}  \hspace{.1in} &  \hspace{.01in} {\sf Age}   \\\hhline{|=|=|=|=|} 
a&Alice&\$100&20\\\hline 
b&Bob&\$250&20\\\hline 
c&Sue&\$300&20\\\hline
d&Alice&\$100&20\\\hline 
e&Bob&\$250&20\\\hline 
f&Sue&\$300&20\\\hline
h&Alice&\$100&30\\\hline 
i&Bob&\$250&30\\\hline 
j&Sue&\$300&30\\\hline
\end{tabular}

\end{footnotesize}

\caption{Example Functorial Data Migrations}
\label{fmd}
\end{mdframed}
\end{figure}


\subsection{Outline}

The rest of this paper is divided into five sections. In section~\ref{cat} we review category theory, and in section~\ref{alg} we review algebraic (multi-sorted equational) logic.  In section~\ref{formal} we describe how we use algebraic theories to define schemas, instances, and the other artifacts necessary to perform data migration and query.   In section~\ref{impl} we describe how our formalism is implemented in the CQL tool.  In section~\ref{pattern} we describe how we use pushouts of schemas and instances to perform data integration and include an extended example.

\subsection{Related Work}

  In this section we describe how our work relates to the functional data model (section~\ref{other}), functional programming (section~\ref{fp}), relational data integration (section~\ref{relrel}), and other work which treats schemas as categories (section~\ref{fdmrel}).  Mathematical related work is discussed in Schultz {\it et al.}~\cite{patrick}.   

\subsubsection{vs The Functional Data Model}
\label{other}

Our data model is simultaneously an extension of, and a restriction of, the functional data model~\cite{Shipman:1981:FDM:319540.319561}.  Both formalisms use functions, rather than relations, over entities and data types as their principal data structure.  The two primary ways the functional data model extends ours is that it allows products of entities (e.g., a function ${\sf Person} \times {\sf Department} \to {\sf String}$), and it allows some non-equational data integrity constraints (e.g., a constraint ${\sf f}(x) = {\sf f}(y) \to x = y$).  These two features can be added to our data model as described in Spivak~\cite{SPIVAK2014}, although doing so alters our data model's properties (e.g., the existence of $\Pi$ cannot in general be guaranteed in the presence of non-equational constraints).  Therefore, our data model can be thought of as extending the fragment of the functional data model where schemas are categories.  Restricting to schemas-as-categories allows us to extend the functional data model with additional operations (e.g., $\Pi$) and to provide strong static reasoning principles (e.g., eliminating the need for run time data integrity checking; see section~\ref{vc}).  The functional data model also includes updates, a topic we are still studying.  The schemas of our data model and the functional data model both induce graphs, so both data models are graph data models in the sense of Angles and Gutierrez~\cite{Angles:2008:SGD:1322432.1322433}.  See Spivak~\cite{Spivak:2012:FDM:2324905.2325108} for a discussion of how our formalism relates to RDF.

\subsubsection{vs Functional Programming}
\label{fp}

Our formalism and functional programming languages both extend equational logic~\cite{Mitchell:1996a}.  As discussed in section~\ref{trans}, many artifacts of our formalism, including schemas and user-defined functions, are ``algebraic data types'' and our instances denote implementations (algebras) of these types.  (Our use of ``algebraic data type'' must be understood in the algebraic specification sense, as a data type specified by a set of equations~\cite{Mitchell:1996a}, rather than in the Haskell/ML sense of a data type specified by products and sums.)  As such, the implementation of our formalism draws heavily on techniques from functional programming: for example, when a set of data integrity constraints forms a confluent and terminating rewrite system, our implementation uses reduction to normal forms to decide equality under the constraints (section~\ref{knuth}).  Our formalism's use of types, rather than terms, to represent sets (e.g., entity sets such as {\sf Person}) is common in expressive type theories such as Coq~\cite{Bertot:2010:ITP:1965123} and contrasts with the sets-as-terms approach used in simply-typed comprehension calculi~\cite{monad}; for a comparison of the two approaches (which we have dubbed QINL and LINQ, respectively), see Schultz {\it et al.}~\cite{qinl}.  Certain categories of schemas and instances are cartesian closed~\cite{BW}, meaning that certain schemas and instances can be defined by expressions in the simply-typed $\lambda$-calculus, and meaning that certain categories of schemas and instances are models of the simply-typed $\lambda$-calculus, but we have yet to find an application of this fact (programming CQL itself using the simply-typed lambda calculus was not as pleasant as the approach developed in this paper).

\subsubsection{vs Relational Data Integration}
\label{relrel}

Our formalism is an alternative to the traditional relational formalisms for both querying data (relational algebra) and migrating / integrating data (``chasing'' embedded dependencies (EDs)~\cite{FKMP05}).  In Spivak \& Wisnesky~\cite{relfound}, we proved that the $\Delta,\Sigma,\Pi$ operations can express any (select, project, cartesian product, union) bag-relational algebra query and gave conditions for the converse to hold.  Our formalism uses purely equational data integrity constraints, which can be captured using first-order EDs, but not all first-order EDs can be captured using only equations.  However, our formalism can be extended in a simple way, described in Spivak~\cite{SPIVAK2014}, to capture all first-order EDs; when this is done, we find that a parallel chase step of an ED on an instance can be expressed using the pushout construction of section~\ref{pattern}, and the CQL tool iterates pushouts to implement chasing of EDs.  At present, we do not understand the relationship between our formalism and second-order EDs~\cite{Fagin:2005:CSM:1114244.1114249}, although our ``uber-flower'' queries (section~\ref{uber}) can be written as second-order EDs of a particular form (section~\ref{eduber}).   For additional discussion about how our formalism relates to EDs and the field of ``model management''~\cite{Melnik:2003:RPP:872757.872782}, see the survey paper Schultz {\it et al.}~\cite{wadt}.

The duality between $\Sigma$ and $\Pi$ in our formalism suggests a ``missing'' operator in relational algebra: the dual to join.  Because a join is a product followed by a filter along an equivalence relation (an equalizer), its dual is a sum (disjoint union) followed by a tuple-merge along an equivalence relation (a co-equalizer).  The tuple-merge operation appears in relational data integration settings~\cite{tuplemerge}, and so does the idea of forming equivalence classes of tuples, in the guise of the chase~\cite{Fagin:2005:CSM:1114244.1114249}, but we have not seen the dual to join explicitly described as such in the relational literature.  

Our formalism defines databases ``intensionally'', as sets of equations, and so in relational parlance our databases are ``deductive databases''~\cite{DBLP:books/aw/AbiteboulHV95}.  As such, some care must be taken when mediating between relational definitions and categorical definitions.  For example, our instances can be inconsistent, in the sense that an instance might prove that $1=2$ for two distinct constant symbols $1$ and $2$, and only consistent instances can be meaningfully translated into relational instances.  In addition, our schemas do not define a set of constants (a ``domain'') that all the instances on that schema share, as is customary in relational data integration~\cite{FKMP05}.  For these and other reasons (mentioned throughout this paper), our work is closer in spirit to traditional logic~\cite{opac-b1104628} than database theory~\cite{Doan:2012:PDI:2401764}.

The pushout data integration pattern (section~\ref{pattern}) is a ``global as view''~\cite{Doan:2012:PDI:2401764} pattern, because the integrated schema  is a function (the pushout) of the source schemas.  But rather than relating the integrated schema to the source schemas by EDs or queries, we use functors.  A pushed-out instance satisfies a universal property similar to that of a universal solution to a set of EDs~\cite{FKMP05}.  Pushouts are investigated for relational data integration purposes in Alagic \& Bernstein~\cite{DBLP:conf/dbpl/AlagicB01}.  In that paper, the authors describe a design pattern for data integration which applies to a large class of formalisms: the so-called {\it institutions}~\cite{Goguen1984}.  Our formalism is an institution, but their work differs from ours in key respects.  First, they are primarily concerned with the $\Delta$ data migration functor (they call our $\Delta$ functor ``Db'' in their paper), because $\Delta$ exists in all institutions.  They recognize that pushouts (what they call ``schema joins'') are a canonical way of obtaining integrated schemas, and that not all institutions have pushouts of schemas (ours does).  Their main theorem is that in any institution the $\Delta$ functor can be used to migrate the data on a pushout schema back to the source schemas.  Our design pattern uses the $\Sigma$ functor to go the other way: pushing data from source schemas to the integrated schema.  See Goguen~\cite{Goguen04informationintegration} for more information about data integration in institutions.  In the more general setting of algebraic specification, pushouts have received considerable attention as a means to integrate specifications~\cite{Blum:1987:ASM:29880.29886}.

\subsubsection{vs The Functorial Data Model}
\label{fdmrel}

Our work is related to a family of data models that treat database schemas as categories or variations thereof~\cite{Johnson2002}~\cite{Fleming02adatabase}~\cite{Spivak:2012:FDM:2324905.2325108}~\cite{SPIVAK2014}~\cite{relfound}~\cite{patrick}.  We refer to these data models as ``functorial data models''.  The original functorial data model~\cite{Fleming02adatabase} treated schemas as finitely presented categories and instances as set-valued functors, but is difficult to use for data integration purposes because most constructions on set-valued functors are only defined up to unique isomorphism, and in the context of data integration, we must distinguish two kinds of values in a database: atomic values such as strings and integers which must be preserved by morphisms (e.g., {\sf Bill}), and meaningless identifiers which need not be preserved (e.g., auto-generated IDs).  (In contexts outside of data integration, such as query, there may not be a need to distinguish two types of values.)  For example, the situation in Figure~\ref{attprob}, which holds in the original functorial data model, is untenable for data integration purposes.

\begin{figure}[h]
\begin{mdframed}
\begin{footnotesize}
\vspace{.2in}
\begin{tabular}{|c|c|c|c|}\hline
 {\sf ID} \ & \  {\sf Name}   & \hspace{.01in}  {\sf Age} & {\sf Salary} \\\hhline{|=|=|=|=|}  
1&Alice&20&\$100\\\hline
2&Bob&20&\$250\\\hline 
3&Sue&30&\$300\\\hline
\end{tabular}
\hspace{.1in}
$\stackrel{\cong}{(\textnormal{good})}$
\hspace{.1in}
\begin{tabular}{|c|c|c|c|}\hline
  {\sf ID}  &   {\sf Name}   &   {\sf Age}  & {\sf Salary} \\\hhline{|=|=|=|=|} 
\color{black}{4}&Alice&20&\$100\\\hline 
\color{black}{5}&Bob&20&\$250\\\hline 
\color{black}{6}&Sue&30&\$300\\\hline
\end{tabular}\hspace{.1in}
$\stackrel{\cong}{(\textnormal{bad})}$\hspace{.1in}
\begin{tabular}{|c|c|c|c|}\hline
  {\sf ID}  &   {\sf Name}  &   {\sf Age}  & {\sf Salary} \\\hhline{|=|=|=|=|}
1&\color{black}{Amy}&20&\$100\\\hline 
2&\color{black}{Bill}&20&\$250\\\hline 
3&\color{black}{Susan}&30&\$300\\\hline
\end{tabular}
\end{footnotesize}
\\
\caption{The Attribute Problem}
\label{attprob}
\end{mdframed}
\end{figure}

Several approaches to this ``attribute problem'' have been proposed, including Johnson {\it et al.}~\cite{Johnson2002} and Spivak \& Wisnesky~\cite{relfound}.  This paper extends the latter  paper by defining database schemas to exist in an ambient computational context called a ``type-side''.  Data values that inhabit types  (e.g., {\sf Bill} : {\sf String}) are preserved by database morphisms, but other values, such as  meaningless identifiers, are not.  As a result, our formalism does not suffer from the attribute problem, and solving the attribute problem was a significant motivation for our work.

Recently, Patterson~\cite{relolog} defined ``relational ologs'', and in so doing generalized the functorial data model (where instances are functors $C \to {\bf Set}$) to relations: Patterson's instances are functors $C \to {\bf Rel}$, where {\bf Rel} is the category of sets and relations.  Interestingly, over finite databases, Patterson's formalism, which is graphical in nature, is equivalent to the extension of our formalism by embedded dependencies (``EDs''; see Section~\ref{relrel}).  The CQL tool's built-in example ``FOAF'' (Friend of a Friend) describes this equivalence.  Independently, Zielinski {\it et al.}~\cite{allegorical} proposed the ``Allegorical Conceptual Data Model'', a similar relational generalization of the functorial data model.

\section{Review of Category Theory}
\label{cat}

In this section we review standard material on category theory~\cite{BW}.  Readers familiar with category theory can safely skim or skip this section.  

\subsection{Categories with Products}
A {\it category} ${\bf C}$ consists of a collection of {\it objects} $Ob({\bf C})$ and a collection of {\it morphisms} $Hom({\bf C})$ between objects.  Each morphism $m$ has a source object $S$ and a target object $T$, which we write as $m : S \to T$.  Every object $X$ has an identity morphism $id_X : X \to X$.  When $X$ is clear from the context we will write $id_X$ as simply $id$.  Two morphisms $f : B \to C$ and $g: A \to B$ may be {\it composed}, written $f \circ g: A \to C$ or $g ; f : A \to C$.  We also write $g.f$ for $g;f$, especially in syntax.  Composition is associative and $id$ is its unit:
$$
f \circ id = f  \ \ \ \ \ \  id \circ f = f \ \ \ \ \ \  f \circ (g \circ h) = (f \circ g) \circ h
$$
\noindent
A morphism $f: X \to Y$ is an {\it isomorphism} when there exists a $g: Y \to X$ such that
$$
f \circ g = id \ \ \ \ \ \  g \circ f = id
$$
Two objects are {\it isomorphic} when there exists an an isomorphism between them.  Example categories include:
\begin{itemize}
\item {\bf Set}, the {\it category of sets}.  The objects of {\bf Set} are sets, and a morphism $f : X \to Y$ is a (total) function from set $X$ to set $Y$.  Given morphisms $f : Y \to Z$ and $g : X \to Y$, the morphism $f \circ g : X \to Z$ is defined as function composition: $(f \circ g)(x) := f(g(x))$.  The isomorphisms of {\bf Set} are bijective functions.  For each object $X$, $id_X$ is the identity function on $X$.
\item Any directed graph generates a category, called the {\it free} category on the graph: its objects are the vertices of the graph, and its morphisms are the paths in the graph.  For each vertex $X$, $id_X$ is the 0-length path $X \to X$.  Composition of morphisms is concatenation of paths, and there are no non-identity isomorphisms.
\end{itemize}

A category {\bf C} is said to {\it have products} when for every pair of objects $X,Y$ in {\bf C}, there exists an object $X \times Y$ in {\bf C}, morphisms $\pi_1 : X \times Y \to X$ and $\pi_2 : X \times Y \to Y$ in {\bf C}, and for every pair of morphisms $f : A \to X, g : A \to Y$ in {\bf C}, there exists a morphism $\langle f, g \rangle : A \to X \times Y$ in {\bf C}, such that
$$
\pi_1 \circ \langle f, g \rangle = f \ \ \ \ \ \pi_2 \circ \langle f, g \rangle = g \ \ \ \ \ 
$$  
and such that for every morphism $h : A \to X \times Y$, 
$$
\langle \pi_1 \circ h, \pi_2 \circ h \rangle = h
$$ 
For example, set-theoretic cartesian product is a product in the category of sets, {\bf Set}.

\subsection{Functors}
A {\it functor} $F: {\bf C} \to {\bf D}$ between two categories ${\bf C}$ and ${\bf D}$ is a mapping of objects of ${\bf C}$ to objects of ${\bf D}$ and morphisms of ${\bf C}$ to morphisms of ${\bf D}$ that preserves identities and composition:
$$
F(f: X \to Y): F(X) \to F(Y) 
\ \ \ \ \ \ 
F(id_X) = id_{F(X)} \ \ \ \ \ \ 
F(f \circ g) = F(f) \circ F(g)
$$
Example functors include:
\begin{itemize}
\item For any category $\bf{C}$, the identity functor $Id : \bf{C} \to \bf{C}$ maps each object and morphism to itself.
\item For any categories ${\bf C}$ and ${\bf D}$ and $D$ an object of ${\bf D}$, there exists a constant functor taking each object $C$ in ${\bf C}$ to $D$ and each morphism in $C$ to $id_D$.
\item  The power set functor $\mathcal{P} : {\bf Set} \to {\bf Set}$ maps each set to its power set and each function $f : X \to Y$ to the function which sends $U \subseteq X$ to its image $f(U) \subseteq Y$.
\item  For each set $A$, there is a product functor $- \times A : {\bf Set} \to {\bf Set}$ mapping each set $X$ to the cartesian product $X \times A$ and each function $f : X \to Y$ to the function $f \times id_A : X \times A \to Y \times A$. 
\end{itemize}

A functor $F : {\bf C} \to {\bf Set}$ for any category $\bf{C}$ is said to be {\it set-valued}.

\subsection{Natural Transformations}
A {\it natural transformation} $\alpha : F \Rightarrow G$ between two functors $F: {\bf C} \to {\bf D}$ and $G: {\bf C} \to {\bf D}$ is a family of morphisms $\alpha_X : F(X) \to G(X)$ in ${\bf D}$, one for each object $X$ in ${\bf C}$, such that for every $f: X \to Y$ in ${\bf C}$,
$$
\alpha_Y \circ F(f) = G(f) \circ \alpha_X 
$$
This equation may conveniently expressed as a {\it commutative diagram}:
\begin{displaymath}
    \xymatrix{
        F(X) \ar[r]^{F(f)} \ar[d]_{\alpha_X} & F(Y) \ar[d]^{\alpha_Y} \\
        G(X) \ar[r]_{G(f)}       & G(Y) }
\end{displaymath}

A natural transformation $\alpha$ is a {\it natural isomorphism} when for every object $X$ in ${\bf C}$, the morphism $\alpha_X$ is an isomorphism in ${\bf D}$.  Example natural transformations include:
\begin{itemize}
\item The identity natural isomorphism $id_F : F \Rightarrow F$ for a functor $F : {\bf C} \to {\bf D}$ is defined as $id_{F_X} : F(X) \to F(X) := id_{F(X)}$.
\item Consider the power set functor $\mathcal{P} : {\bf Set} \to {\bf Set}$.  There is a natural transformation $sng : Id_{\bf Set} \Rightarrow \mathcal{P}$ that maps every set $X$ to the singleton set $\{ X \}$ (i.e., $sng_X : X \to \mathcal{P}(X)$), and there is a natural transformation $union : \mathcal{P} \circ \mathcal{P} \Rightarrow \mathcal{P}$ that maps a set of sets $\{ X_1, \ldots, X_n\}$ to its $n$-ary union $X_1 \cup \ldots \cup X_n$ (i.e., $union_X : \mathcal{P}(\mathcal{P}(X)) \to \mathcal{P}(X)$). 
\end{itemize}


\subsection{Adjunctions}
An {\it adjunction} between categories ${\bf C}$ and ${\bf D}$ consists of
a functor $F : {\bf D} \to {\bf C}$ called the left adjoint,
a functor $G : {\bf C} \to {\bf D}$ called the right adjoint,
a natural transformation $\epsilon  : F \circ G \Rightarrow Id_{\bf C}$ called the counit, and
a natural transformation $\eta  : Id_{\bf D} \Rightarrow G \circ F$ called the unit,
such that for every object $X$ in ${\bf C}$ and $Y$ in {\bf D}, the following equations hold:
$$
id_{F(Y)} = \epsilon_{F(Y)} \circ F(\eta_Y) \ \ \ \ \ \ id_{G(X)} = G(\epsilon_X) \circ \eta_{G(X)}
$$
Consequently, the set of morphisms $F(Y) \to X$ is bijective with the set of morphisms $Y \to G(X)$.  Example adjunctions include:
\begin{itemize}
\item Let $A$ be a set and consider the product functor $- \times A : {\bf Set} \to {\bf Set}$.  The exponential functor $-^A : {\bf Set} \to {\bf Set}$, which maps each set $X$ to the set of functions from $A$ to $X$ (written $X^A$), is right adjoint to $- \times A$.  Intuitively, this is because the set of functions $X \times Y \to Z$ is bijective with the set of functions $X \to Z^Y$.  
\item Consider the category of groups and group homomorphisms, ${\bf Grp}$.  The functor $\mathit{free} : {\bf Set} \to {\bf Grp}$, which maps each set $X$ to the free group generated by $X$, and the functor $\mathit{forget} : {\bf Grp} \to {\bf Set}$ which maps each group  to its underlying set, are adjoint. Intuitively, maps from the free group $\mathit{free}(X)$ to a group $Y$ correspond precisely to maps from the set $X$ to the set $\mathit{forget}(Y)$: each homomorphism from $\mathit{free(X)}$ to $Y$ is fully determined by its action on generators.
\end{itemize}


\section{Review of Multi-sorted Equational Logic}
\label{alg}

In this section we review standard material on multi-sorted equational logic, following the exposition in Mitchell~\cite{Mitchell:1996a}.  Theories in multi-sorted equational logic are also called ``algebraic theories'', as well as ``Lawvere theories'' and ``product theories''.  We will use these phrases interchangeably.  For a category-theoretic study of such theories, see Adamek  {\it et al.}~\cite{Adamek.Rosicky.Vitale:2011a}.  Readers familiar with equational logic can safely skim or skip this section.  We will write ``theory'' instead of ``multi-sorted equational theory'' in this section.  

\subsection{Syntax}

In this section we define the syntax of theories.  Readers may wish to refer to the example theory about strings and natural numbers in Figure~\ref{Type} while reading this section.

 A {\it signature} $Sig$ consists of:
\begin{enumerate}
\item A set $Sorts$ whose elements are called {\it sorts},
\item A set $Symbols$ of pairs $(f, s_1 \times \ldots \times s_k \to s)$ with $s_1, \ldots, s_k , s \in Sorts$ and no $f$ occurring in two distinct pairs.  We write $f \taking X$ instead of $(f, X) \in Symbols$.  When $k = 0$, we may call $f$ a {\it constant symbol} and write $f \taking s$ instead of $ f : \ \to s$.  Otherwise, we may call $f$ a {\it function symbol}.  
\end{enumerate}
We assume we have some countably infinite set $\{ v_1, v_2, \dots \}$, whose elements we call {\it variables} and which are assumed to be distinct from any sort or symbol we ever consider.  A {\it context} $\Gamma$ is defined as a finite set of variable-sort pairs, with no variable given more than one sort:
$$
\Gamma := \{ v_1 : s_1, \ldots, v_k : s_k \}
$$
When the sorts $s_1 , \ldots, s_k$ can be inferred, we may write a context as $\{ v_1, \ldots, v_k \}$.  We may write $\{ v_1 : s, \ldots, v_k : s \}$ as $\{ v_1, \ldots, v_k : s\}$.  We may write $\Gamma \cup \{ v : s \}$ as $\Gamma, v:s$.  We inductively define the set $Terms^s(Sig, \Gamma)$ of {\it terms} of sort $s$ over signature $Sig$ and context $\Gamma$ as:
\begin{enumerate}
\item $x \in Terms^s(Sig, \Gamma)$, if $x : s \in \Gamma$, 
\item $f(t_1, \ldots, t_k)  \in Terms^s(Sig, \Gamma)$, if $f \taking s_1 \times \ldots \times s_k \to s$ and $t_i \in Terms^{s_i}(Sig, \Gamma)$ for $i = 1, \ldots , k$. When $k=0$, we may write $f$ for $f()$.  When $k=1$, we may write $t_1.f$ instead of $f(t_1)$.  When $k=2$, we may write $t_1 \ f \ t_2$ instead of $f(t_1,t_2)$.
\end{enumerate}
We refer to $Terms^s(Sig, \emptyset)$ as the set of {\it ground} terms of sort $s$.  We will write $Terms(Sig, \Gamma)$ for the set of all terms in context $\Gamma$, i.e., $\bigcup_s Terms^s(Sig, \Gamma)$. Substitution of a term $t$ for a variable $v$ in a term $e$ is written as $e[v \mapsto t]$ and is recursively defined as usual:
$$
v [v \mapsto t] = t \ \ \ \ \ 
v' [v \mapsto t] = v' \ (v \neq v') \ \ \ \ 
f(t_1, \ldots, t_n)[v \mapsto t] = f(t_1[v \mapsto t], \ldots, t_n[v \mapsto t])
$$
We will only make use of substitutions that are sort-preserving; i.e., to consider $e[v \mapsto t]$, we require $e \in Terms(Sig, \Gamma)$ for some $\Gamma$ such that $v : s \in \Gamma$ and $t \in Terms^s(Sig, \Gamma)$.  To indicate a simultaneous substitution for many variables on a term $e$ we will write e.g., $e[v_2 \mapsto t_2, v_1 \mapsto t_1]$.  To indicate a sequential substitution for many variables on a term $e$ we will write $e [v_1 \mapsto t_1] \circ [v_2 \mapsto t_2]$, meaning  $(e[v_1 \mapsto t_1]) [v_2 \mapsto t_2]$.

An {\it equation} over $Sig$ is a formula $\forall \Gamma. \ t_1 = t_2 : s$ with $t_1, t_2 \in Terms^s(Sig,\Gamma)$; we will omit the $: s$ when doing so will not lead to confusion.  A {\it  theory} $Th$ is a pair of a signature and a set of equations over that signature.  In this paper, we will make use of a theory we call $Type$, which is displayed in Figure~\ref{Type}.  Additional axioms, such as the associativity of $+$, can be added to $Type$, but doing so does not impact the examples in this paper.  
\begin{figure}[h]
\begin{mdframed}
$$
Sorts := \{ {\sf Nat}, \ {\sf Char}, \ {\sf String} \}
$$
$$
Symbols := \{ {\sf zero} \taking {\sf Nat}, \ {\sf succ} \taking {\sf Nat} \to {\sf Nat}, \ {\sf A} \taking {\sf Char}, {\sf B} \taking {\sf Char}, \ldots , {\sf Z} : {\sf Char},
$$
$$
  {\sf nil}\taking{\sf String}, \ {\sf cons} \taking{\sf Char} \times {\sf String} \to {\sf String}, \ {\sf length} \taking {\sf String} \to {\sf Nat}, {\sf +} : {\sf String} \times {\sf String} \to {\sf String} \}
$$
$$
Equations := \{ {\sf length}({\sf nil}) = {\sf zero}, \ \ \ \forall c\taking{\sf Char}, s\taking{\sf String}. \ {\sf length}({\sf cons}(c,s)) = {\sf succ}({\sf length}(s)) \}  
$$
\\
We will abbreviate {\sf zero} as {\sf 0} and ${\sf succ}^n({\sf zero})$ as {\sf n}.  We will write e.g. {\sf Bill} for 
$${\sf cons}({\sf B}, {\sf cons}({\sf I}, {\sf cons}({\sf L}, {\sf cons}({\sf L}, {\sf nil})))).$$

\caption{The Multi-sorted Equational Theory $Type$}
\label{Type}

\end{mdframed}
\end{figure}

 Associated with a theory $Th$ is a binary relation between (not necessarily ground) terms, called {\it provable equality}.  We write $Th \vdash \forall \Gamma. \ t = t' : s$ to indicate that the theory $Th$ proves that terms $t, t' \in Terms^s(Sig, \Gamma)$ are equal according to the usual rules of multi-sorted equational logic.  From these rules it follows that provable equality is the smallest equivalence relation on terms that is a congruence, is closed under substitution, is closed under adding variables to contexts, and contains the equations of $Th$.  Provable equality is semi-decidable in general~\cite{Bachmair:1989a}, but some theories are decidable.  Formally, $Th \vdash$ is defined by the inference rules in Figure~\ref{eqlogic}.
 
 \begin{figure}[h]
 \begin{mdframed}
\setgroup{g1}
\irule{}{t \in Terms^s(Sig,\Gamma)}{Th \vdash \forall \Gamma. \ t = t : s}
\irule{}{Th \vdash \forall \Gamma. \ t = t' : s}{Th \vdash \forall \Gamma. \ t' = t : s}
\irule{}{Th \vdash \forall \Gamma. \ t = t' : s \\ Th \vdash \forall \Gamma. \ t' = t'' : s}{Th \vdash \forall \Gamma. \ t = t'' : s}
\irule{}{Th \vdash \forall \Gamma. \ t = t'  : s\\ v \notin \Gamma}{Th \vdash \forall \Gamma, v:s'. \ t = t' : s}
\irule{}{Th \vdash \forall \Gamma, v:s. \ t = t' : s' \\ Th \vdash \forall \Gamma. \ e = e' : s}{Th \vdash \forall \Gamma. \ t[v \mapsto e] = t'[v \mapsto e'] : s'}
\showrules{g1} 
\caption{Inference Rules for Multi-sorted Equational Logic}
\label{eqlogic}
\end{mdframed}
\end{figure}
%
%
%

A {\it morphism of signatures} $F \taking Sig_1 \to Sig_2$ consists of:
\begin{itemize}
\item a function $F$ from sorts in $Sig_1$ to sorts in $Sig_2$, and
\item a function $F$ from function symbols $f \taking s_1 \times \ldots \times s_n \to s$ in $Sig_1$ to terms in 
$$Terms^{F(s)}(Sig_2, \{ v_1 \taking F(s_1), \ldots, v_n \taking F(s_n) \} ).$$
To clearly indicate the context $\{ v_1, \ldots, v_n \} $, the function $F(f)$ may be written in ``$\lambda$ notation'', i.e. as $F(f) = \lambda v_1, \ldots, v_n . g(v_1, \ldots, v_n)$ for some term $g$, where the $\lambda$ is omitted if $n=0$.
\end{itemize}

For example, let $Sig_1$ consist of two sorts, $a, b$, and one function symbol, $f \taking a \to b$, and let $Sig_2$ consist of one sort, $c$, and one function symbol, $g\taking c \to c$.  There are countably infinitely many morphisms $F \taking Sig_1 \to Sig_2$, one of which is defined as $F(a) := c$, $F(b) := c$, and $F(f) := \lambda v \taking c . \ g(g(v))$.  In the literature on algebraic specification, our definition of signature morphism is called a ``derived signature morphism''~\cite{Mossakowski2015}.

The function $F$ taking function symbols to terms can be extended to take terms to terms:
$$
F(v) = v \ \ \ \ \ F(f(t_1, \ldots, t_n)) = F(f)[v_1 \mapsto F(t_1), \ldots, v_n \mapsto F(t_n)]
$$
As before, when we are defining the action of a specific $F$ on a specific $f \taking s_1 \times \ldots \times s_n \to s$, to make clear the variables we are using, we may write $F(f) := \lambda v_1 , \ldots, v_n. \ \varphi$, where $\varphi$ may contain $v_1, \ldots, v_n$.   A {\it morphism of theories} $F : Th_1 \to Th_2$ is a morphism of signatures that preserves provable equality of terms:
$$
Th_1 \vdash \forall v_1 : s_1, \ldots v_n : s_n. \ t_1 = t_2 : s \ \ \ \Rightarrow \ \ \ Th_2 \vdash \forall v_1 : F(s_1), \ldots, v_n : F(s_n) . \ F(t_1) = F(t_2) : F(s)
$$
In the theory $Type$ (Figure~\ref{Type}), any permutation of {\sf A, B, \ldots Z} induces a morphism $Type \to Type$, for example.  Although morphisms of signatures are commonly used in the categorical approach to logic~\cite{Adamek.Rosicky.Vitale:2011a}, such morphisms do not appear to be as commonly used in the traditional set-theoretic approach to logic.   Checking that a morphism of signatures is a morphism of theories reduces to checking provable equality of terms and hence is semi-decidable.

{\bf Remark.} Multi-sorted equational logic differs from single-sorted logic by allowing empty sorts (sorts that have no ground terms).  Empty sorts are required by our formalism; without them, we could not express empty entities.  For the theoretical development, this difference between multi-sorted and single-sorted logic can be safely ignored.  But the fact that many algorithms are based on single-sorted logic means that care is required when implementing our formalism.  For example, certain theorem proving methods based on Knuth-Bendix completion~\cite{Knuth:1970a} require a ground term of every sort.  

{\bf Categorical Remark.} From a theory $Th$ we form a cartesian multi-category~\cite{BW} $ \llparenthesis Th \rrparenthesis$ as follows.  The objects of $\llparenthesis Th \rrparenthesis$ are the sorts of $Th$.  The elements of the hom-set $s_1, \ldots, s_k \to s$ of $\llparenthesis Th \rrparenthesis$ are equivalence classes of terms of sort $s$ in context $\{ v_1\taking s_1, \ldots, v_k \taking s_k \}$, modulo the provable equality relation $Th \vdash$.  Composition is defined by substitution.  A morphism of theories $F : Th_1 \to Th_2$ denotes a functor $\llparenthesis F \rrparenthesis : \llparenthesis Th_1 \rrparenthesis \to \llparenthesis Th_2 \rrparenthesis$.  Although cartesian multi-categories are the most direct categorical semantics for theories, in many cases it is technically more convenient to work with product categories instead.  Every cartesian multi-category generates a product category, and  we often conflate the multi-category just described with the product category it generates, as is usually done in the categorical algebraic theories literature.  For details, see Schultz {\it et al.}~\cite{patrick}.  

\subsection{Semantics}
 \label{nf}
An {\it algebra} $A$ over a signature $Sig$ consists of:
\begin{itemize}
\item a set of {\it carriers} $A(s)$ for each sort $s$, and
\item a function $A(f) : A(s_1) \times \ldots \times A(s_k) \to A(s)$ for each symbol $f : s_1 \times \ldots \times s_k \to s$.
\end{itemize}
Let $\Gamma := \{ v_1 : s_1 , \ldots , v_n : s_n \}$ be a context.  An $A$-{\it environment} $\eta$ for $\Gamma$ associates each $v_i$ with an element of $A(s_i)$.  The meaning of a term in $Terms(Sig, \Gamma)$ relative to $A$-environment $\eta$ for $\Gamma$ is recursively defined as:
$$
A \llbracket v\rrbracket \eta = \eta(v) \ \ \ \ \ \ \ A \llbracket f(t_1, \ldots, t_n) \rrbracket \eta = A(f)(A\llbracket t_1 \rrbracket \eta , \ldots, A \llbracket t_i \rrbracket \eta)
$$
An algebra $A$ over a signature $Sig$ is a {\it model} of a theory $Th$ on $Sig$ when $Th \vdash \forall \Gamma. \ t = t' : s$ implies $A\llbracket t \rrbracket\eta = A\llbracket t' \rrbracket\eta$ for all terms $t, t' \in Terms^s(Sig, \Gamma)$ and $A$-environments $\eta$ for $\Gamma$.  Deduction in multi-sorted equational logic is sound and complete: two terms $t, t'$ are provably equal in a theory $Th$ if and only if $t$ and $t'$ denote the same element in every model of $Th$.  One model of the theory $Type$ (Figure~\ref{Type}) has carriers consisting of the natural numbers, the 26 character English alphabet, and all strings over the English alphabet.  Another model of $Type$ uses natural numbers modulo four as the carrier for ${\sf Nat}$.  
 
From a signature $Sig$ we form its {\it term algebra} $\llbracket Sig \rrbracket$, a process called {\it saturation}, as follows.  The carrier set $\llbracket Sig \rrbracket(s)$ is defined as the set of ground terms of sort $s$.  The function $\llbracket Sig \rrbracket(f)$ for $f : s_1 \times \ldots \times s_k \to s$ is defined as the function $t_1 , \ldots t_n \mapsto f(t_1, \ldots, t_n)$.  From a theory $Th$ on $Sig$ we define its {\it term model} $\llbracket Th \rrbracket$ to be the quotient of $\llbracket Sig \rrbracket$ by the equivalence relation $Th \vdash$.  In other words, the carrier set $\llbracket Th \rrbracket(s)$ is defined as the set of equivalence classes of ground terms of sort $s$ that are provably equal under $Th$.  The function $\llbracket Th \rrbracket(f)$ is $\llbracket Sig \rrbracket (f)$ lifted to operate on equivalence classes of terms.  To represent $\llbracket Th \rrbracket$ on a computer, or to write down $\llbracket Th \rrbracket$ succinctly, we must choose a {\it representative} for each equivalence class of terms; this detail can be safely ignored for now, but we will return to it in the implementation of our formalism (Section~\ref{impl}).  When we must choose representatives for $\llbracket Th \rrbracket$, we will write $nf_{Th}(e)$ to indicate the unique $e' \in \llbracket Th \rrbracket$ such that $Th \vdash e = e'$ (i.e., the {\it normal form} for $e$ in $Th$).  For example, the term model of the algebraic theory $Type$ (Figure~\ref{Type}) is shown in Figure~\ref{termmodel}.

\begin{figure}[h]
\begin{mdframed}
$$\llbracket Type \rrbracket({\sf Nat}) = \{{\sf 0}, {\sf 1}, {\sf 2}, \ldots \} $$
$$\llbracket Type \rrbracket({\sf Char}) = \{{\sf A}, {\sf B}, {\sf C}, \ldots \}$$
$$\llbracket Type \rrbracket({\sf String}) = \{{\sf nil}, {\sf A}, {\sf B}, \ldots, {\sf AA}, {\sf AB}, \ldots \}$$
$$\llbracket Type \rrbracket({\sf succ}) = \{({\sf 0}, {\sf 1}), ({\sf 1}, {\sf 2}), \ldots \}$$
$$\ldots$$
\caption{The Term Model $\llbracket Type \rrbracket$ of Theory $Type$ (Figure~\ref{Type})}
\label{termmodel}
\end{mdframed}
\end{figure}
A {\it morphism of algebras} $h : A \to B$ on a signature $Sig$ is a family of functions $h(s) : A(s) \to B(s)$ indexed by sorts $s$ such that:
$$
h(s)(A(f)(a_1, \ldots, a_n)) = B(f)(h(s_1)(a_1), \ldots, h(s_n)(a_n)) 
$$
for every symbol $f : s_1 \times \ldots \times s_n \to s$ and $a_i \in A(s_i)$.  We may abbreviate $h(s)(a)$ as $h(a)$ when $s$ can be inferred.  The term algebras for a signature $Sig$ are initial among all $Sig$-algebras: there is a unique morphism from the term algebra to any other $Sig$-algebra.  Similarly, the term models are initial among all models.  It is because of initiality that in many applications of equational logic to functional programming, such as algebraic datatypes~\cite{Mitchell:1996a}, the intended meaning of a theory is its term model.

 {\bf Categorical Remark.} Models of a theory $Th$ correspond to functors $Th \to {\bf Set}$, and the term model construction yields an initial such model.  That is, an algebraic theory $Th$ denotes a cartesian multi-category, $\llparenthesis Th \rrparenthesis$, and the term model construction yields a functor $\llparenthesis Th \rrparenthesis \to {\bf Set}$.  At the risk of confusion, we will write also write $\llparenthesis Th \rrparenthesis$ for the functor $\llparenthesis Th \rrparenthesis \to {\bf Set}$; hence, we have $\llparenthesis Th \rrparenthesis : \llparenthesis Th \rrparenthesis \to {\bf Set}$.  Morphisms between models correspond to natural transformations.  A morphism of theories $F : Th_1 \to Th_2$ induces a functor, $\llparenthesis F \rrparenthesis : \llparenthesis Th_1 \rrparenthesis \to \llparenthesis Th_2 \rrparenthesis$ between the cartesian multi-categories $\llparenthesis Th_1 \rrparenthesis$ and $\llparenthesis Th_2 \rrparenthesis$, as well as a natural transformation between the set-valued functors $\llparenthesis Th_1 \rrparenthesis$ and $\llparenthesis Th_2 \rrparenthesis$.
\section{An Equational Formalism for Functorial Data Migration}
\label{formal} 
In this section we describe how to use multi-sorted equational theories to define schemas and instances, and how to migrate data from one schema to another.  To summarize, we proceed as follows; each of these steps is described in detail in this section.  First, we fix an arbitrary multi-sorted equational theory $Ty$ to serve as an ambient {\it type-side} or ``background theory'' against which we will develop our formalism.  We say that the sorts in $Ty$ are {\it types}.  For example, we may define $Ty$ to contain a sort ${\sf Nat}$; function symbols $0,1,+,\times$; and equations such as $0+x=x$.  A {\it schema} $S$ is an equational theory that extends $Ty$ with new sorts (which we call {\it entities}), for example ${\sf Person}$; unary function symbols between entities (which we call {\it foreign keys}) and from entities to types (which we call {\it attributes}), for example, ${\sf father} : {\sf Person} \to {\sf Person}$ and ${\sf age} : {\sf Person} \to {\sf Nat}$; and additional equations.  An instance $I$ on $S$ is an equational theory that extends $S$ with 0-ary constant symbols (which we call {\it generators}), such as ${\sf Bill}$ and ${\sf Bob}$; as well as additional ground equations, such as ${\sf father}({\sf Bill})={\sf Bob}$.  The intended meaning of $I$ is its term model (i.e., the canonical model built from $I$-equivalence classes of terms).  Morphisms of schemas and instances are defined as theory morphisms; i.e., mappings of sorts to sorts and function symbols to (open) terms that preserve entailment: $h : C \to D$ exactly when $C \vdash p = q$ implies $D \vdash h(p) = h(q)$.  Our ``uber-flower'' query language is based on a generalization of for-let-where-return (flwr) notation, and generalizes the idea from relational database theory that conjunctive queries can be represented as ``frozen'' instances~\cite{DBLP:books/aw/AbiteboulHV95}.

\subsection{Type-sides, Schemas, Instances, Mappings, and Transforms}
\label{trans}

Our formalism begins by fixing a specific multi-sorted equational theory $Ty$ which will be called the {\it type-side} of the formalism.  The sorts in $Ty$ are called {\it types}.  The type-side is meant to represent the computational context within which our formalism will be deployed.  For example, a type-side for SQL would contain sorts such as VARCHAR and INTEGER and functions such as LENGTH : VARCHAR $\to$ INTEGER, as well as any user-defined scalar functions we wish to consider (e.g., squaring a number); a type-side for SK combinator calculus would contain a sort $o$, constants $S, K : o$, a function symbol $\cdot : o \to o$, and equations $K\cdot x \cdot y = x$ and $S \cdot x \cdot y \cdot z = x \cdot z \cdot (y \cdot z)$.  From a database point of view, choosing a particular multi-sorted equational theory $Ty$ can be thought of as choosing the set of built-in and user-defined types and functions that can appear in schemas and queries.

Simply-typed, first-order functional programs can be expressed using multi-sorted equational theories, and using a functional program as a type-side is a best-case scenario for the automated reasoning required to implement our formalism; see section~\ref{impl} for details.  We will abbreviate ``multi-sorted equational theory'' as ``theory'' in this section.

A {\it schema} $S$ on type-side $Ty$ is a theory extending $Ty$.  If $s$ is a sort in $S$ but not in $Ty$, we say that $s$ is an {\it entity}; otherwise, that $s$ is a {\it type}.  (Note that although we say ``entity'', the synonyms ``entity type'' and ``entity set'' are also common in database literature.) $S$ must meet the following conditions:
\begin{enumerate}
\item If $\forall \Gamma. \ t_1 = t_2$ is in $S$ but not in $Ty$, then $\Gamma = \{ v : s \}$ where $s$ is an entity.
\item If $f \taking s_1 \times \ldots \times s_n \to s$ is in $S$ but not $Ty$, then $n = 1$ and $s_1$ is an entity.  If $s$ is an entity we say that $f$ is a {\it foreign key}; otherwise, we say that $f$ is an {\it attribute}.
\end{enumerate} 
In other words, every equation in a schema will have one of two forms: 
$\forall v : t. \ v.p = v.p' : t'$, where $t$ and $t'$ are entities and $p$ and $p'$ are sequences of foreign keys; or some combination of type-side functions applied to attributes, for example $\forall v : t. \ v.p_1.att_1 + v.p_2.att_2 = v.att : t' $ where $t$ is an entity and $t'$ is a type.  Due to these restrictions, $S$ admits three sub-theories: the {\it type-side} of $S$, namely, $Ty$; the {\it entity-side} of $S$, namely, the restriction of $S$ to entities (written $S_E$); and the {\it attribute-side} of $S$, namely, the restriction of $S$ to attributes (written $S_A$).  We can also consider the entities and attributes together ($S_{EA}$), and the attributes and type-side together ($S_{AT}$).  A morphism of schemas, or {\it schema mapping}, $S_1 \to  S_2$ on type-side $Ty$ is a morphism of theories $S_1 \to S_2$ that is the identity on $Ty$.  An example schema $Emp$ on type-side $Type$ (Figure~\ref{Type}) is shown in Figure~\ref{Emp}.  We may draw the entity and attribute part of the schema in a graphical notation, with every sort represented as a dot, and the foreign keys and attributes represented as edges.  

\begin{figure}[h]
\begin{mdframed}
\vspace{-.05in}
$$
Sorts := \{ {\sf Emp}, \ {\sf Dept} \}
$$
$$
Symbols := \{ {\sf mgr} \taking {\sf Emp} \to {\sf Emp}, \ {\sf wrk} \taking {\sf Emp} \to {\sf Dept}, \ {\sf secr} \taking {\sf Dept} \to {\sf Emp}  \},
$$
$$
{\sf dname} \taking {\sf Dept} \to {\sf String}, \ {\sf ename} \taking {\sf Emp} \to {\sf String}
$$
$$
Equations := \{ \forall v. \ v.{\sf mgr};{\sf wrk} = v;{\sf wrk}, \ \ \ 
\forall v. \ v = v;{\sf secr};{\sf wrk}, \ \ \  \forall v. \ v;{\sf mgr};{\sf mgr} = v;{\sf mgr}
 \}  
$$ 

$$
\xymatrix@=9pt{&\LTO{{\sf Emp}}\ar@<.5ex>[rrrrr]^{\sf wrk}\ar@(l,u)[]+<0pt,13pt>^{\sf mgr}\ar[dddr]^{\sf ename}&&&&&\LTO{{\sf Dept}}\ar@<.5ex>[lllll]^{\sf secr}\ar[dddllll]^{\sf dname}\\\\\\&&\DTO{\sf String}&~&~&~&}
$$
 
\caption{Schema $Emp$, on Type-side $Type$ (Figure~\ref{Type})}
\label{Emp}
\end{mdframed}
\end{figure}

In schema $Emp$ (Figure~\ref{Emp}), {\sf Emp} and {\sf Dept} are sorts (entities) of employees and departments, respectively; {\sf mgr} takes an employee to their manager; {\sf secr} takes a department to its secretary; and {\sf wrk} takes an employee to the department they work in.  The equations are data integrity constraints saying that managers work in the same department as their employees, that secretaries work for the department they are the secretary for, and that the management hierarchy is two levels deep (this constraint ensures that $\llbracket Emp_E \rrbracket$ is finite, a condition useful for our examples but not required by our implementation; see section~\ref{impl}). 

The first restriction on schemas (bullet 1 in the above list) rules out products of entities (for example, using a symbol ${\sf CommonBoss} : {\sf Emp} \times {\sf Emp} \to {\sf Emp}$), and the second restriction on schemas  (bullet 2) rules out the use of types as domains (for example, using a symbol ${\sf EmpNumber} : {\sf Nat} \to {\sf Emp}$).  These restrictions are necessary to guarantee the existence of a right adjoint $\Pi$ to $\Delta$, which we use to model product and filter operations.

An {\it instance} $I$ on schema $S$ is a theory extending $S$, meeting the following conditions:
\begin{enumerate}
\item If $s$ is a sort in $I$, then $s$ is a sort in $S$.
\item If $\forall \Gamma. \ t_1 = t_2$ is in $I$ but not $S$, then $\Gamma = \emptyset$.
\item If $f : s_1 \times \ldots \times s_n \to s$ is in $I$ but not $S$, then $n = 0$.  We say $f$ is a {\it generator} of sort $s$.  
\end{enumerate} 

That is, an instance only adds $0$-ary symbols and ground equations to its schema.  Mirroring a similar practice in database theory, we use the phrase {\it skolem term} to refer to a term in an instance that is not provably equal to a term entirely from the type-side, and whose sort is a type.  Although skolem terms are very natural in database theory, skolem terms wreak havoc in the theory of algebraic datatypes, where their existence typically implies an error in a datatype specification that causes computation to get ``stuck''~\cite{Mitchell:1996a}.  

Similarly to how schemas admit sub-theories for entity, attribute, and type-sides, an instance $I$ contains sub-theories for entities ($I_E$), attributes ($I_A$), and types ($I_T$).  Note that $I_T$ may not be the ambient type-side $Ty$, because $I$ can declare new constant symbols whose sorts are types (so-called skolem variables), as well as additional equations; for example, ${\sf infinity} : {\sf Nat}$ and ${\sf succ}({\sf infinity})={\sf infinity}$.  A morphism of instances, or {\it transform}, $h : I_1 \to I_2$ is a morphism of theories $I_1 \to I_2$ that is the identity on $S$, and the requirement of identity on $S$ rules out the ``attribute problem'' from Figure~\ref{attprob}.   

The intended meaning of an instance $I$ is its term model, $\llbracket I \rrbracket$.  In practice, the term model $\llbracket I \rrbracket$ will often have an infinite type-side, but $\llbracket I_{EA} \rrbracket$ will be finite.  Therefore, our implementation computes $\llbracket I_{EA} \rrbracket$, as well as an instance $talg(I)$ called the {\it type-algebra}\footnote{Technically, it is $\llbracket talg(I) \rrbracket$ that is a $Ty$-algebra, and $talg(I)$ presents this algebra.  But we will almost never be interested in $\llbracket talg(I) \rrbracket$, so to save space we will refer to the equational theory $talg(I)$ as $I$'s type-algebra.} for $I$.  The type-algebra $talg(I)$ is an instance on the empty schema over $I_T$.  For every attribute $att : s \to t$ in $S$, and every term $e \in \llbracket I_E \rrbracket(s)$, the type-algebra contains a generator $e;att : t$.  We call these generators {\it observables} because they correspond to type-valued observations one can make about an entity.  Observables have the form $e;fk_1; \ldots ; fk_n. att$; i.e., have a 0-ary constant symbol as a head, followed by a possibly empty list of foreign keys, followed by an attribute.  We define the function $trans : Terms^t(I, \emptyset) \to talg(I)$, for every type (non-entity) sort $t$, as:
$$
trans(e;fk_1; \ldots ; fk_n; att) := nf_I(e;fk_1; \ldots ; fk_n);att \ \textnormal{for observables} 
$$
$$
trans(f(e_1, \ldots, e_n)) := f(trans(e_1), \ldots, trans(e_n)) \ \textnormal{otherwise}
$$
By $nf_I(x)$ we mean the normal form for $x$ in $\llbracket I \rrbracket$; see section~\ref{nf}.  The equations for $talg(I)$ are the images under $trans$ of the (necessarily ground) equations of $I$ but not $S$ and all substitution instances of the equations at types in $S$ but not $Ty$. Note that  $talg(I)$ does not present $\llbracket I_T \rrbracket$ (the restriction of $I$ to types), rather, $talg(I)$ presents $\llbracket I \rrbracket_T$ (the skolem terms of $\llbracket I \rrbracket$ and their relationships).  

We visually present term models using a set of tables, with one table per entity, with an ID column corresponding to the carrier set.  Sometimes, we will present the type-algebra as well.  An instance on the $Emp$ schema, and its denotation, are shown in Figure~$\ref{Inst}$.


\begin{figure}[t]
\begin{mdframed}
$$
Symbols := \{ {\sf a}, {\sf b}, {\sf c} \taking {\sf Emp}, \ {\sf m}, {\sf s}  \taking {\sf Dept} \} 
$$
$$
Eqs := \{ 
 {\sf a};{\sf ename} = {\sf Al}, \ \
 {\sf c};{\sf ename} = {\sf Carl}, \ \
 {\sf m};{\sf dname} = {\sf Math}, \ \
 $$
 $$
\ \ \ \ \  \ \ {\sf a};{\sf wrk} = {\sf m}, \ \ {\sf b};{\sf wrk} = {\sf m}, \ \ {\sf s};{\sf secr} = {\sf c}, \ \ {\sf m};{\sf secr} = {\sf b}  \}  
$$  
$$
\begin{tabular}{|c|c|c|}  
\multicolumn{3}{c}{{\sf Dept}}  \vspace{.1in} \\\hline  
\hspace{.1in} {\sf  ID} \hspace{.1in} & \hspace{.1in}{\sf dname} \hspace{.1in}& \hspace{.1in} {\sf secr} \hspace{.1in} \\ \hhline{|=|=|=|}
{\sf m} & {\sf Math} & {\sf b}  \\ \hline
{\sf s} & {\sf s;dname} & {\sf c} \\ \hline
\end{tabular}
 \ \ \ \ \ \ \ \ \ \ \ \ \ 
\begin{tabular}{|c|c|c|c|}
\multicolumn{4}{c}{{\sf Emp}} \vspace{.1in} \\ \hline
\hspace{.2in} {\sf  ID} \hspace{.2in} & \hspace{.2in} {\sf ename} \hspace{.2in} & \hspace{.2in} {\sf mgr} \hspace{.2in} & \hspace{.02in} {\sf wrk} \hspace{.02in} \\ \hhline{|=|=|=|=|}
{\sf a} & {\sf Al} & {\sf a};{\sf mgr} & {\sf m}  \\ \hline
{\sf b} & {\sf b.ename} & {\sf b;mgr} & {\sf m} \\  \hline
{\sf c} & {\sf Carl} &{\sf c;mgr}  & {\sf s} \\ \hline
{\sf a;mgr} & {\sf a;mgr;ename} & {\sf a;mgr} & {\sf m} \\ \hline
{\sf b;mgr} & {\sf b;mgr;ename} & {\sf b;mgr} & {\sf m} \\ \hline
{\sf c;mgr} & {\sf c;mgr;ename} & {\sf c;mgr} & {\sf s} \\ \hline
\end{tabular}
$$

The type-algebra extends $Type$ with:
$$
Symbols := \{ 
 {\sf m};{\sf dname}, \ \  {\sf s;dname}, \ \  {\sf a};{\sf ename}, \ \ {\sf b};{\sf ename}, \ \ {\sf c};{\sf ename},
 $$
 $$
\ \ \ \ \ \ \ \ \ \  {\sf a};{\sf mgr};{\sf ename}, \ \  {\sf b};{\sf mgr};{\sf ename}, \ \  {\sf c};{\sf mgr}.{\sf ename} ; {\sf String}
 \} 
$$
$$
Eqs := \{ 
 {\sf a};{\sf ename} = {\sf Al}, \ \
 {\sf c};{\sf ename} = {\sf Carl}, \ \
 {\sf m};{\sf dname} = {\sf Math} \}
 $$
\caption{Instance $Inst$ on Schema $Emp$ (Figure~\ref{Emp})}
\label{Inst}
\end{mdframed}
\end{figure}


In many cases we would like for an instance $I$ to be a {\it conservative extension} of its schema $S$, meaning that for all terms $t, t' \in Terms^s(S, \Gamma)$, $I \vdash \forall \Gamma. \ t = t' : s$ if and only if $S \vdash \forall \Gamma . \ t = t' : s$.  (Similarly, we may also want schemas to conservatively extend their type-sides.)  For example, $Emp \nvdash {\sf Al} = {\sf Carl} : {\sf String}$, but there is an $Emp$-instance $I$ for which $I \vdash {\sf Al} = {\sf Carl} : {\sf String}$.   In the context of ``deductive databases''~\cite{DBLP:books/aw/AbiteboulHV95} (databases that are represented intensionally, as theories, rather than extensionally, as tables) such as our formalism, non-conservativity is usually regarded as non-desirable~\cite{Ghilardi06didi}, although nothing in our formalism requires conservativity.  Checking for conservativity is decidable for the description logic underlying OWL~\cite{Ghilardi06didi}, but not decidable for multi-sorted equational logic (and hence our formalism), and not decidable for the formalism of embedded dependencies~\cite{FKMP05} that underlies much work on relational data integration (the chase fails when conservativity may be violated).  In section~\ref{consalg} we give a simple algorithm that soundly approximates conservativity.  Note that the $\Delta$ and $\Pi$ migration functors preserve the conservative extension property, but $\Sigma$ does not; hence, one may want to be careful when using $\Sigma$.  (More pedantically, $\Delta$ and $\Pi$ preserve type-algebras, but $\Sigma$ does not.)

 {\bf Remark.} There is a precise sense in which our definition of transform corresponds to the definition of database homomorphism in relational database theory.  Recall~\cite{DBLP:books/aw/AbiteboulHV95} that in database theory a schema is a triple ($dom$, $null$, $R$), where $dom$ is a set (called the {\it domain}), $null$ is a set disjoint from $dom$ (called the {\it labeled nulls}), and $R$ is a set of relation names and arities; an instance $I$ is a set of relations over $dom \cup null$, indexed by $R$, of appropriate arity; and a  homomorphism $h : I_1 \to I_2$ is a function $h : dom \cup null \to dom \cup null$ that is constant on $dom$ and such that $(c_1, \ldots, c_n) \in I_1(R)$ implies $(h(c_1), \ldots, h(c_n)) \in I_2(R)$ for every $R$ of arity $n$.  If we interpret a term model $\llbracket I \rrbracket$ as a relational instance by considering every skolem term in $\llbracket I \rrbracket$ to be a labeled null and every non-skolem term to be a domain value, then a transform of instances in our formalism induces a homomorphism of the encoded relational instances.  In this encoding, $dom$ is playing the role of a free (equation-less), discrete (function-less) type-side.
 
{\bf Categorical Remark}.  By forgetting the entity/attribute distinction, we can consider a schema $S$ as a single algebraic theory, $\hat{S}$; the cartesian multi-category $\llparenthesis \hat{S} \rrparenthesis$ is called the {\it collage} of $S$.  A schema mapping $F : S \to T$ is then a functor between collages $\llparenthesis \hat{S} \rrparenthesis \to \llparenthesis \hat{T} \rrparenthesis$ that is the identity on $Ty$.  More pedantically, a schema $S$ is a profunctor $\llparenthesis S \rrparenthesis : \llparenthesis S_E \rrparenthesis^{op} \times \llparenthesis Ty \rrparenthesis \to {\bf Set}$ which preserves products in $\llparenthesis Ty \rrparenthesis$.  The observables from an entity $e \in S_E$ to type $ty \in Ty$ are given by $\llparenthesis S \rrparenthesis(e, ty)$.  A schema mapping $F : S \to T$ denotes a functor $\llparenthesis F_E \rrparenthesis : \llparenthesis S_E \rrparenthesis \to \llparenthesis T_E \rrparenthesis$ and a natural transformation $\llparenthesis S \rrparenthesis \Rightarrow \llparenthesis T \rrparenthesis \circ (\llparenthesis F_E \rrparenthesis^{op} \times id) $:
\vspace{-.05in}
$$
\xymatrix@=20pt{
\llparenthesis S_E \rrparenthesis^{op} \times \llparenthesis Ty \rrparenthesis \ar[d]_{\llparenthesis F_E \rrparenthesis^{op} \times id} \ar[r]^{ \ \ \ \ \ \ \llparenthesis S \rrparenthesis} & {\bf Set} \\
 \llparenthesis T_E \rrparenthesis^{op} \times \llparenthesis Ty \rrparenthesis \ar[ru]^{\Downarrow}_{\llparenthesis T \rrparenthesis} & 
 }
$$
\subsection{Functorial Data Migration}
\label{fdmsec}

We are now in a position to define the data migration functors.  We first fix a type-side (multi-sorted equational theory), $Ty$.  The following are proved in Schultz  {\it et al.}~\cite{patrick}:

\begin{itemize}

\item The schemas on $Ty$ and their mappings form a category.  

\item The instances on a schema $S$ and their transforms form a category, $S\iinst$. 

\item The models of $S$ and their homomorphisms obtained by applying $\llbracket \rrbracket$ to $S\iinst$ form a category, $\llbracket S\iinst \rrbracket$, which is equivalent to, but not equal to, $S\iinst$.

\item A schema mapping $F : S \to T$ induces a unique functor $\Sigma_F : S\iinst \to T\iinst$ defined by substitution, $\Sigma_F(I) := F(I)$, with a right adjoint, $\Delta_F : T\iinst \to S\iinst$, which itself has a right adjoint, $\Pi_F : S\iinst \to T\iinst$. 

\item A schema mapping $F : S \to T$ induces a unique functor $\llbracket \Delta_F \rrbracket : \llbracket T\iinst \rrbracket \to \llbracket S\iinst \rrbracket$ defined by composition, $\llbracket \Delta_F \rrbracket(I) := I \circ F$, with a left adjoint, $\llbracket \Sigma_F  \rrbracket : \llbracket S\iinst \rrbracket \to \llbracket T\iinst \rrbracket$, and a right adjoint $\llbracket \Pi_F \rrbracket : \llbracket S\iinst \rrbracket \to \llbracket T\iinst \rrbracket$.

\end{itemize}

Although $\Sigma_F$ and $\llbracket \Delta_F \rrbracket$ are canonically defined, their adjoints are only defined up to unique isomorphism.  The canonically defined  migration functors enjoy properties that the other data migration functors do not, such as $\Sigma_F(\Sigma_G(I)) = \Sigma_{F \circ G}(I)$ and $\llbracket \Delta_F \rrbracket( \llbracket \Delta_G \rrbracket (I)) = \llbracket \Delta_{F \circ G} \rrbracket(I)$ (for the other functors, these are not equalities, but unique isomorphisms).  

It is possible to give explicit formulae to define the three data migration functors $\Delta, \Sigma, \Pi$ \cite{patrick}.  However, we have found that it is more convenient to work with two derived data migration functors, $\Delta \circ \Pi$ and $\Delta \circ \Sigma$, which we describe in the next section.  Therefore, we now simply describe examples of $\Delta,\Sigma,\Pi$ in Figures~\ref{fmd},~\ref{fkmex}, and~\ref{fkmex2}.  Because these examples display instances as tables, rather than equational theories, we are actually illustrating $\llbracket \Delta \rrbracket, \llbracket \Sigma \rrbracket, \llbracket \Pi \rrbracket$.
Figures~\ref{fmd} and~\ref{fkmex} shows a schema mapping $F$ which takes two distinct source entities, {\sf N1} and {\sf N2}, to the target entity {\sf N}.  The $\llbracket \Delta_F \rrbracket$ functor projects in the opposite direction of $F$: it projects columns from the single table for {\sf N} to two separate tables for {\sf N1} and {\sf N2}, similar to {\tt FROM N AS N1} and {\tt FROM N AS N2} in SQL.  When there is a foreign key between {\sf N1} and {\sf N2}, the $\llbracket \Delta_F \rrbracket$ functor populates it so that {\sf N} can be recovered by joining {\sf N1} and {\sf N2}.  The $\llbracket \Pi_F \rrbracket$ functor takes the cartesian product of {\sf N1} and {\sf N2} when there is no foreign key between {\sf N1} and {\sf N2}, and joins {\sf N1} and {\sf N2} along the foreign key when there is.  The $\llbracket \Sigma_F \rrbracket$ functor disjointly unions {\sf N1} and {\sf N2}; because {\sf N1} and {\sf N2} are not union compatible (have different columns), $\llbracket \Sigma_F \rrbracket$ creates null values.  When there is a foreign key between {\sf N1} and {\sf N2}, $\llbracket \Sigma_F \rrbracket$ merges~\cite{tuplemerge} the tuples that are related by the foreign key, resulting in a join.  As these two examples illustrate, $\Delta$ can be thought of as ``foreign-key aware'' projection, $\Pi$ can be thought of as a product followed by a filter (which can result in a join), and $\Sigma$ can be thought of as a (not necessarily union compatible) disjoint union followed by a potentially recursive merge (which can also result in a join).
   
Figure~\ref{fkmex2} shows a traditional ``data exchange setting''~\cite{FKMP05} where data on a source schema about amphibians must be migrated onto a target schema about animals, where the target schema contains a data integrity constraint enforcing that each amphibian is only counted as a single animal.  The schema mapping $F$ is an inclusion, and $\llbracket \Sigma_F \rrbracket$ has precisely the desired data exchange semantics.

\begin{figure}[t]
\begin{mdframed}
\hspace{2.1in}
\parbox{1.2in}{\fbox{\xymatrix@=8pt{
& \LTO{String}\\
\LTO{N1} \ar[ur]^{\sf name} \ar[dr]_{\sf salary} \ar[rr]^{\sf f} & & \LTO{N2}\ar[dl]^{\sf age}\\
&  \DTO{Nat}}} }
$\Too{\ \ \ \  F \ \ \ \ }$ \hspace{.05in}
\parbox{1.2in}{\fbox{\xymatrix@=8pt{
\LTO{String}\\
\LTO{N}\ar[u]^{\sf name} \ar@/^/[d]^{\sf age} \ar@/_/[d]_{\sf salary}\\
\DTO{Nat}}}}

\vspace{.05in}

\begin{footnotesize}
\hspace{.07in}
\begin{tabular}{|c|c|c|c|}
\multicolumn{4}{c}{{\sf N1}} \vspace{.01in} \\\hline
\hspace{.01in} {\sf ID} \hspace{.01in} & \hspace{.06in} {\sf name} \hspace{.06in} & \hspace{.06in} {\sf salary} \hspace{.06in} & \hspace{.1in} {\sf f} \hspace{.1in}  \\\hhline{|=|=|=|=|}
1&Alice&\$100 & 1 \\\hline 
2&Bob&\$250 & 2 \\\hline 
3&Sue&\$300 & 3 \\\hline
\end{tabular}
\begin{tabular}{|c|c|}
\multicolumn{2}{c}{{\sf N2}} \vspace{.01in} \\\hline
\hspace{.01in} {\sf ID} \hspace{.01in} &  \hspace{.1in} {\sf age}  \hspace{.1in} \\\hhline{|=|=|} 
1&20\\\hline 
2&20\\\hline 
3&30\\\hline
\end{tabular}
$\overset{\Fromm{\llbracket \Delta_F \rrbracket }}{\Too{\llbracket \Pi_F \rrbracket, \llbracket \Sigma_F \rrbracket }}$
\begin{tabular}{|c|c|c|c|}
\multicolumn{4}{c}{{\sf N}} \vspace{.01in} \\\hline
\hspace{.01in} {\sf ID} \hspace{.01in} &  \hspace{.01in} {\sf name}  \hspace{.01in} &  \hspace{.01in} {\sf salary}  \hspace{.01in} &  \hspace{.01in} {\sf age}  \hspace{.01in} \\\hhline{|=|=|=|=|} 
1&Alice&\$100&20\\\hline 
2&Bob&\$250&20\\\hline 
3&Sue&\$300&30\\\hline
\end{tabular}
\end{footnotesize}

\caption{Example Functorial Data Migrations, with Foreign Keys}
\label{fkmex}
\end{mdframed}

\begin{mdframed}
\vspace{-.15in}
\begin{centering}
\begin{align*} 
\parbox{1.5in}{\fbox{\xymatrix@=9pt{
 &  \LTO{LandAnimal}  \ar[dl]_{\sf name} & \\
  \LTO{String}  &  \LTO{Amphibian} \ar[u]_{\sf is1}  \ar[d]^{\sf is2}  &     \\
& \DTO{WaterAnimal} \ar[ul]^{\sf name'} &
}} }
\hspace{.1in}
\Too{F}
\hspace{.1in}
\parbox{2.3in}{\fbox{\xymatrix@=9pt{
 &  \LTO{LandAnimal}  \ar[dl]_{\sf name} \ar[dr]^{\sf is3} & \\
  \LTO{String}  &  \LTO{Amphibian}  \ar[u]_{\sf is1}  \ar[d]^{\sf is2}  & \LTO{Animal}    \\
& \DTO{WaterAnimal} \ar[ul]^{\sf name'} \ar[ur]_{\sf is4} & \\
& \forall a. \ a.{\sf is1}.{\sf is3} = a.{\sf is2}.{\sf is4} & \\
} 
 }}
\end{align*}
\end{centering}

\vspace{.05in}

\begin{footnotesize}
\hspace{.4in}
\begin{tabular}{|c|c|}
\multicolumn{2}{c}{{\sf LandAnimal}} \vspace{.01in} \\\hline
{\sf ID}  & {\sf Name}  \\\hhline{|=|=|}
\color{black}{1}& \color{black}{{frog}} \\\hline 
\color{black}{2}& \color{black}{ cow}  \\\hline
\end{tabular}
\hspace{.1in}
\begin{tabular}{|c|c|}
\multicolumn{2}{c}{{\sf WaterAnimal}} \vspace{.01in} \\\hline
{\sf ID}  &  {\sf Name}   \\\hhline{|=|=|}
\color{black}{1}& \color{black}{{ frog}} \\\hline 
\color{black}{3}& \color{black}{ fish}  \\\hline
\end{tabular}
\hspace{.1in}
$\Too{\llbracket \Sigma_F \rrbracket}$
\hspace{.1in}
\begin{tabular}{|c|c|c|}
\multicolumn{3}{c}{{\sf LandAnimal}} \vspace{.01in} \\\hline
 {\sf ID} \hspace{.01in} & \hspace{.06in} {\sf name} \hspace{.06in} & {\sf is3}  \\\hhline{|=|=|=|} 
\color{black}{1}& \color{black}{{ frog}} & \color{black}{1} \\\hline 
\color{black}{2}& \color{black}{ cow}  & \color{black}{2} \\\hline
\end{tabular}
\hspace{.1in}
\begin{tabular}{|c|c|c|}
\multicolumn{3}{c}{{\sf WaterAnimal}} \vspace{.01in} \\\hline
\hspace{.01in} {\sf ID} \hspace{.01in} & \hspace{.06in} {\sf name'} \hspace{.06in} &  {\sf is4}  \\\hhline{|=|=|=|}
\color{black}{1}& \color{black}{{ frog}} & \color{black}{1} \\\hline 
\color{black}{3}& \color{black}{ fish}  & \color{black}{3} \\\hline
\end{tabular} 
 \\ 
 \hspace*{.4in}
\begin{tabular}{|c|c|c|}
\multicolumn{3}{c}{{\sf Amphibian}} \vspace{.01in} \\\hline
\hspace{.01in} {\sf ID} \hspace{.01in} & \hspace{.06in} {\sf is1} \hspace{.06in}  & \hspace{.06in} {\sf is2} \hspace{.06in}\\\hhline{|=|=|=|}
\color{black}{1} & \color{black}{1} & \color{black}{1}  \\\hline
\end{tabular}
\hspace{2in}
\begin{tabular}{|c|c|c|} 
\multicolumn{3}{c}{{\sf Amphibian}} \vspace{.01in}  \\\hline
\hspace{.01in} {\sf ID} \hspace{.01in} & \hspace{.06in} {\sf is1} \hspace{.06in}  & \hspace{.06in} {\sf is2} \hspace{.06in}\\\hhline{|=|=|=|}
\color{black}{1} & \color{black}{1} & \color{black}{1}  \\\hline
\end{tabular}
  \hspace{.1in}
\begin{tabular}{|c|}
\multicolumn{1}{c}{{\sf Animal}} \vspace{.01in} \\\hline
\hspace{.01in} {\sf ID} \hspace{.01in} \\\hhline{|=|}
\color{black}{1} \\\hline
\color{black}{2} \\\hline
\color{black}{3} \\\hline
\end{tabular}

\end{footnotesize}

\vspace{-.2in}

\caption{An Example $\Sigma$ Data Migration, with Path Equalities}
\label{fkmex2}
\end{mdframed}

\end{figure}


\subsection{Uber-flower Queries}
\label{uber}

It is possible to form a query language directly from schema mappings.  This is the approach of Spivak \& Wisnesky~\cite{relfound}, where a query is defined to be a triple of schema mappings $(F, G, H)$ denoting $\llbracket \Sigma_F \rrbracket \circ \llbracket \Pi_G \rrbracket \circ \llbracket \Delta_H \rrbracket$.  Suitable conditions on $F, G, H$ guarantee closure under composition, computability using relational algebra, and other properties desirable in a query language.  In practice, however, we found this query language to be challenging to program.  Having to specify entire schema mappings is onerous; it is difficult to know how to use the data migration functors to accomplish any particular task without a thorough understanding of category theory; and as a kind of ``join all'', $\Pi$ is expensive to compute.  Hence, in Schultz  {\it et al.}~\cite{patrick} we developed a new syntax, which we call {\it uber-flower} syntax because it generalizes flwr (for-let-where-return) syntax (a.k.a select-from-where syntax, a.k.a. comprehension syntax~\cite{monad}).  We have found uber-flower syntax to be more concise, easier to program, and easier to implement than the language based on triples of schema mappings in Spivak \& Wisnesky~\cite{relfound}. 

An uber-flower $Q : S \to T$, where $S$ and $T$ are schemas on the same type-side, induces a data migration $eval(Q) : S\iinst \to T\iinst \cong \Delta_G \circ \Pi_F$ and an adjoint data migration $coeval(Q) : T\iinst \to S\iinst \cong \Delta_F \circ \Sigma_G$ for some $X, \ F : S \to X, \ G : T \to X$.  In fact, all data migrations of the form $\Delta \circ \Pi$ can be expressed as the eval of an uber-flower, and all migrations of the form  $\Delta \circ \Sigma$ can be expressed as the coeval of an uber-flower.  In sections~\ref{querytrans} and \ref{ubertrans} we describe the correspondence between uber-flowers and data migration functors in detail.   In the remainder of this section we describe uber-flowers, but defer a description of how to (co-)evaluate them to sections~\ref{evaluber} and~\ref{coevaluber}. 
A {\it tableau}~\cite{DBLP:books/aw/AbiteboulHV95} over a schema $S$ is a pair of:
\begin{itemize}
\item a context over $S$, called the {\it for} clause, $fr$ and
\item a set of quantifier-free equations between terms in $Terms(S,fr)$, called the {\it where} clause $wh$.
\end{itemize}
Associated with a tableau over $S$ is a canonical $S$-instance, the so-called ``frozen'' instance~\cite{DBLP:books/aw/AbiteboulHV95}.  In our formalism, a tableau trivially becomes an instance by the validity-preserving {\it Herbrandization} process  (the dual of the satisfiability-preserving {\it Skolemization} process) which ``freezes'' variables into fresh constant symbols.  For example, we can consider the tableau $(\{ v_1: {\sf Emp}, v_2 : {\sf Dept} \}, v_1;{\sf wrk} = v_2, v_1;{\sf ename} = {\sf Peter})$ to be an $Emp$-instance with generators ${\sf v_1}$ and ${\sf v_2}$ and equivalent equations.  In this paper, we may silently pass between a tableau and its frozen instance.

An {\it uber-flower} $S \to T$ consists of, for each entity $t \in T$:
\begin{itemize}

\item a tableau $(fr_t, wh_t)$ over $S$ and, 
\item for each attribute $att : t \to t' \in T$, a term $[att]$ in $Terms^{t'}(S, fr_t)$, called the {\it return} clause for $att$, and
\item for each foreign key $fk : t \to t' \in T$, a transform $[fk]$ from the tableau for $t'$ to the tableau for $t$ (note the reversed direction), called the {\it keys} clause for $fk$,

\item such that an equality-preservation condition holds.  We defer a description of this condition until section~\ref{vc}. 

\end{itemize}

We prefer to use $fr_t, wh_t, [att], [fk]$ notation when discussing a single uber-flower.  When we are discussing many queries $Q_1, \ldots, Q_n$, we will write $Q_k(t), Q_k(att), Q_k(fk)$ to indicate $(fr_t, wh_t)$, $[att]$, and $[fk]$, respectively, for $Q_k$.  

We usually require that the for clauses in an uber-flower only bind variables to entities, not to types (e.g., $v : {\sf Person}$ is allowed, but $v : {\sf Nat}$ is not).  While not strictly necessary, there are two reasons for preferring this restriction.  First, in practice, types will almost always be infinite, so the data migrations induced by a non-restricted uber-flower would often return infinite instances.  Second, the restriction ensures that the induced data migrations are {\it domain independent}~\cite{DBLP:books/aw/AbiteboulHV95}, allowing some evaluations of uber-flowers to be computed using relational algebra~\cite{relfound}.  Semantically, this restriction means that evaluations of uber-flowers correspond to migrations of the form $\Delta_G \circ \Pi_F$, where $F$ is surjective on attributes, a condition described in section~\ref{querytrans}.


In Figure~\ref{query} we present an uber-flower, from our $Emp$ schema to our $Emp$ schema (Figure~\ref{Emp}), which when evaluated makes each employee their own boss and appends their old boss's name to their name.  Note that {\sf Dept} is copied over unchanged; only {\sf Emp} changes.  

{\bf Remark.}  Because boolean algebra can be equationally axiomatized, evaluation of uber-flowers can express queries that might not be considered conjunctive in certain relational settings.  For example, when our type-side contains boolean algebra (Figure~\ref{bool}), evaluation of uber-flowers can express queries such as $Q(R) := \{ x \in R \ | \ P(x) \vee \neg P'(x) = \top \}$.  In addition, instances can contain skolem terms of type {\sf Bool}, implying the existence of truth values besides $\top, \bot$.    Similarly, evaluation of uber-flowers can express non-computable queries whenever a type-side contains an equational theory that is Turing-complete.  In such cases, the type-side will not have a decidable equality relation and the theorem proving methods employed by the implementation (section~\ref{impl}) will diverge or fail.

\begin{figure}[h]
\begin{mdframed}
$$
\top, \bot : {\sf Bool} \ \ \ \ \ \ \ \neg : {\sf Bool} \to {\sf Bool} \ \ \ \ \ \ \ \wedge, \vee : {\sf Bool} \times {\sf Bool} \to {\sf Bool}
$$
\begin{center}
\begin{tabular}{ccc}
$x \vee y = y \vee x$ & & $x \wedge y = y \wedge x$ \\
$x \vee (y \vee z) = (x \vee y) \vee z$ & & $x \wedge (y \wedge z) = (x \wedge y) \wedge z$  \\
$x \wedge (y \vee z) = (x \wedge y) \vee (x \wedge z)$ & $ \ \ \ \ \ \ \ \ \ \ $ & $x \vee (y \wedge z) = (x \vee y) \wedge (x \vee z)$ \\
$x \vee \neg x = \top$ & & $x \wedge \neg x = \bot$ \\
$x \vee \bot = x$ &  &  $x \wedge \top = x$ 
\end{tabular}
\end{center}\vspace{-.1in}
\caption{An Equational Axiomatization of Boolean Algebra}
\label{bool}
\end{mdframed}
\vspace{-.2in}
\end{figure}

{\bf Categorical Remark.}  An uber-flower is syntax for a structure which has several equivalent formulations. One is induced by a cospan of schemas (section~\ref{ubertrans}).  Another is a certain kind of profunctor between schemas: let $Q : S \to T$ be an uber-flower on type-side $Ty$.  Then $Q$ denotes a {\it bimodule}~\cite{patrick}, i.e., a functor $\llparenthesis Q \rrparenthesis : \llparenthesis \hat{T} \rrparenthesis^{op} \to \llparenthesis S\iinst \rrparenthesis$ where $\llparenthesis Q \rrparenthesis(t) = y(t)$ for all $t \in \llparenthesis Ty \rrparenthesis$, where $y : \llparenthesis \hat{S} \rrparenthesis^{op} \to \llparenthesis S\iinst \rrparenthesis$ is the Yoneda embedding.  See the errata for an update on this line of thinking.


\subsubsection{Verification Conditions for Uber-flower Well-formedness}
\label{vc}

Let $Q : S \to T$ be an uber-flower.  To verify that $Q$ is well-formed (pro-functorial), for every equation in $T$ we must verify an induced set of equations on $S$.  There are two kinds of equations in $T$ that we must consider, both of which require the notion of {\it path} to describe.  A {\it path} in $T$, namely $p : t_0 \to t_n := fk_1 . \ldots . fk_n$, is a sequence of foreign keys, $fk_n : t_{n-1} \to t_n$, and we write $[p]$ to indicate the substitution $[fk_1] \circ  \ldots \circ [fk_n]$ taking each $v : s \in fr_{t_n}$ to some term in $Terms^s(S, fr_{t_0})$ (when $n=0$ this is the identity substitution on $fr_{t_0}$).   One kind of equation to verify is an equality between entities of paths with a shared head variable in a singleton entity context:
$$
\forall v : t. \ v;p = v;p' : t'
$$
For each variable $u : s \in fr_{t'}$, we must verify the following:
$$
S \cup (fr_t, wh_t) \vdash u[p] = u[p'] : t'
$$
The other kind of equation to verify is an equality between types of arbitrary terms where paths share head variables in a singleton entity context, for example:
$$
\forall v : t. \ c + v;p';att' = v;p;att : t'
$$
Here we must check, for example:
$$ 
S \cup (fr_t, wh_t) \vdash c + [att'][p'] = [att][p] : t'
$$
Figure~\ref{query} shows an uber-flower and its verification conditions.  In that figure $\varphi \rightsquigarrow \psi$ means that equation $\varphi \in T$ generates verification condition $\psi$.  Rather than simply give the verification conditions, the figure illustrates how the verification conditions are obtained.  For example, the first verification condition, 
$$
\forall v. \ v;{\sf mgr;wrk} = v;{\sf wrk} : {\sf Dept}
\rightsquigarrow
Emp \cup {\sf e} : {\sf Emp} \vdash \ {\sf d}[{\sf d} \mapsto {\sf e;}{\sf wrk}][{\sf e} \mapsto {\sf e}] = 
{\sf d}[{\sf d} \mapsto {\sf e};{\sf wrk}] : {\sf Dept}
$$
means that the equation $eq := \forall v. \ v;{\sf mgr;wrk} = v;{\sf wrk} : {\sf Dept}$ induces the (tautological) verification condition 
$Emp \cup {\sf e} : {\sf Emp} \vdash  {\sf e;}{\sf wrk} = {\sf e};{\sf wrk} : {\sf Dept}$ by starting with a substitution  $eq' := {\sf d};{\sf mgr;wrk} = {\sf d};{\sf wrk} : {\sf Dept}$ of $eq$, and then applying the keys clauses (substitutions) of the uber-flower, $[wrk] = [{\sf d} \mapsto {\sf e;}{\sf wrk}]$ and $[mgr] = [{\sf e} \mapsto {\sf e}]$, to $eq'$.  The notation $Emp \ \cup$ means that we can use the equations from the $Emp$ schema (Figure~\ref{Emp}) to prove the desired equality.

From a practical standpoint, well-formed queries are guaranteed to only materialize instances which obey their data integrity constraints, so runtime checking of data integrity constraints are not needed.  Mathematically a query must be well-formed to even be considered a query.

 \begin{figure}[h]
\begin{mdframed}
\begin{verbatim}

Dept = for d:Dept
       where
       return (dname -> d;dname)
       keys (secr -> [e -> d;sec])

Emp = for e:Emp
      where             
      return (ename -> e;ename +  e;mgr;ename) 
      keys (mgr -> [e -> e], wrk -> [d -> e;wrk]) 

\end{verbatim}
\vspace{-.2in}
$$
\xymatrix@=9pt{&\LTO{{\sf Emp}}\ar@<.5ex>[rrrrr]^{\sf wrk}\ar@(l,u)[]+<0pt,13pt>^{\sf mgr}\ar[dddr]^{\sf ename}&&&&&\LTO{{\sf Dept}}\ar@<.5ex>[lllll]^{\sf secr}\ar[dddllll]^{\sf dname}\\\\\\&&\DTO{\sf String}&~&~&~&}
$$
 \vspace{-.15in}
Verification conditions:

$$
\forall v. \ v;{\sf mgr;wrk} = v;{\sf wrk} : {\sf Dept}
\rightsquigarrow
Emp \cup {\sf e} : {\sf Emp} \vdash \ {\sf d}[{\sf d} \mapsto {\sf e;}{\sf wrk}][{\sf e} \mapsto {\sf e}] = 
{\sf d}[{\sf d} \mapsto {\sf e};{\sf wrk}] : {\sf Dept}
$$
$$
\forall v. \ v = v;{\sf secr};{\sf wrk} : {\sf Dept} \
\rightsquigarrow \
Emp \cup {\sf d}:{\sf Dept} \vdash {\sf d} = {\sf d}[{\sf d} \mapsto {\sf e};{\sf wrk}][{\sf e} \mapsto {\sf d}.{\sf secr}] : {\sf Dept}
$$
$$
\forall v. \ v;{\sf mgr.mgr} = v;{\sf mgr} : {\sf Emp} \
\rightsquigarrow \ 
Emp \cup {\sf e} : {\sf Emp} \vdash {\sf e}[{\sf e} \mapsto {\sf e}][{\sf e} \mapsto {\sf e}] = {\sf e}[{\sf e} \mapsto {\sf e}] : {\sf Emp}
$$
 
\caption{The Uber-flower $Promote : Emp \to Emp$}
\label{query}
\end{mdframed}
\end{figure}

\subsubsection{Morphisms of Uber-flowers}


Let $Q_1, Q_2 : S \to T$ be uber-flowers.  For every foreign key $fk : t \to t'$, we have transforms $Q_1(f) : fr^1_{t'} \to fr^1_t$, $Q_2(f) : fr^2_{t'} \to fr^2_t$.  A morphism $h : Q_1 \to Q_2$ is, for each entity $t \in T$, a transform of frozen instances $h(t) : fr^1_t \to fr^2_t$, such that for every foreign key $fk : t \to t' \in T$, and every $v' \in fr^1_{t'}$,
$$
S \cup wh^2_t \vdash v'[h(t')][Q_2(f)] = v'[Q_1(f)][h(t)]
$$
and for every attribute $att : t \to t' \in T$, 
$$
S \cup wh^2_t \vdash Q_2(att) = Q_1(att)[h(t)]
$$

The morphism $h$ induces a transform $ eval(h) : eval(Q_2)(I) \to eval(Q_1)(I)$ for each $S$-instance $I$ (in this way, $eval$s of uber-flowers are similar to relational conjunctive queries, for which $Q \to Q'$ implies $\forall J, Q'(J) \subseteq Q(J)$) and a transform $coeval(h) : coeval(Q_1)(J) \to coeval(Q_2)(J)$ for each $T$-instance $J$, described in sections~\ref{xxx6} and~\ref{xxx7}, respectively.

\subsubsection{Composing Uber-flowers}
\label{composing}


Let $Q_1 : A \to B$ and $Q_2 : B \to C$ be uber-flowers on the same type-side.  We define the uber-flower $Q_2 \circ Q_1 : A \to C$ as follows.  For every entity $c \in C$,
\begin{itemize}
\item We define (the frozen $A$-instance of) the for and where clause at $c$ of $Q_2 \circ Q_1$ to be the frozen $A$-instance that is obtained by applying $coeval(Q_1) : B\iinst \to A\iinst$ to the frozen $B$-instance for $c$ in $Q_2$:
$$
(Q_2 \circ Q_1)(c) := coeval(Q_1)(Q_2(c))
$$
\item Similarly, the transform associated with a foreign key $fk : c \to c' \in C$ is:
 $$
(Q_2 \circ Q_1)(fk) := coeval(Q_1)(Q_2(fk))
$$
\item To define the term associated with an attribute $att : c \to c' \in C$ we first define the instance $y(c')$ (a so-called ``representable''~\cite{BW} instance) to be the $B$-instance with a single generator $v : c'$.  We then define a transform / substitution from $y(c')$ to the frozen $B$-instance for $c$ in $Q_2$, namely, $h : y(c') \to Q_2(c) := v \mapsto Q_2(att)$.  Finally, we define:
$$
(Q_2 \circ Q_1)(att) := coeval(Q_1)(h)(v)
$$
\end{itemize}

The above definition of query composition, which mathematically correct, is algorithmically incomplete in the sense that the composed queries described above may have infinitely many variables even when there is an equivalent query with finitely many variables.  To obtain a complete algorithm, we perform the algebraic analog of ``query/view unfolding''~\cite{Garcia-Molina:2008:DSC:1450931}.  For every entity $c \in C$,
\begin{itemize}
\item We define (the frozen $A$-instance of) the for and where clause at $c$ of $Q_2 \circ Q_1$ to be the frozen $A$-instance that is obtained by ``union unfolding'' each generating variable $(v:b)$ of $Q_2(c)$ into the entire $B$-instance $Q_1(b)$, plus some extra equations $X$ that we will describe momentarily:
$$
(Q_2 \circ Q_1)(c) := X \ \cup \ \sqcup_{(v:b) \in Q_2(c)} Q_1(b) 
$$
Suppose we have $b_1.p = b_2.q : b$ in $Q_2(c)$, for some paths $p : B_1 \to b$, $q : B_2 \to b$.  Then $Q_1(p) : Q_1(b) \to Q_1(B_1)$ and $Q_1(q) : Q_1(b) \to Q_1(B_2)$.  Define $h_1 : Q_1(B_1) \to \sqcup_{(v:b) \in Q_2(c)} Q_1(b)$ as $h_1(x):=(b_1:B_1,x)$, and define $h_2 : Q_1(B_2) \to \sqcup_{(v:b) \in Q_2(c)} Q_1(b)$ as $h_2(x):=(b_2:B_2,x)$.  Then
$$\{ (Q_1(p) ; h_1)(v) = (Q_1(q) ; h_2)(v) \ | \ \forall (v : b) \in Q_1(b) \} \in X$$ 

Note that we have implicitly constructed a function $H$ taking a term of entity $b$ in $Q_2(c)$ and a $(v : b') \in Q_1(b)$ to a term in $(Q_2 \circ Q_1)(c)$.  This function $H$ will be used to define the foreign keys clause of our composed query, and will be used to define a function $G$ taking a term of type $t$ in $Q_2(c)$ to a term in $(Q_2 \circ Q_1)(c)$ as follows.  On typeside symbols, define
$
 G(f(a_1, \ldots, a_n)) := f(G(a_1), \ldots, G(a_n))
$
. On attributes $att : b \to t$, we have $Q_1(att)$ is a term in $Q_1(b)$ of the form $g.p$ for some $g : b' \in Q_1(b)$ and path $p : b' \to t$.  We define
$
G(e;att) := Q_1(att)[g \mapsto H(e, g:b')]   
$
.  Finally, we use $G$ to add equations-at-types to $X$, similar to how we used $H$ to add equations-at-entities to $X$.  Given $e = e' : t$ in $Q_2(c)$ we add $G(e) = G(e')$ to $X$.

\item We construct the transform associated with a foreign key $fk : c \to c' \in C$ as follows.  Let $(v:b) \in Q_2(c')$ and $(u:b') \in Q_1(b)$ be given.  We have $Q_2(fk) : Q_2(c') \to Q_2(c)$.  Hence $Q_2(fk)(v)$ yields a term of sort $b$ in $Q_2(c)$.  We apply $H$ as above to $Q_2(fk)(v)$ and $u:b' \in Q_1(b)$ to get the result.   Symbolically:
$$
(Q_2 \circ Q_1)(fk : c \to c')((v:b) \in Q_2(c'), (u:b') \in Q_1(b)) := H(Q_2(fk)(v), u)
$$

\item We construct the term associated with attribute $att : c \to t$ as follows.  We have $Q_2(att)$ is a term in $Q_2(c)$ and we define
$$
(Q_2 \circ Q_1)(att : c \to t) := G(Q_2(att))
$$
where $G$ is defined as above.

\end{itemize}

\begin{figure}
\begin{verbatim} 
schema S = literal : ty {
    entities s s0
    foreign_keys ss : s0 -> s    
    attributes att : s -> Integer     }

schema T = literal : ty {
    entities t1 t2 t3
    foreign_keys
        f : t1 -> t3  g : t2 -> t3  l : t1 -> t1  h : t2 -> t1
    path_equations  t1 = l    
    attributes   att1 : t1 -> Integer      att2 : t2 -> Integer }

schema U = literal : ty {
    entities  u11 u12 u21 u22 u3 u3x
    foreign_keys
        f1 : u11 -> u3       g1 : u21 -> u3
        f2 : u12 -> u3       g2 : u22 -> u3
        f1x: u11 -> u3x     h2 : u21 -> u11
        f2x: u12 -> u3x h3 : u22 -> u12
        h  : u21 -> u11.   l  : u11 -> u11
     path_equations  l.l = l            
    attributes
        att1 : u11 -> Integer       att2 : u21 -> Integer
        att2x: u12 -> Integer      att3 : u3 -> Integer     }

query qUT = literal : U -> T {
    entities
        t1 -> {from u11:u11 u12:u12
               where u11.l.l.f1 = u12.f2   neg(u11.l.l.f1.att3) = neg(u12.f2.att3)
               return att1 -> neg(u12.att2x)}
        t2 -> {from u21:u21 u22:u22 
               where u21.g1 = u22.g2  u21.h2.l.l.f1 = u22.h3.f2
               return att2 -> neg(u21.att2)}
        t3 -> {from u3:u3 u3x:u3x}

    foreign_keys
        f -> {u3 -> u11.f1  u3x -> u12.f2x}
        g -> {u3 -> u21.g1  u3x -> u21.h.f1x}
        l -> {u11 -> u11  u12 -> u12} 
        h -> {u11 -> u21.h2  u12 -> u22.h3}
}

query qTS = literal : T -> S {
    entities
        s -> {from t1:t1 t2:t2  
        where t1.f = t2.g       t1.att1=t2.att2 
        return att -> t1.l.att1}
        s0 -> { from x:t2 
        where x.h.f = x.g     x.h.att1=x.att2 }
    foreign_keys ss -> {t1 -> x.h t2 -> x }  }
    \end{verbatim}
\caption{Example of Query Composition, Input}
    \end{figure}
    
\begin{figure}
\begin{footnotesize}
\begin{verbatim} 
query qUS = literal : U -> S { 
s -> {from (t1,u12) : u12  (t1,u11) : u11   (t2,u21) : u21   (t2,u22) : u22
         where 
                  (t1,u12).att2x.neg = (t2,u21).att2.neg
                  (t1,u11).f1 = (t2,u21).g1
                  (t1,u11).l.l.f1 = (t1,u12).f2
	                  .f2x = (t2,u21).h.f1x
                  (t2,u21).h2.l.l.f1 = (t2,u22).h3.f2
                  (t2,u21).g1 = (t2,u22).g2
                  (t1,u11).l.l.f1.att3.neg = (t1,u12).f2.att3.neg
        return att -> (t1,u12).att2x.neg }

s0 -> {from (x,u21) : u21.  (x,u22) : u22
        where 
                  (x,u22).h3.att2x.neg = (x,u21).att2.neg
                  (x,u21).h2.l.l.f1 = (x,u22).h3.f2
                  (x,u21).h2.f1 = (x,u21).g1
                  (x,u21).g1 = (x,u22).g2
                  (x,u22).h3.f2x = (x,u21).h.f1x
        foreign_keys 
                  ss -> { (t1,u12) -> (x,u22).h3
                          (t1,u11) -> (x,u21).h2
                          (t2,u21) -> (x,u21)
                          (t2,u22) -> (x,u22) }
   \end{verbatim}
\end{footnotesize} \vspace{-.2in}
\caption{Example of Query Composition, Output}
\end{figure}


\subsubsection{Converting Data Migrations to Uber-flowers}
\label{querytrans}


Let $F : S \to T$ be a schema mapping on a type-side $Ty$.  In this section we define two uber-flowers, $Q_F : S \to T$ and $Q^F : T \to S$ such that:
$$
eval(Q^F) \cong \Delta_F \ \ \ \ \ coeval(Q^F) \cong \Sigma_F \ \ \ \ \ eval(Q_F) \cong \Pi_F \ \ \ \ \ coeval(Q_F) \cong \Delta_F
$$
 We now describe $Q^F$ and $Q_F$:
\begin{itemize}

\item The for clause for $Q^F : T \to S$ at entity $s \in S$ is defined to have a single variable, $v_s : F(s)$, and $Q^F$ has an empty where clause.  For each foreign key $fk : s \to s' \in S$, $F(fk)$ is ($\alpha$-equivalent to) a term in $Terms^{F(s')}(T, \{v_s:F(s)\})$ and we define $Q^F(fk) : Q^F(s') \to Q^F(s)$ to be the transform $v_{s'} \mapsto F(fk)$. For each attribute $att : s \to s' \in S$, $s'$ is a type and $F(att)$ is ($\alpha$-equivalent to) a term in $Terms^{s'}(T, \{v_s:F(s)\})$ and we define $Q^F(att)$ to be $F(att)$.   For example, the $Q^F$ that corresponds to Figure~\ref{fkmex} is:
\begin{verbatim}
N1 := for vN1 : N,
      return name -> vN1;name, salary -> vN1;salary
      keys f -> [vN2 -> vN1;f]
          
N2 := for vN2 : N,
      return age -> vN2;age          
\end{verbatim}

\item The frozen instance (for/where clause) for $Q_F : S \to T$ at entity $t \in T$ is defined to be $\Delta_F(y(t))$, where $y(t)$ is the instance with a single generator $\{ v_t : t \}$.  For each foreign key $fk : t \to t' \in T$, we define the transform $Q_F(fk) : Q_F(t') \to Q_F(t)$ to be $\Delta_F(v_{t'} \mapsto v_t.fk)$.  For each attribute $att : t \to t' \in T$, $v_t ; att \in Terms^{t'}(T, \{ v_t : t \})$ and $trans(v_t;att) \in talg(\{ v_t : t \})$.  Since $\Delta$ preserves type algebras, we have $trans(v_t;att) \in talg(\Delta_F(\{ v_t : t \}))$, and hence we can define $Q_F(att)$ to be $trans(v_t;att)$.  

If $T$ has attributes, then the $Q_F$ constructed above will contain a for clause that binds variables to types.  As discussed previously, such ``domain dependent''~\cite{DBLP:books/aw/AbiteboulHV95} queries are undesirable in practice.   To obtain an equivalent query that does not bind variables to types, it is necessary to replace each such Skolem variable with an equivalent non-Skolem expression, which exists whenever $F$ is {\it surjective on attributes}.  Formally, $F$ is surjective on attributes (a semi-decidable condition) when for every attribute $att : t \to t' \in T$, there exists an entity $s \in S$ such that $F(s) = t$ and there exists an $e \in Terms^{t'}(T, \{ v : s \})$ such that $T \vdash \forall v : F(s). \ F(e)(v) = v;att$.  For example, the $Q_F$ that corresponds to Figure~\ref{fkmex} is (note that we write `$x$' to indicate an $x \in talg(y(N))$):
\begin{verbatim}
y(N) = vN : N

N := for vN1 : N1, vN2 : N2,
       'vN.name' : String, 'vN.salary' : Int, 'vN.age' : Int
     where vN1.f = vN2, vN1.name='vN.name', 
      vN1.salary = 'vN.salary', vN2.age = 'vN.age'
     return name -> 'vN.name', salary -> 'vN.salary', age -> 'vN.age'
     
simplified N := for vN1 : N1, vN2 : N2 
     where vN1.f = vN2
     return name -> vN1;name, salary -> vN2;salary, age -> vN2;age
        
\end{verbatim}

\end{itemize}


\subsubsection{Converting Uber-flowers to Data Migrations}
\label{ubertrans}

An uber-flower $Q : S \to T$, where $S$ and $T$ are schemas on the same type-side $Ty$, induces a data migration $eval(Q) : S\iinst \to T\iinst \cong \Delta_G \circ \Pi_F$ and adjoint data migration $coeval(Q) : T\iinst \to S\iinst \cong \Delta_F \circ \Sigma_G$ for some $X, \ F : S \to X, \ G : T \to X$.  In this section, we construct $X$, $F$, and $G$.  First, we define a schema $X$ such that $S \subseteq X$ and we define $F : S \hookrightarrow X$ to be the inclusion mapping.  We start with:
$$
En(X) :=En(S) \sqcup En(T) \ \ \ \ \ Att(X) := Att(S) \ \ \ \ \ 
Fk(X) \subseteq Fk(S) \sqcup Fk(T)  
$$
Then, for each entity $t \in T$ and each $v : s$ in $fr_t$ (the frozen instance for $t$ in $Q$), we add a foreign key to $X$:
$$
 (v, s, t) : t \to s \ \in Fk(X)
$$
Let us write $\sigma_x$ for the substitution $[v_k \mapsto x.(v_k, s_k, t) , \ \forall v_k:s_k \in fr_t]$.   For each equation $e = e' \in wh_t$ we add an equation to $X$:
$$
\forall x : t. \ e\sigma_x = e'\sigma_x \ \ \in Eq(X)
$$
and for each foreign key $fk : t \to t'$ and for each $v' : s' \in fr_{t'}$, we add an equation to $X$:
\begin{equation}
\label{eq:one}
\forall x : t. \ x.fk.(v',s',t') = v'[fk]\sigma_x   \ \ \in Eq(X)
\end{equation}
This almost completes the schema $X$, but we will need to add the equations of $T$, suitably translated, to $X$.  To do so, we must first define $G : T \to X$ to be the identity on entities and foreign keys, and on attributes we define:
$$
G(att : t \to t') := \forall x:t. \ [att]\sigma_x
$$
Finally, to complete $X$ we add the images of $T$'s equations under $G$ to $X$.  

For example, the schema $X$ for the uber-flower $Promote$ (Figure~\ref{query}) is shown in Figure~\ref{promote2}.  Rather than simply give the equations of the schema $X$, the figure illustrates how the equations conditions are obtained.  For example, the first equation, 
$$
\forall x. \ x;{\sf wrk};d = d[d \mapsto e;{\sf wrk}][e \mapsto x;e] \equiv x;e;{\sf wrk}
$$
means that schema $X$ contains the equation $\forall x. \ x;{\sf wrk};d = x;e;{\sf wrk}$, which was obtained from foreign key ${\sf wrk} : {\sf Emp} \to {\sf Dept}$ and for-bound variable $d:{\sf Dept}$ by equation \ref{eq:one}.

\subsubsection{ED syntax for Uber-flowers}
\label{eduber}

Intriguingly, the intermediate schema and schema mappings that are created when translating uber-flowers into data migrations, as described in the previous section, suggest an alternative syntax for uber-flowers that resembles the syntax of second-order embedded dependencies (EDs)~\cite{Fagin:2005:CSM:1114244.1114249}.  The uber-flower $Promote$ (Figure~\ref{query}) is shown as a data migration in Figure~\ref{promote2}, and we can express the intermediate schema and mapping in Figure~\ref{promote2} using the following second-order ED:
$$
\exists e : {\sf Emp}_{src} \to {\sf Emp}_{dst}, \ \exists d : {\sf Dept}_{src} \to {\sf Dept}_{dst}. 
$$
$$
 \forall x. \ d({\sf wrk}_{src}(x)) = {\sf wrk}_{dst}(e(x)) \ \wedge \ \forall x. \ e({\sf secr}_{src}(x)) = {\sf secr}_{dst}(d(x)) \ \wedge \ \forall x. \ e({\sf mgr}_{src}(x)) = e(x)
$$
We do not understand how our formalism relates to second-order EDs, but the implementation of our formalism in the CQL tool allows users to input uber-flowers using the above second-order ED syntax.
\begin{figure}[h]
\begin{mdframed}
\begin{center}
\begin{tabular}{p{2.6in}cc}
 \vspace{-.4in}

$$\forall x. \ x;{\sf wrk};d = d[d \mapsto e;{\sf wrk}][e \mapsto x;e] \equiv x;e;{\sf wrk}$$

$$
\forall x. \ x;{\sf secr};e = e[e\mapsto d;{\sf secr}][d \mapsto x;d] \equiv x;d;{\sf secr}
$$

$$
\forall x. \ x;{\sf mgr};e = e[e \mapsto e][e \mapsto x;e] \equiv x;e
$$

$$
 \forall v. \ v;{\sf mgr};{\sf wrk} = v;{\sf wrk}
 $$
 $$
 \ldots
 $$

 & &
$$
\xymatrix@=9pt{
& \ar[ddd]^{e} \LTO{{\sf Emp}}\ar@<.5ex>[rrrrr]^{\sf wrk}\ar@(l,u)[]+<0pt,13pt>^{\sf mgr}&&&&&\LTO{{\sf Dept}}\ar@<.5ex>[lllll]^{\sf secr} \ar[ddd]^{d}\\\\\\
&\LTO{{\sf Emp}}\ar@<.5ex>[rrrrr]^{\sf wrk}\ar@(l,u)[]+<0pt,13pt>^{\sf mgr}\ar[dddr]^{\sf ename}&&&&&\LTO{{\sf Dept}}\ar@<.5ex>[lllll]^{\sf secr}\ar[dddllll]^{\sf dname}\\\\\\
&&\DTO{\sf String}&~&~&~&
}
$$
\end{tabular}
\end{center}
\caption{Uber-flower $Promote$ (Figure~\ref{query}) as a Data Migration}
\label{promote2}
\end{mdframed}

\end{figure}

\section{Implementation: the CQL tool}
\label{impl}

We have implemented our formalism in the open-source CQL tool, which can be downloaded at \url{http://categoricaldata.net}.  In this section we discuss certain implementation issues that arise in negotiating between syntax and semantics, and provide algorithms for key parts of the implementation: deciding equality in equational theories, saturating theories into term models, checking conservativity of equational theories, (co-)evaluating queries, and (co-)pivoting (converting instances into schemas).  Even though the goal of the CQL tool is merely to prove that algebraic data integration can be done, we close with a discussion about the CQL tool's performance.

\begin{figure}[h]
\begin{center}
  \includegraphics[width=5.0in]{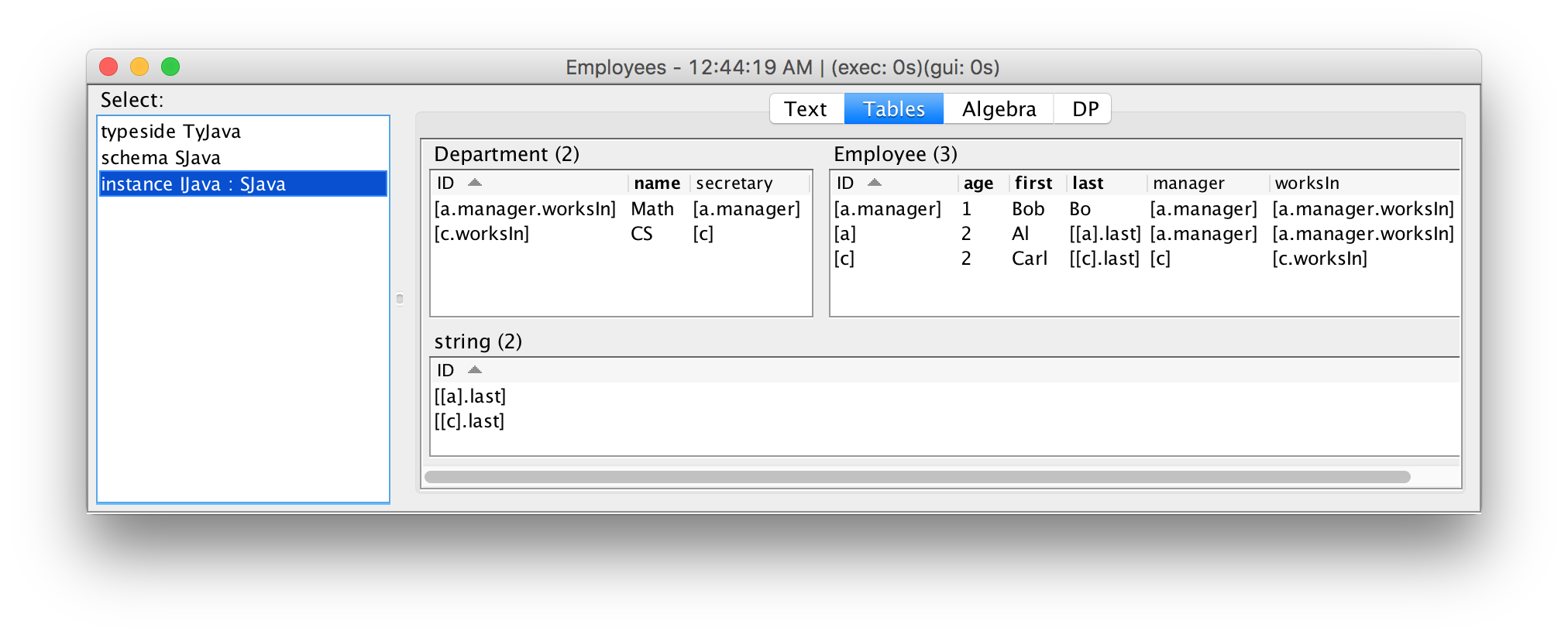}
  \vspace*{-.45in}
  \end{center}
  \caption{The CQL Tool Displaying an Instance}
  \label{fig:OPL_pic}
\end{figure}

\subsection{Deciding Equality in Equational Theories}
\label{knuth}

Many constructions involving equational theories, including uber-flower (co-)evaluation, depend on having a decision procedure for provable equality in the theory.  A decidable equational theory is said to have a decidable {\it word problem}.  The word problem is obviously semi-decidable: to prove if two terms (words) $p$ and $q$ are equal under equations $E$, we can
systematically enumerate all of the (usually infinite) consequences of $E$ until we find $p = q$.
However, if $p$ and $q$ are not equal, then this enumeration will never stop. In practice, not only
is enumeration computationally infeasible, but for uber-flower (co-)evaluation, we
require a true decision procedure: an algorithm which, when given $p$ and $q$ as input, will always
terminate with ``equal'' or ``not equal''.  Hence, we must look to efficient, but incomplete, automated theorem proving techniques to decide word problems.  

The CQL tool provides a built-in theorem prover based on Knuth-Bendix completion~\cite{Knuth:1970a}: from a set of equations $E$, it attempts to construct a system of rewrite rules (oriented equations), $R$, such
that $p$ and $q$ are equal under $E$ if and only if $p$ and $q$ rewrite to syntactically equal
terms (so-called \emph{normal forms}) under $R$.  We demonstrate this with an example.  Consider the equational theory of groups, on the left, in Figure~\ref{bendix}. Knuth-Bendix completion yields
the rewrite system on the right in Figure~\ref{bendix}. To see how these rewrite rules are used to decide the word problem, consider the two terms $(a^{-1}
\ast a) \ast (b \ast b^{-1})$ and $b \ast ((a \ast b)^{-1} \ast a)$.  Both of these terms rewrite
to $1$ under the above rewrite rules; hence, we conclude that they are provably equal. In contrast, the two
terms $1 \ast (a \ast b)$ and $b \ast (1 \ast a)$ rewrite to $a \ast b$ and $b \ast a$,
respectively, which are not syntactically the same; hence, we conclude that they are not provably equal.

\begin{figure}[h]
\begin{mdframed} 
\begin{center}
  
    \begin{tabular}{@{}c@{\hspace{3em}}c@{}c@{}}
    Equations & Rewrite Rules & \\
    $1 \ast x = x$ & $1 \ast x \rightsquigarrow x$ & $x \ast 1 \rightsquigarrow x$  \\
    $x^{-1} \ast x = 1$ & $x^{-1} \ast x \rightsquigarrow 1$ & $(x^{-1})^{-1} \rightsquigarrow x$ \\
    $(x \ast y) \ast z = x \ast (y \ast z)$ & $(x \ast y) \ast z \rightsquigarrow x \ast (y \ast z)$ &  $x \ast x^{-1} \rightsquigarrow 1$\\
    & $x^{-1} \ast (x \ast y) \rightsquigarrow y$ & $x \ast (x^{-1} \ast y) \rightsquigarrow y$ \\
    & $1^{-1} \rightsquigarrow 1$ & $(x \ast y)^{-1} \rightsquigarrow y^{-1} \ast x^{-1}$ \\

  \end{tabular}
  \caption{Knuth-Bendix Completion for Group Theory}
  \label{bendix}
\end{center}
\end{mdframed}
\end{figure}

The details of how the Knuth-Bendix algorithm works are beyond the scope of this paper.  However, we
make several remarks.  First, Knuth and Bendix's original algorithm~\cite{Knuth:1970a} can fail even
when a rewrite system to decide a word problem exists; for this reason, we use the more modern,
``unfailing'' variant of Knuth-Bendix completion~\cite{Bachmair:1989a}.  Second, first-order, simply-typed
functional programs are equational theories that are already complete in the sense of Knuth-Bendix.  Third, specialized Knuth-Bendix algorithms exist for particular kinds of theories; a particular algorithm~\cite{doi:10.1137/0214073} for theories where all function symbols are 0-ary or unary, such as for the entity and attribute parts of our schemas, works well in practice.

\subsection{Saturating Theories into Term Models}
\label{satsec}

Many constructions involving equational theories, including uber-flower (co-)evaluation, depend on having a procedure, called {\it saturation}, for constructing finite term models from theories.  This process is semi-computable: there are algorithms that will construct a finite term model if it exists, but diverge if no finite term model exists.  The CQL tool has two different methods for saturating theories: theories where all function symbols are 0-ary or unary can be saturated using an algorithm for computing Left-Kan extensions~\cite{Bush2003107}, and arbitrary theories can be saturated by using a decision procedure for the theory's word problem as follows.  Let $Th$ be an equational theory, and define the {\it size} of a term in $Th$ to be the height of the term's abstract syntax tree; for example, $\sf{max}(x\sf{;sal},x\sf{;mgr;sal})$ has size of three.  We construct $\llbracket Th \rrbracket$ in stages: first, we find all not provably equal terms of size $0$ in $Th$; call this $\llbracket Th \rrbracket^0$. Then, we add to
$\llbracket Th \rrbracket^0$ all not provably  equal terms of size $1$ that are not provably equal to a term in $\llbracket Th \rrbracket^0$; call this
$\llbracket Th \rrbracket^1$.  We iterate this procedure, obtaining a sequence $\llbracket Th \rrbracket^0, \llbracket Th \rrbracket^1, \ldots$.  If $\llbracket Th \rrbracket$ is indeed finite, then there will exist some $n$ such that $\llbracket Th \rrbracket^n = \llbracket Th \rrbracket^{n+1} =
\llbracket Th \rrbracket$ and we can stop. Otherwise, our attempt to construct $\llbracket Th \rrbracket$ will run forever: it is not
decidable whether a given theory $Th$ has a finite term model. 

Note that the model $\llbracket Th \rrbracket$ computed using the above procedure is technically not the canonical term model for the theory; rather, we have constructed a model that is isomorphic to the canonical term model by choosing representatives for equivalence classes of terms under the provable equality relation.  Depending on how we enumerate terms, we can end up with different models.  

Saturation is used for constructing tables from instances to display to the user, and for (co-)evaluating queries on instances.  In general, the type-side $Ty$ of an instance $I$ will be infinite, so we cannot saturate the equational theory of the instance directly (i.e., $\llbracket I \rrbracket$ is often infinite).  For example, if the type-side of $I$ is the free group on one generator ${\sf a}$, then $\llbracket I \rrbracket$ will contain ${\sf a}$, ${\sf a} \ast {\sf a}$, ${\sf a} \ast {\sf a} \ast {\sf a}$, and so on. Hence, as described in section~\ref{formal}, the CQL tool computes the term model for only the entity and attribute part of $I$ (namely, $\llbracket I_{EA} \rrbracket$), along with an instance (equational theory) called the type-algebra of $I$ (namely, $talg(I)$).  The pair $(\llbracket I_{EA} \rrbracket, talg(I))$ is sufficient for all of CQL's purposes.

\begin{figure}[h]
\begin{centering}$$
  \includegraphics[width=4.5in]{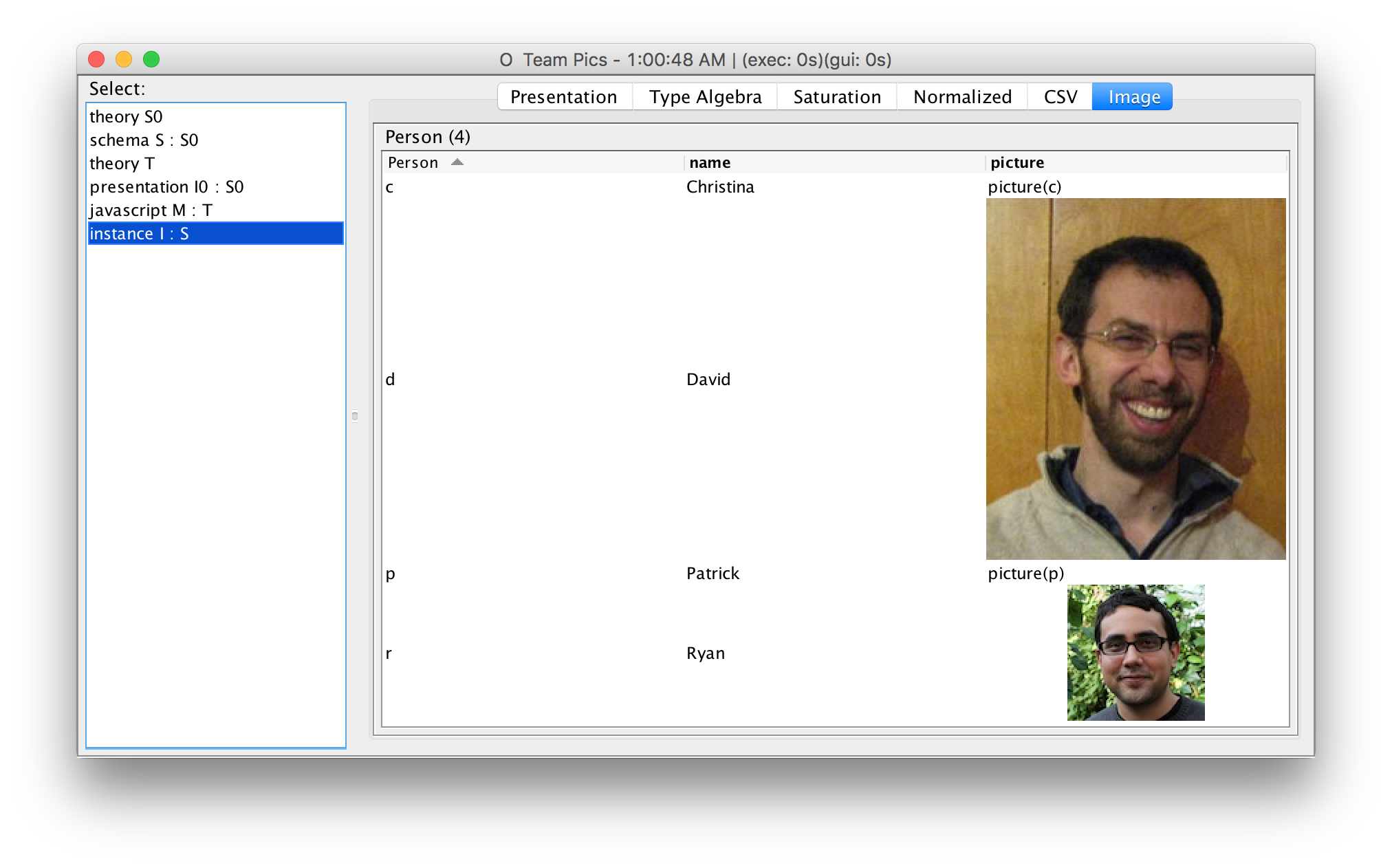}$$
  \end{centering}
  \vspace{-.2in}
  \caption{The CQL Tool Displaying an Instance with a Computational Type-side}
  \label{images}
\end{figure}

The CQL tool supports an experimental feature that we call ``computational type-sides''.  The mathematics behind this feature have not been fully worked out, but it provides a mechanism to connect CQL to other programming languages.  An $\mathcal{L}${\it -valued model} of $talg(I)$ is similar to a (set-valued) model of $talg(I)$, except that instead of providing a carrier set for each sort in $talg(I)$,  an $\mathcal{L}$-valued model provides a type in $\mathcal{L}$, and instead of providing a function for each symbol in $talg(I)$, an $\mathcal{L}$-valued model provides an expression in $\mathcal{L}$.  For example, if $\mathcal{L} =$ Java, then we can interpret {\sf String} as {\sf java.lang.String}, {\sf Nat} as {\sf java.lang.Integer}, $+ : {\sf String} \to {\sf String}$ as ${\sf java.lang.String.append}$, etc.  (Note that our $\mathcal{L}$-models are on $talg(I)$, not $Ty$; so an $\mathcal{L}$-model must provide a meaning for the skolem terms in $talg(I)$, which can be tricky.) Given an $\mathcal{L}$-model $M$, we can take the image of $\llbracket I_{EA} \rrbracket$ under $M$ by replacing each term $talg(t) \ni t \in \llbracket I_{EA} \rrbracket$ with the value of $t$ in $M$.  We write this as $M(\llbracket I_{EA} \rrbracket)$.  The pair $(M(\llbracket I_{EA} \rrbracket), M)$ can be used by the CQL tool in many situations where $(\llbracket I_{EA} \rrbracket, talg(A))$ is expected; for example, displaying instances (see Figure~\ref{images}), and (co-)evaluating uber-flowers.  Formalizing computational type-sides is an important area for future work.

\vspace{-.2in}
\subsection{Deciding that a Theory Conservatively Extends Another}
\label{consalg}

As described in Section~\ref{formal}, we may want instances to conservatively extend their schemas, where a theory $Th_2$ conservatively extends $Th_1$ when $Th_1 \vdash \forall \Gamma. \ t = t' : s$ iff $Th_2 \vdash \forall \Gamma. \ t = t' : s$ for all  $t, t' \in Terms^s(Th_1, \Gamma)$ for every $\Gamma$.  Conservativity in equational logic is not decidable, and the only system we are aware of that automates conservativity checks in a language at least as expressive as equational logic is CCC~\cite{LuethEA04}.  In this section, we give a simple algorithm that soundly but incompletely checks that a theory $Th_2$ conservatively extends a theory $Th_1$ by showing that $Th_2$ freely extends $Th_1$. 

Let $Th_1$ be an equational theory, and let $Th_2$ extend $Th_1$ with new sorts, symbols, and equations.  We will simplify the presentation of $Th_2$ by repeatedly looking for equations of the form $g = t$, where $g$ is a generator (0-ary symbol) of $Th_2$ but not of $Th_1$, and $t$ does not contain $g$; we then substitute $g \mapsto t$ in $Th_2$.  If after no more substitutions are possible, all equations in $Th_2$ are either reflexive or provable in $Th_1$, then $Th_2$ is conservative (actually, free) over $Th_1$.  For example, we can show that the theory: 
$$\{{\sf infinity} : {\sf Nat}, \ {\sf undef} : {\sf Nat}, \ {\sf infinity}={\sf 0}, \ {\sf undef}={\sf 1}, \ {\sf infinity}={\sf undef}\}$$ is not conservative over $Type$ (Figure~\ref{Type}), because the simplification process yields the non-reflexive equation ${\sf 0} = {\sf 1}$, which is not provable in $Type$.  However, the algorithm is far from complete. The theory:
$$\{{\sf +} : {\sf Nat \times Nat} \to {\sf Nat}, {\sf infinity} : {\sf Nat}, \ {\sf undef} : {\sf Nat}, {\sf infinity}{\sf +}{\sf 1} = {\sf undef}{\sf +}{\sf 1} \}$$ does not pass our check, even though it is a conservative extension of $Type$.  Developing a better conservativity checker is an important area for future work, lest we inadvertently ``damage our ontologies''~\cite{Ghilardi06didi}.  The process of repeatedly substituting $g \mapsto t$, where $g$ is a generator in an instance's type-algebra and $t$ is a type-side term is also used by the CQL tool to simplify the display of tables by biasing the tables to display e.g., {\sf 45} instead of e.g. {\sf age.bill} when {\sf 45} and {\sf age.bill} are provably equal. 

\subsection{Evaluating Uber-flowers}
\label{evaluber}

Although it is possible to evaluate an uber-flower by translation into a data migration of the form $\Delta \circ \Pi$, we have found that in practice it is faster to evaluate such queries directly, using algorithms which extend existing join algorithms from relational query processing.  In this section, we describe such an algorithm which intuitively extends the most basic join algorithm, ``nested loops join''~\cite{Garcia-Molina:2008:DSC:1450931}.   

   Let $Q : S \to T$ be an uber-flower and let $I$ be an $S$-instance.  We now describe how to compute the instance (theory) $eval(Q)(I)$.  First, we copy the generators and equations of the type-algebra $talg(I)$ into $eval(Q)(I)$.  Then, for every target entity $t \in T$, we perform the following:

\begin{itemize}

\item We define the generators of entity $t$ in $eval(Q)(I)$ to be those $\llbracket I_{EA} \rrbracket$ environments for $fr_t$ which satisfy $wh_t$.  Formally, we represent these environments as ground substitutions $fr_t \to Terms(I,\emptyset)$ and define, where $fr_t := \{ \overrightarrow{v_i : s_i} \}$: 
$$
eval(Q)(I)(t) := \{ [\overrightarrow{v_i \mapsto e_i}] \ | \ I \vdash eq[\overrightarrow{v_i \mapsto e_i}] \ ,\  \forall eq \in wh_t \ , \forall e_i \in \llbracket I_{EA} \rrbracket (s_i) \}
$$

\item For each attribute $att : t \to t' \in T'$, we have a term $[att] \in Terms^{t'}(S, fr_t)$ from the return clause for $t$.  For every substitution $\sigma \in eval(Q)(t)$, we have $[att]\sigma \in Terms^{t'}(S, \emptyset)$, and we add: 
$$\sigma . att = trans([att]\sigma) \ \in eval(Q)(I)$$  
The reason that $trans([att]\sigma) \in Terms^{t'}(eval(Q)(I), \emptyset)$ is because $trans([att]\sigma) \in Terms^{t'}(talg(I), \emptyset)$ and $talg(I) \subseteq eval(Q)(I)$.  See \ref{trans} for the definition of $trans$.  

\item For each foreign key $fk : t \to t' \in T$, we have a transform from the frozen instance for $t'$ to the frozen instance for $t$ from the keys clause for $t$, which can be thought of as a substitution $[fk] : fr_{t'} \to Terms(S, fr_t)$.   For every substitution $\sigma : fr_{t} \to Terms(S, \emptyset) \in eval(Q)(t)$, we add the equation: 
$$\sigma . fk = \sigma \circ [fk]  \ \in eval(Q)(I) $$

We know that $\sigma \circ [fk] \in eval(Q)(I)(t')$ because $[fk]$ is a transform, not an arbitrary substitution.  

\end{itemize}

Note that in order to build the instance $eval(Q)(I)$, we have effectively constructed the term model $\llbracket eval(Q)(I)_{EA} \rrbracket$, and then ``de-saturated'' it into an equational theory.  The description above is implemented in the CQL tool by a simple nested loop join algorithm: for each target entity $t$, to find those $\llbracket I_{EA} \rrbracket$ environments for $fr_t := \{ \overrightarrow{v_i : s_i} \}$ satisfying $wh_t$, compute a temporary table $\tau := \Pi_i \llbracket I_{EA} \rrbracket(s_i)$ and then filter $\tau$ by $wh_t$, using provable equality in $I$.   The CQL tool contains additional implementations of query evaluation based on more sophisticated algorithms such as hash-join~\cite{Garcia-Molina:2008:DSC:1450931}, but we do not describe these algorithms here.

To make the above description concrete, we will now evaluate the uber-flower $Promote : Emp \to Emp$ from Figure \ref{query} on the instance $Inst$ from Figure \ref{Inst}, which in turn is on the schema $Emp$ from Figure~\ref{Emp} on the type-side $Type$ from Figure~\ref{Type}.  Our goal is to compute the instance (equational theory) $eval(Promote, Inst)$.  We start by copying $talg(Inst)$ into  $eval(Promote, Inst)$.  Next, we process the tableau.  We start with target entity ${\sf Dept} \in T$.  The from and where clauses give us a set of substitutions $\{ [d \mapsto {\sf m}], [d \mapsto {\sf s}] \}$, which are the generators of $eval(Promote, Inst)$ at entity {\sf Dept}.  The return clause adds equations $[d \mapsto {\sf m}].{\sf dname} = {\sf m.dname}$ and $[d \mapsto {\sf s}].{\sf dname} = {\sf s.dname}$; note that {\sf s.dname} is one of the generators from $talg(Inst)$, and ${\sf s}$ will not be a term in $eval(Promote, Inst)$.  The keys clause for ${\sf secr} : {\sf Dept} \to {\sf Emp}$ adds equations $[d \mapsto {\sf m}].{\sf secr} = [e \mapsto {\sf b}]$ and $[d \mapsto {\sf s}].{\sf secr} = [e \mapsto {\sf c}]$; we have not added $[e \mapsto {\sf b}]$ and $[e \mapsto {\sf b}]$ to $eval(Promote,Inst)$ yet but we will momentarily.  Note that so far, we have simply copied the table {\sf Dept} from $Inst$ to $eval(Promote, Inst)$, up to isomorphism.  We next consider the target entity ${\sf Emp} \in T$.  The from and where clause give us a set of substitutions $\{ [e \mapsto {\sf a}], [e \mapsto {\sf b}], [e \mapsto {\sf c}], [e \mapsto {\sf a;mgr}], [e \mapsto {\sf b;mgr}], [e \mapsto {\sf c.mgr}] \}$, which are the generators of $eval(Promote, Inst)$ at entity {\sf Emp}.  The return clause adds equations such as $[e \mapsto {\sf a}].{\sf ename} = {\sf a.ename} + {\sf a;mgr;ename}$, where ${\sf a;ename}$ and ${\sf a;mgr;ename}$ are generators in $talg(Inst)$.  The keys clause for ${\sf mgr} : {\sf Emp} \to {\sf Emp}$ adds equations such as $[e \mapsto {\sf a}];{\sf mgr} = [e \mapsto {\sf a}]$, and the keys clause for ${\sf wrk} : {\sf Emp} \to {\sf Dept}$ adds equations such as $[e \mapsto {\sf a}];{\sf wrk} = [d \mapsto m]$ (we added $[d \mapsto {\sf m}]$ to $eval(Promote, Inst)$ when processing the target entity ${\sf Emp}$).  The entire instance is displayed in Figure~\ref{evald}.

\begin{figure}[h]
\begin{mdframed}
$$
Generators := \{ 
 {\sf m}.{\sf dname}, \ {\sf s.dname}, \ {\sf a}.{\sf ename}, \ {\sf b}.{\sf ename}, \ {\sf c}.{\sf ename} : {\sf String},
 $$
 $$
\ \ \ \ \ \ \ \ \ \  \ \ \ \ {\sf a}.{\sf mgr}.{\sf ename}, \ {\sf b}.{\sf mgr}.{\sf ename}, \ {\sf c}.{\sf mgr}.{\sf ename} : {\sf String},
$$
$$
[d \mapsto {\sf m}], \ [d \mapsto {\sf s}] : {\sf Dept},$$
$$ \ \ \ \ \ \ \ \ \ \ \ \ \ \ \ \ \ \ \ \ \ \ \ \ \ \ \ \ \ \ \ \ \ \ \ \ \ \ \ \ \ \ \ \ \ [e \mapsto {\sf a}],\  [e \mapsto {\sf b}], \ [e \mapsto {\sf c}], \ [e \mapsto {\sf a.mgr}], \ [e \mapsto {\sf b.mgr}], \ [e \mapsto {\sf c.mgr}] : {\sf Emp} \}
$$

$$
Eqs := \{ 
 {\sf a}.{\sf ename} = {\sf Al}, \ \
 {\sf c}.{\sf ename} = {\sf Carl}, \ \
 {\sf m}.{\sf dname} = {\sf Math}, 
 $$
$$
\ \ \ \ \ \ \ \ \ \ \ \ \ \ \ \ \ \ [d \mapsto {\sf m}];{\sf secr} = [e \mapsto {\sf b}], \
[d \mapsto {\sf s}];{\sf secr} = [e \mapsto {\sf c}], \
 [d \mapsto {\sf m}];{\sf dname} = {\sf Math}, \ \ldots \}
$$

\vspace{.2in}
\begin{footnotesize}
\begin{tabular}{|c|c|c|}  
\multicolumn{3}{c}{{\sf Dept}}  \vspace{.1in} \\  \hline
\hspace{.1in} {\sf  ID} \hspace{.1in} & \hspace{.1in}{\sf dname} \hspace{.1in}& \hspace{.1in} {\sf secr} \hspace{.1in} \\ \hhline{|=|=|=|}
$[d \mapsto {\sf m}]$ & {\sf Math} & $[e \mapsto {\sf b}]$  \\ \hline
$[d \mapsto {\sf s}]$ & {\sf s;dname} & $[e \mapsto {\sf c}]$ \\ \hline
\end{tabular}
 \ \ \ \ \ \ 
\begin{tabular}{|c|c|c|c|}
\multicolumn{4}{c}{{\sf Emp}} \vspace{.1in} \\ \hline
\hspace{.2in} {\sf  ID} \hspace{.2in} & \hspace{.6in} {\sf ename} \hspace{.6in} & \hspace{.2in} {\sf mgr} \hspace{.2in} & \hspace{.1in} {\sf wrk} \hspace{.1in} \\ \hhline{|=|=|=|=|}
$[e \mapsto {\sf a}]$& {\sf Al} + {\sf a.mgr.ename} & $[e \mapsto {\sf a}]$ & $[d \mapsto {\sf m}]$  \\ \hline
$[e \mapsto {\sf b}]$& {\sf b.ename} + {\sf b.mgr.ename} & $[e \mapsto {\sf b}]$ & $[d \mapsto {\sf m}]$ \\  \hline
$[e \mapsto {\sf c}]$& {\sf Carl} + {\sf c.mgr.ename} & $[e \mapsto {\sf c}]$  & $[d \mapsto {\sf s}]$ \\ \hline
$[e \mapsto {\sf a}.{\sf mgr}]$ & {\sf a.mgr.ename} +  {\sf a.mgr.ename}&$[e \mapsto {\sf a}.{\sf mgr}]$ &  $[d \mapsto {\sf m}]$  \\ \hline
$[e \mapsto {\sf b}.{\sf mgr}]$ & {\sf b.mgr.ename} +  {\sf b.mgr.ename}& $[e \mapsto {\sf b}.{\sf mgr}]$ &  $[d \mapsto {\sf m}]$  \\ \hline
$[e \mapsto {\sf c}.{\sf mgr}]$ & {\sf c.mgr.ename} +  {\sf c.mgr.ename}& $[e \mapsto {\sf c}.{\sf mgr}]$ &  $[d \mapsto {\sf s}]$  \\ \hline
\end{tabular}
\end{footnotesize}

\vspace{.1in}
\caption{Evaluation of Uber-flower $Promote$ (Figure~\ref{query}) on $Inst$ (Figure~\ref{Inst})}
\label{evald}
\end{mdframed}
\end{figure}

\subsubsection{Evaluating Uber-flowers on Transforms}
\label{xxx5}

Let $Q : S \to T$ be an uber-flower and let $h : I \to J$ be a transform of $S$-instances $I$ and $J$.  Our goal is to define the transform $eval(Q)(h) : eval(Q)(I) \to eval(Q)(J)$.  For each target entity $t \in T$, consider the generators in $eval(Q)(I)(t)$: they will be the ground substitutions $fr_t \to Terms(S, \emptyset)$ satisfying $wh_t$.  We will map each such substitution to a substitution (generator) in $eval(Q)(J)(t)$.  Given such a substitution $\sigma := [v_0 : s_0 \mapsto e_0 , \ldots, v_n : s_n \mapsto e_n]$, where $e_i \in Terms^{s_i}(S, \emptyset)$, we define: 
$$eval(Q)(\sigma) :=  [v_0 : s_0 \mapsto nf_{J_{E}}(h(e_1)) , \ldots, v_n : s_n \mapsto nf_{J_{E}}(h(e_n))$$

 In general, $h(e_i)$ need not appear in $\llbracket J_E \rrbracket$, so we must use $nf$ (section~\ref{nf}) to find the normal form of $h(e_i)$ in $J_E$. 
 

\subsubsection{Evaluating Morphisms of Uber-flowers}
\label{xxx6}

  If $Q_1, Q_2 : S \to T$ are uber-flowers, a morphism $h : Q_1 \to Q_2$ is, for each entity $t \in T$, a morphism from the frozen instance for $t$ in $Q_1$ to the frozen instance for $t$ in $Q_2$, and it induces a  transform $eval(h) : eval(Q_2)(I) \to eval(Q_1)(I)$ for every $S$-instance $I$; we now show how to compute $eval(h)$.  Let $t \in T$ be an entity and $fr^1_t := \{v_1, \ldots, v_n\}$ be the for clause for $t$ in $Q_1$.  The generators of $eval(Q_2)(I)$ are substitutions $\sigma : fr^2_t \to \llbracket I_E \rrbracket$, and
  $$
  eval(h)(\sigma) :=  [v_1 \mapsto nf_{I_E}(h(v_1)\sigma), \ldots v_n \mapsto nf_{I_E}(h(v_n)\sigma)] 
$$
In the above we must use $nf$ (section~\ref{nf}) to find appropriate normal forms.
\vspace{-.1in}
\subsection{Co-Evaluating Uber-flowers}
\label{coevaluber}

Although it is possible to co-evaluate an uber-flower by translation into a data migration of the form $\Delta \circ \Sigma$, we have implemented co-evaluation directly.  Let $Q : S \to T$ be an uber-flower and let $J$ be a $T$-instance.  We are not aware of any algorithm in relational database theory that is similar to $coeval(Q)$; intuitively, $coeval(Q)(J)$ products the frozen instances of $Q$ with the input instance $J$ and equates the resulting pairs based on either the frozen part or the input part.  We now describe how to compute the $S$-instance (theory) $coeval(Q)(J)$.  First, we copy the generators and equations of the type-algebra $talg(J)$ into $coeval(Q)(J)$.  We define $coeval(Q)(I)$ to be the smallest theory such that, for every target entity $t \in T$, where $fr_t := \{ v_1 : s_1, \ldots , v_n : s_n \}$,

\begin{itemize}

\item $\forall (v : s) \in fr_t,$ and $\forall j \in \llbracket J \rrbracket(t),$ 
$$
  (v, j) : s \  \in \ coeval(Q)(J)  
$$
\item $\forall e=e'  \in wh_t,$ and $\forall j \in \llbracket J \rrbracket(t),$
$$  
   (e=e') [v_1 \mapsto (v_1, j), \ldots, v_n \mapsto (v_n, j) ]  \  \ \in \ coeval(Q)(J)
$$

\item $\forall att : t \to t' \in T$, and $\forall j \in \llbracket J \rrbracket(t),$
$$ 
  trans(\llbracket J \rrbracket(att)(j)) = [att][v_1 \mapsto (v_1, j), \ldots, v_n \mapsto (v_n, j) ]  \  \ \in coeval(Q)(J)
$$
recall that $[att] \in Terms^{t'}(S, fr_t)$ is the return clause for attribute $att$ and $trans : Terms^{t'}(J, \emptyset) \to talg(J)$ is defined in section~\ref{trans}.

\item $\forall fk : t \to t' \in T,$ and $\forall j \in \llbracket J \rrbracket(t)$, and $\forall (v' : s') \in fr_{t'},$ 
$$
  (v', \llbracket J \rrbracket(fk)(j)) = v'[fk][v_1 \mapsto (v_1, j), \ldots, v_n \mapsto (v_n, j) ] \ \in coeval(Q)(J)
$$
\noindent
recall that the substitution $[fk] : fr_{t'} \to Terms(S, fr_t)$ is the keys clause for $fk$. 
\end{itemize}

The co-evaluation of the uber-flower $Promote : Emp \to Emp$ from Figure~\ref{query} on the instance $Inst$ from Figure~\ref{Inst} is in fact isomorphic to the evaluation of $promote$ (Figure~\ref{evald}); the reason is that evaluation and co-evaluation of $Promote$ are semantically both projections ($\Delta$-only operations).


\subsubsection{Co-Evaluating Uber-flowers on Transforms}


Let $Q : S \to T$ be an uber-flower and let $h : I \to J$ be a transform of $T$-instances $I$ and $J$.  Our goal is to define the transform $coeval(Q)(h) : coeval(Q)(I) \to coeval(Q)(J)$.  For each entity $t \in T$, and for every $(v : s) \in fr_t,$ and for every $j \in \llbracket J_E \rrbracket(t),$ we define:
$$
coeval(Q)(  (v, j) ) := (v, nf_{J_E}(h(j)))
$$
 As was the case for evaluation of uber-flowers on transforms, we must use $nf$ (section~\ref{nf}) to find appropriate normal forms.
\subsubsection{Co-Evaluating Morphisms of Uber-flowers}
\label{xxx7}

If $Q_1, Q_2 : S \to T$ are uber-flowers, a morphism $h : Q_1 \to Q_2$ is, for each entity $t \in T$, a morphism from the frozen instance for $t$ in $Q_1$ to the frozen instance for $t$ in $Q_2$, and it induces a transform $coeval(h) : coeval(Q_1)(J) \to coeval(Q_2)(J)$ for every $T$-instance $J$; in this section, we compute $coeval(h)$.  Let $t \in T$ be an entity. The generators of $coeval(Q_1)(J)$ are pairs $(v, j)$ with $v : s \in fr^{Q_2}_t$ and $j \in \llbracket J_E \rrbracket(t)$.  Define:
$$
coeval(h)((v, j)) := (v', j).fk_1.\ldots fk_n   \ \ \ \textnormal{where} \  h(v) := v'.fk_1 . \ldots . fk_n 
$$

\subsection{The Unit and Co-Unit of the Co-Eval , Eval Adjunction}

Let $Q : S \to T$ be an uber-flower.  Then $coeval(Q)$ is left adjoint to $eval(Q$), i.e., $coeval(Q) \dashv eval(Q)$.  This means that the set of morphisms $coeval(Q)(I) \to J$ is isomorphic to the set of morphisms $I \to eval(Q)(J)$ for every $I, J$.  The {\it unit} and {\it co-unit} of the adjunction, defined here, describe this isomorphism.  Let $I$ be a $S$-instance.  The component at $I$ of the co-unit transform $\epsilon_I : coeval(Q)(eval(Q)(I)) \to I$ is defined as:
$$
\epsilon_I((v_k, [v_1  \mapsto e_1, \ldots , v_n  \mapsto e_2]) ) := e_k, \ \forall k \in \{ 1 , \ldots, n \}  
$$
Let $J$ be a $T$-instance.  The component at $I$ of the unit transform $\eta_J : J \to eval(Q)(coeval(Q), J)$ is:
$$
\eta_J(j) := [v_1 \mapsto (v_1, nf_{J_E}(j)), \ldots , v_n \mapsto (v_n, nf_{J_E}(j))]
$$ 
\subsection{Converting Instances to Schemas}
\label{ppivot}
Let $S$ be a schema on type-side $Ty$ and let $I$ be an $S$-instance.  We now describe how to convert $I$ into a schema, written $\int I$, an operation we call ``pivoting''.  Then, we describe an alternative way to convert $I$ into a schema, written $\oint I$, an operation we call ``co-pivoting''.  We indicate (co-)pivots using integral signs because related categorical constructions, such as the Grothendieck construction or co-ends~\cite{BW} are often indicated using integral signs; however, here we are {\it defining} the schemas $\int I$ and $\oint I$, and our integral notation is not meant to indicate any existing construction.

 The {\it pivot} of $I$ is defined to be a schema $\int I$ on $Ty$, a mapping $F : \int I \to S$, and a $\int I$-instance $J$, such that $\Sigma_F(J) = I$, defined as follows.  First, we copy $talg(I)$ into $J$. Then for every entity $s \in S$, and every  $ i \in \llbracket I_E \rrbracket(s)$, we define:
$$
  i : s  \in \int I
  \ \ \ \ \ \ 
F(i) := s
\ \ \ \ \ \ 
i : s \in J
$$
 and for every attribute $att : s \to s' \in S$,
$$
  (i, att) : i \to s' \in \int I
\ \ \ \ \ \ 
F((i, att)) := att
\ \ \ \ \ \ \
i.(i, att) = trans( \llbracket I_{EA} \rrbracket(att)(i) ) \in J
$$
where $trans$ is defined in section~\ref{trans}, and for every foreign key $fk : s \to s' \in S$,
$$
  (i, fk) : i \to \llbracket I_E \rrbracket(fk)(i) \in \int I
\ \ \ \ \ \ 
F((i, fk)) := fk
\ \ \ \ \ \ \
i.(i, fk) = \llbracket I_E \rrbracket(fk)(i)\in J
$$
In addition, for each generator $g := e  .  fk_1  .  \ldots .  fk_n .  att $ in $talg(I)$, we have a term $[g] \in Terms(J, \emptyset)$ defined as:
$$
nf_{I_E}(e) \ \ . \ \ (nf_{I_E}(e), fk_1) \ \ . \ \  (nf_{I_E}(e.fk_1), fk_2) \ \ . \ \ \ldots \ \ . \ \ (nf_{I_E}(e.fk_1., \ldots fk_n), att)  
$$
and we add $g = [g]$ to $J$.  See Figure~\ref{pivot} for an example.  

The {\it co-pivot} of $I$ is defined to be a schema $\oint I$ on $Ty$ that extends $S$ and an inclusion mapping $F : S \hookrightarrow \oint I$.  First, we add a single entity $\star$ to $ \oint I$, and for every generator $g : s \in talg(I)$, an attribute $g_A$: 
$$
\star \in \oint I  \ \ \ \ \ \  \ g_A : \star \to s  \ \in \oint I
$$
Then for every entity $s \in S$, and $ i \in \llbracket I_E \rrbracket(s)$ we add a foreign key:
$$
\ \ \ \ \ \   
i_E : \star \to s  \in \oint I
  \ \ \ \ \ \ 
F(s) := s
$$
and additionally, for every attribute $att : s \to s'$, and every foreign key $fk : s \to s''$, we add:
$$
\forall x : \star. \ x. i_E. att = x.trans(i.att)_A
\ \ \ \ \ \ 
\forall x : \star. \ x.i_E . fk = x.nf_{I_E}(i.fk)_E
 \ \ \ \ \in \oint I
$$
where $trans$ is defined in section~\ref{trans}.  See Figure~\ref{copivot} for an example.  Note that in this figure, we include an instance $J := \Delta_F(I)$ to make the duality of co-pivoting and pivoting explicit, and that although $J$ includes the tables from $I$, we do not display these tables in order to make the figure smaller.

\subsection{Performance}

In this section we give preliminary performance results for the June 2017 version of the CQL tool.  Some of these experiments reference schemas and constructions (e.g., colimits) that are not introduced until the next section (Section~\ref{pattern}).  In practice, the time required to check schema mappings and uber-flowers for well-formedness (Section~
\ref{vc}) is negligible, so in this section we focus on the scalability of the saturation procedure (Section~
\ref{satsec}; used for $\Sigma$) and uber-flower evaluation (Section~\ref{evaluber}; used for $\Delta, \Pi$).

When interpreting the following results, it is important to consider several caveats.  First, the CQL tool is naively implemented; for example, it uses  string-valued variables rather than a more sophisticated variable representation such as De Bruijn indices~\cite{Mitchell:1996a}.  Second, the CQL tool deviates somewhat from the formalism in this paper for reasons of efficiency and expediency.  Third, there are other evaluation strategies for CQL besides the one that the CQL tool uses; for example, implementation via SQL generation~\cite{relfound} (for $\Delta$ and $\Pi$) or using the ``chase'' algorithm~\cite{FKMP05} (for $\Sigma$) -- in practice, these strategies may be necessary to get practical performance from our algebraic approach to data integration.  

As previously mentioned the CQL tool ships with many different automated theorem provers.  In this section, we refer to three of these provers by name:

\begin{itemize}

\item {\it Thue}. This prover implements a Knuth-Bendix completion algorithm specialized to equational theories where all symbols are 0 or 1-ary~\cite{doi:10.1137/0214073}.

\item {\it Congruence}. This prover implements a congruence closure algorithm~\cite{Nelson:1980:FDP:322186.322198} to decide equational theories without quantifiers.

\item {\it Program}. This prover orients equations into size-reducing rewrite rules and checks for ``weak orthogonality''~\cite{Baader:1998:TR:280474}.  Such theories are first-order functional programs and are decided by rewriting terms into normal forms. 

\end{itemize}

The experiments in this section are synthetic.  CQL includes the ability to construct random instances: let $gens(s)$ indicate a desired set of generators for each entity $s$ in some schema.  For each attribute or foreign key $f : s \to s'$ and for each generator $g \in gens(s)$ the random instance contains an equation $f(g) = g'$, where $g'$ is a (uniformly) randomly chosen element of $gens(s')$.  These random instances thus have a special, ``dense'' form, and as such performance tests on them may not indicate real-world performance. 
\begin{figure}[t]
\begin{mdframed}
\begin{footnotesize}

\begin{tabular}{ccc}
\parbox{1.5in}{
\fbox{ $S :=$ \hspace{.3in}
\xymatrix@=25pt{
\LTO{String} \\
  \LTO{Person} \ar[d]_{\sf livesIn}  \ar[u]^{\sf name} \\
   \LTO{Home}  \ar[d]_{\sf size} \\
   \LTO{Nat}
} \hspace{.3in} }}
& $\Fromm{F}$ &
\hspace{-.7in}
\parbox{1.5in}{
\fbox{ $\int I :=$ \hspace{.4in}
\xymatrix@=25pt{
\LTO{String} & \\
 \ar[u]^{\sf name} \LTO{a} \ar[d]_{\sf livesIn}  & \LTO{b} \ar[ul]_{\sf name} \ar[dl]^{\sf livesIn}  \\ 
  \LTO{ h } \ar[d]_{\sf size} \\
   \LTO{Nat} 
} \hspace{.4in} }}
\vspace{.1in} \\
$I :=$
\begin{tabular}[t]{|c|c|}
\multicolumn{2}{c}{{\sf Home}} \vspace{.03in} \\\hline 
{\sf  ID}& \hspace{.02in} {\sf size} \hspace{.02in}   \\\hhline{|=|=|}
{\sf h} & {\sf 4} \\ \hline 
\end{tabular}
\hspace{.1in}
\begin{tabular}[t]{|c|c|c|}
\multicolumn{3}{c}{{\sf Person}}\vspace{.03in} \\\hline 
{\sf  ID}& \hspace{.02in} {\sf livesIn} \hspace{.02in} & {\sf name} \\\hhline{|=|=|=|} 
{\sf a} & {\sf h} & {\sf Alice} \\\hline 
{\sf b} & {\sf h} & {\sf Bill} \\ \hline 
\end{tabular}
& $ \ \ \ \Fromm{\llbracket \Sigma_F \rrbracket} \ \ \ $ &
$J :=$
\begin{tabular}[t]{|c|c|}
\multicolumn{2}{c}{{\sf h}}\vspace{.03in} \\\hline 
{\sf ID}& \hspace{.02in}{\sf size}\hspace{.02in}  \\\hhline{|=|=|}
{\sf h} & {\sf 4} \\\hline 
\end{tabular}
\hspace{.1in}
 \begin{tabular}[t]{|c|c|c|}
\multicolumn{3}{c}{{\sf a}}\vspace{.03in} \\\hline 
{\sf  ID}&\hspace{.02in} {\sf name}\hspace{.02in} & {\sf livesIn} \\\hhline{|=|=|=|}
{\sf a}  & {\sf Adam} & {\sf h } \\\hline 
\end{tabular}
\\ & &
\begin{tabular}[t]{|c|c|c|}
\multicolumn{3}{c}{{\sf b}}\vspace{.03in} \\\hline
{\sf  ID}& \hspace{.02in}{\sf name} \hspace{.02in}& {\sf livesIn} \\\hhline{|=|=|=|}
{\sf b} & {\sf Bill} & {\sf h} \\\hline 
\end{tabular}
\end{tabular}

\end{footnotesize}
\vspace{.1in}
\caption{Example of a Pivot}
\label{pivot}
\end{mdframed}
\end{figure}

\begin{figure}[b]
\begin{mdframed}
\begin{footnotesize}

\begin{tabular}{ccc}
\parbox{1.5in}{
\fbox{ $S :=$ \hspace{.3in}
\xymatrix@=25pt{
\LTO{String} \\
  \LTO{Person} \ar[d]_{\sf livesIn}  \ar[u]^{\sf name} \\
   \LTO{Home}  \ar[d]_{\sf size} \\
   \LTO{Nat}
} \hspace{.3in} }}
& \hspace{.3in} $\Too{F}$ &
\hspace{.3in}
\parbox{1.5in}{
\fbox{ $\oint I :=$ \hspace{.2in}
\xymatrix@=25pt{
\LTO{String} & \\
 \ar[u]^{\sf name} \LTO{Person} \ar[d]_{\sf livesIn}  & \LTO{\star} \ar@/^/[l]^{\sf a} \ar@/_/[l]_{\sf b} \ar[ld]^{\sf h}  \\ 
  \LTO{ Home } \ar[d]_{\sf size} \\
   \LTO{Nat} 
} \hspace{.2in} }}
 \vspace{.2in} \\ 
$I :=$
\begin{tabular}[t]{|c|c|}
\multicolumn{2}{c}{{\sf Home}} \vspace{.03in} \\\hline 
{\sf  ID}& \hspace{.02in} {\sf size} \hspace{.02in}   \\\hhline{|=|=|}
{\sf h} & {\sf 4} \\\hline 
\end{tabular}
\hspace{.1in}
\begin{tabular}[t]{|c|c|c|}
\multicolumn{3}{c}{{\sf Person}}\vspace{.03in} \\\hline 
{\sf  ID}& \hspace{.02in} {\sf livesIn} \hspace{.02in} & {\sf name} \\\hhline{|=|=|=|} 
{\sf a} & {\sf h} & {\sf Alice} \\\hline 
{\sf b} & {\sf h} & {\sf Bill} \\\hline 
\end{tabular}
& $ \ \ \ \ \ \ \ \ \  \ \ \Fromm{\llbracket \Delta_F \rrbracket} \ \ \ $ &
$J := I \ \cup \ $
\begin{tabular}[t]{|c|c|c|c|}
\multicolumn{4}{c}{$\star$}\vspace{.03in} \\\hline 
{\sf ID} \hspace{.02in} & \hspace{.04in}{\sf a}\hspace{.04in} &  \hspace{.04in}{\sf b}\hspace{.04in} &  \hspace{.04in}{\sf h}\hspace{.04in}  \\\hhline {|=|=|=|=|}
$\star$ & {\sf a} & {\sf b} & {\sf h}  \\\hline 
\end{tabular}

\end{tabular}

\end{footnotesize}
\vspace{.1in}
\caption{Example of a Co-Pivot}
\label{copivot}
\end{mdframed}
\end{figure}

The experiments in this section refer to two kinds of tasks: 
\begin{itemize}
\item {\it Saturation}.  That is, building a term model / initial algebra from an equational theory {\it by first constructing a decision procedure for the theory} and then enumerating terms (Section~\ref{satsec}).  Saturation arises following a $\Sigma$ or pushout operation (see Section~\ref{pattern}); data that is imported from e.g., a SQL database is already saturated.  An examination of Java profiling info suggests that the time to construct a decision procedure dominates the time required to build the term model.

\item {\it Query Evaluation}.  That is, building an instance by evaluating a query on another instance (Section~\ref{evaluber}).  Although this task does involve using a decision procedure, it does not require constructing a new decision procedure from scratch (a decision procedure for the output instance of a $\Delta$ or a $\Pi$ or an $eval$ can be easily obtained from a decision procedure for the input instance).  Hence, performance on this task is tantamount to the performance of the naive nested loops join~\cite{Garcia-Molina:2008:DSC:1450931} algorithm described in Section~\ref{evaluber}.  We have also implemented a faster hash join~\cite{Garcia-Molina:2008:DSC:1450931} algorithm, but it is out of scope for this paper.
\end{itemize}

The results in this section were obtained on an Intel Core i7-6770HQ CPU at 2.60GHz.  All entries are averages: we occasionally saw  $2\times$ speed ups or slow downs from run to run; we are not sure if this is due to CQL's inherent non-determinism (e.g., the order in which sets are traversed) or some artifact of the Java virtual machine upon which CQL runs.

\begin{itemize}

\item {\it Pharma Data Integration Example. }  This example uses the ``Pharma Colim'' example which is built-in to the CQL IDE and described in Section~\ref{pattern}.  It shows the amount of time, in seconds, required to saturate a random instance on a given number of rows on the pushout schema in Figure~\ref{fig1}.  The experiment is repeated for three different provers: Thue, Congruence, and Program.   We find that the overhead of the Thue algorithm compared to congruence closure is small, and that orienting equations according to size and checking for weak orthogonality is the fastest theorem proving method by a large margin.

\item {\it Finance Data Integration Example. }  This example uses the ``Finance Colim'' example which is built-in to the CQL IDE.  It shows the amount of time, in seconds, required to saturate a random instance on a given number of rows on a schema containing 9 entities, 8 foreign keys, 52 attributes, and 12 equations.  It also includes an additional schema (9 entities, 16 foreign keys, 50 attributes, 16 equations) and measures the time it takes to evaluate an uber-flower between the two schemas.  We find that saturation time dominates query evaluation time.

\item {\it Pullback Example.}  This example uses the ``Pullback'' example which is built-in to the CQL IDE.  It defines a schema that is a co-span $\cdot \to \cdot \leftarrow \cdot$ and a schema that is a commutative square.  It saturates a random instance on the co-span schema and then evaluates a query that joins the two root entities into the square schema.  We find that saturation time exceeds query evaluation time, but not nearly as much as with the above finance example.

\end{itemize}

\begin{figure}
\begin{footnotesize}
\begin{center}

\begin{tabular}{r | r | r | r}  
	Rows & $\ \ \ $ Thue &	$\ \ \ $Program & 	$\ \ \ $Congruence \\ \hline 
	500	& 6	& 1	& 4 \\ \hline
	1000 &	41	& 10 &	34 \\ \hline
	1500	& 134&	32	&116 \\ \hline
	2000	&314&	78&	252 \\ \hline
	2500 &	552	& 156&	504 \\ \hline
	3000	&987	&189&725 \\ \hline
	3500	&1451	&286	&1270 \\ \hline
	4000	&2236	&446	&1773
\end{tabular} 
\parbox{3.5in}{\includegraphics[width=3.5in]{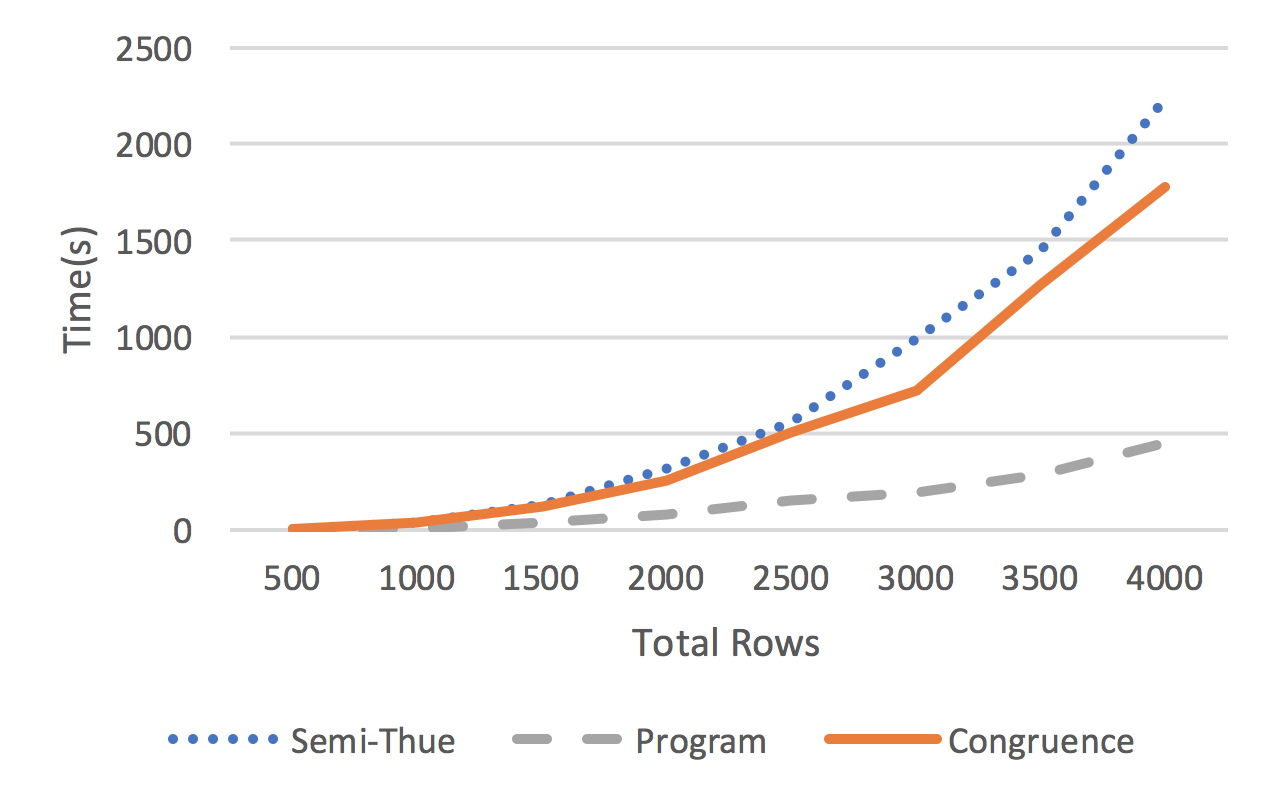}} 
\end{center}
\end{footnotesize}


\caption{Time (seconds) to Saturate a Random Instance on the Integrated Schema of the Pharma Example (Figure~\ref{fig2}), for Three  Theorem Provers. }
\label{fig_perf_pharma}
\end{figure}

\begin{figure}
\begin{center}
\begin{footnotesize}
\begin{tabular}{r | r | r }
	Rows & $\ \ \ $ Saturation &	 $\ \ \ $Query Evaluation \\ \hline 
90	&	0.26	&	0.03 \\ \hline
450	&	9.6	&	1.2 \\ \hline
810	&	57.5	&	5.7 \\ \hline
1170	&	263.3&		19.6 \\ \hline
1530	&	1150	&	38
\end{tabular}
\end{footnotesize}
\parbox{3.5in}{\includegraphics[width=3.5in]{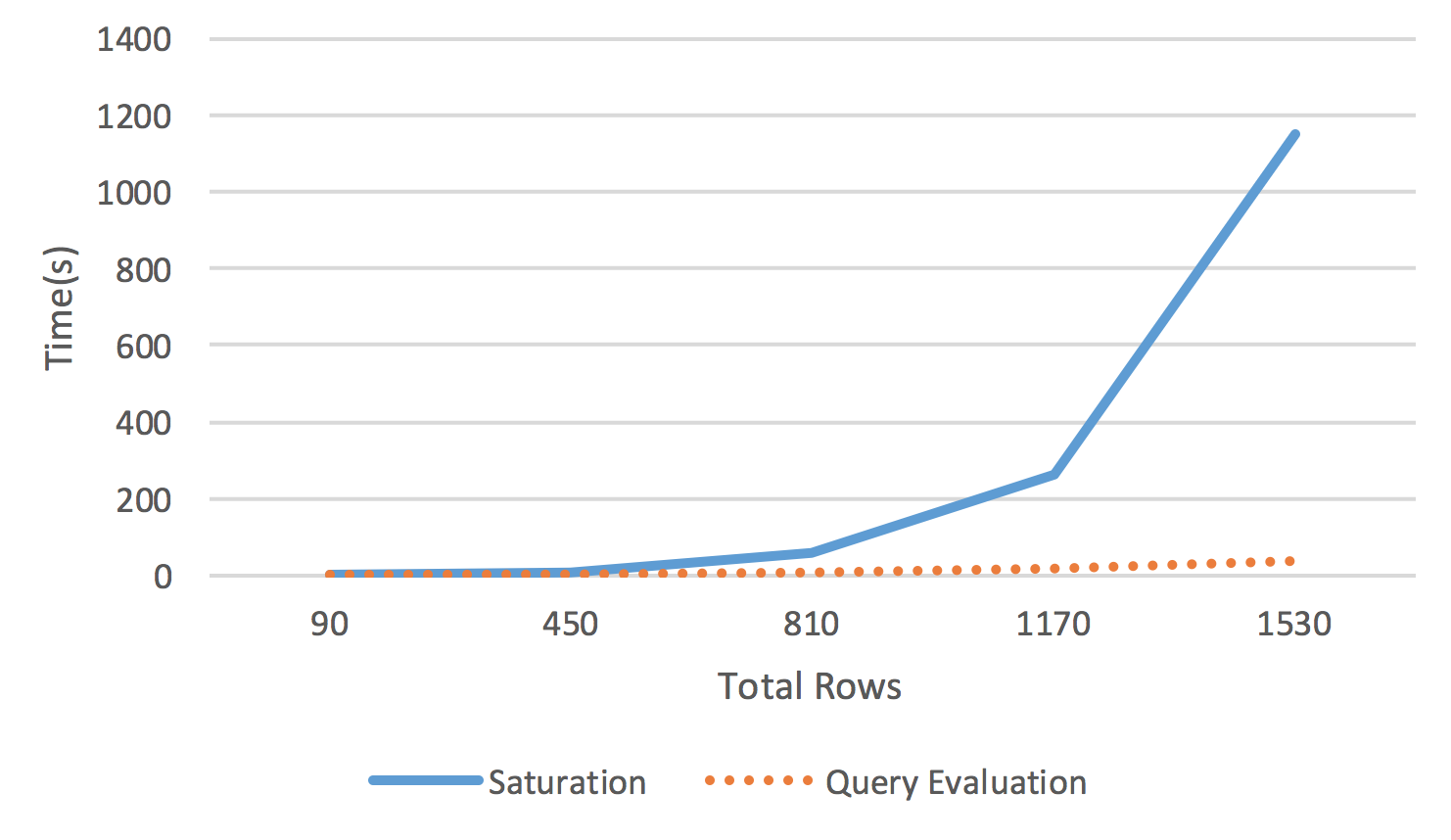}}
\end{center}

\caption{Time (seconds) to Saturate and Query a Random Instance on the Colimit Schema of the Finance Integration Example using the Thue Prover. }
\end{figure}

\begin{figure}
\begin{center}
\begin{footnotesize}
\begin{tabular}{r | r | r }
	Rows & $\ \ \ $ Saturation  &	 $\ \ \ $Query Evaluation \\ \hline 
300	&	1.2	&	0.2 \\ \hline
600	&	5.6	&	1.1 \\ \hline
900	&	19	&	3.6 \\ \hline
1200	&	45.3	&	8.88 \\ \hline
1500	&	84.3	&	16 \\ \hline
1800	&	147.6&	28 \\ \hline
2100	&	256	&	45.1
\end{tabular}
\end{footnotesize}
\parbox{3.5in}{\includegraphics[width=3.5in]{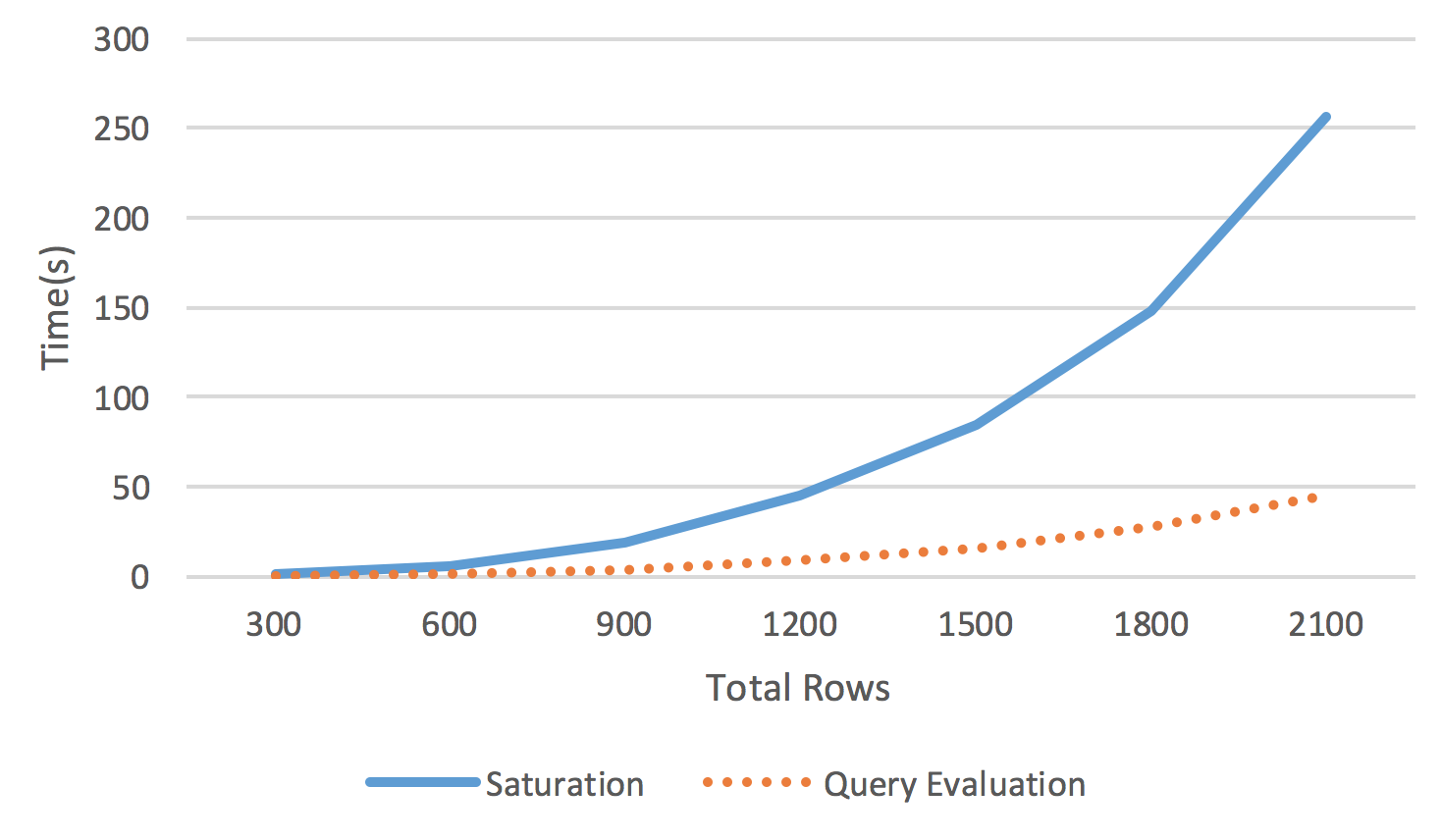}}
\end{center}

\caption{Time (seconds) to Saturate and Query a Random Instance on the Pullback Schema Using the Thue Prover. }
\label{fig_perf_pullback}
\end{figure}

\section{A Pushout Design Pattern for Algebraic Data Integration}
\label{pattern} 

In this section we describe a design pattern for integrating two instances on two different schemas, relative to an overlap schema and an overlap instance, using the formalism defined in this paper.  The overlap schema is meant to capture the schema elements common to the two input schemas (e.g., that {\sf Patient} and {\sf Person} should be identified), and the overlap instance is meant to capture the instance data that should be identified (e.g., that {\sf Pete} and {\sf Peter} are the same person).  

Although the $\Sigma,\Delta,\Pi$ data migration functors are sufficient to express queries and data migrations, as unary operations they are insufficient to express data integrations, which involve many schemas and instances and their relationships~\cite{Doan:2012:PDI:2401764}.  So, we need to define additional operations on our formalism as we develop our pattern.  In particular, we will define {\it pushouts}~\cite{BW} of schemas and instances and use pushouts as the basis of our pattern.  The idea of using pushouts to integrate data is not new and was for example discussed in Goguen~\cite{Goguen04informationintegration}; our goal here is to express this pattern using our formalism.  With pushouts defined, we describe our pattern at an abstract level, and then we describe a medical records example that uses the pattern.  This example is built into the CQL tool as the ``Pharma Colim'' example.

Pushouts have a dual, called a pullback, obtained by reversing the arrows in the pushout diagram (Figure~\ref{pushout}).   Exploring applications of pullbacks to data integration, as well as finding other useful design patterns for algebraic data integration, are important areas for future work. 

 \subsection{Pushouts of Schemas and Instances}
 
Let ${\bf C}$ be a category and $F_1 :  S \to S_1$ and $F_2 : S \to S_2$ be morphisms in ${\bf C}$.  A {\it pushout} of $F_1,F_2$ is any pair $G_1 : S_1 \to T$ and $G_2 : S_2 \to T$ such that $G_2 \circ F_2 = G_1 \circ F_1$, with the property that for any other pair $G_1' : S_1 \to T'$ and $G_2' : S_2 \to T'$ for which $G'_2 \circ F_2 = G'_1 \circ F_1$, there exists a unique $t : T \to T'$ such that $G_1' = t \circ G_1$ and $G_2' = t \circ G_2$, as shown in Figure~\ref{pushout}.

 \begin{figure}[h]
 \begin{mdframed}
 \begin{align*}
  \xymatrix@C+2em@R+2em{
   S \ar[r]^{F_1}  \ar[d]_{F_2} & S_1 \ar[d]^{G_1} \ar@/^10pt/[rdd]^{G_1'}   & \\
   S_2 \ar[r]_{G_2} \ar@/_10pt/[rrd]_{G_2'} & T \ar@{-->}[dr]^t & \\
    & & T' \\
   }  \end{align*}
  
   \caption{Pushouts}
   \label{pushout}
   \end{mdframed}
  \end{figure}
Our formalism admits pushouts of schemas and instances.  Let $S := (Ens, Symbols, Eqs)$, $S_1 := (Ens_1, Symbols_1, Eqs_1)$ and $S_2 := (Ens_2, Symbols_2, Eqs_2)$ be schemas on some type-side, where $En$ indicates entities, $Symbols$ indicates foreign keys and attributes, and $Eqs$ indicates schema, but not type-side, equations.  Let $F_1 : S \to S_1$ and $F_2 : S \to S_2$ be schema mappings.  The pushout schema $T$ is defined with entities:
$$
Ens_T := (Ens_1 \sqcup Ens_2) / \sim
$$
where $\sqcup$ means disjoint union, $\sim$ is the least equivalence relation such that $F_1(e) \sim F_2(e)$ for every entity $e \in S$, and $/$ means set-theoretic quotient.  We define further that: 
$$
Symbols_T := Symbols_1 \ \sqcup \ Symbols_2
$$
$$
Eqs_T := Eqs_1 \ \sqcup \ Eqs_2 \ \sqcup $$
$$\{ v_1 : F_1(s_1), \ldots, v_n : F_1(s_n) . \ F_1(e) = F_2(e) : F_1(s) \ | \ e : s_1 \times \ldots \times s_n \to s \in Symbols_S \}
$$
and the schema mappings $G_1$ and $G_2$ inject each entity into its equivalence class under $\sim$ and inject each symbol appropriately. 

 Pushouts of instances are slightly easier to define than pushouts of schemas.  Let $S := (Gens, Eqs)$, $S_1 := (Gens_1, Eqs_1)$ and $S_2 := (Gens_2, Eqs_2)$ be instances on some schema, where $Gens$ indicates generators and $Eqs$ indicates instance, but not schema, equations.  Let $F_1 : S \to S_1$ and $F_2 : S \to S_2$ be transforms.  The pushout instance $T$ is:
$$
Gens_T := Gens_1 \ \sqcup \ Gens_2
$$
$$
Eqs_T := Eqs_1 \ \sqcup \ Eqs_2 \ \sqcup \ \{ F_1(e) = F_2(e) : s \ | \ e : s \in Gens_S \}
$$
and the transforms $G_1, G_2$ are inclusions.  
  
The pushout schemas and instances defined in this section are canonical, but the price for canonicity is that their underlying equational theories tend to be highly redundant (i.e., have many symbols that are provably equal to each other).  These canonical pushout schemas and instances can be simplified, and in fact CQL can perform simplification, but the simplification process is necessarily non-canonical (a fact which complicates the use of algebraic specification techniques in general~\cite{rabe}).  In our extended medical example (Figure~\ref{fig1}), we will use a simplified non-canonical pushout schema.  

\vspace{-.1in}
 \subsection{Overview of the Pattern}
  
Given input schemas $S_1$, $S_2$, an overlap schema $S$, and mappings $F_1, F_2$ as such: 
$$S_1 \stackrel{F_1}{\leftarrow} S \stackrel{F_2}{\rightarrow} S_2$$
we propose to use their pushout: 
$$S_1 \stackrel{G_1}{\to} T \stackrel{G_2}{\leftarrow} S_2$$ 
as the integrated schema.  Given input $S_1$-instance $I_1$, $S_2$-instance $I_2$, overlap $S$-instance $I$ and transforms $h_1 \taking \Sigma_{F_1}(I) \to I_1$ and $h_2 : \Sigma_{F_2}(I) \to I_2$,  we propose the pushout of:  
$$\Sigma_{G_1}(I_1) \stackrel{\Sigma_{G_1}(h_1)}{\leftarrow} \bigl( \Sigma_{G_1 \circ F_1}(I) = \Sigma_{G_2 \circ F_2}(I)  \bigr)\stackrel{\Sigma_{G_2}(h_2)}{\to} \Sigma_{G_2}(I_2)$$ 
as the integrated $T$-instance.  

Because pushouts are initial among the solutions to our design pattern, our integrated instance is the ``best possible'' solution in the sense that if there is another solution  to our pattern, then there will be a unique transform from our solution to the other solution.  In functional programming terminology, this means our solution has ``no junk'' (extra data that should not appear) and ``no noise'' (missing data that should appear)~\cite{Mitchell:1996a}.  Initial solutions also appear in the theory of relational data integration, where the chase constructs weakly initial solutions to data integration problems~\cite{FKMP05}.


\subsection{An Example of the Pattern}


As usual for our formalism, we begin by fixing a type-side.  We choose the $Type$ type-side from Figure~\ref{Type}.  Then, given two source schemas $S_1$, $S_2$, an overlap schema $S$, and mappings $F_1, F_2$ as input, our goal is to construct a pushout schema $T$ and mappings $G_1, G_2$, as shown in Figure~\ref{fig1}.  In that figure's graphical notation, an attribute $\bullet_A \to_{att} \bullet_{\sf String}$ is rendered as $\bullet_A - \circ_{att}$.  Next, given input $S_1$-instance $I_1$, $S_2$-instance $I_2$, overlap $S$-instance $I$ and morphisms $h_1 : \Sigma_{F_1}(I) \to I_1$ and $h_2 : \Sigma_{F_2}(I) \to I_2$, our goal is to construct a pushout $T$-instance $J$ and morphisms $j_1, j_2$, as shown in Figure~\ref{fig2}; note that in this figure, by $K \to_{(F, h)} L$ we mean that $h : \Sigma_F(K) \to L$, and that we use $italic$ font for generators, ${\sf sans} \ {\sf serif}$ font for sorts and symbols, and regular font for terms in the type-side.

Our example involves integrating two different patient records databases.  In $S_1$, the ``observations'' done on a patient  have types, such as heart rate and blood pressure.  In $S_2$, the observations still have types, but via ``methods'' (e.g., patient self-report, by a nurse, by a doctor, etc; for brevity, we have omitted attributes for the names of these methods).  Another difference between schemas is that $S_1$ assigns each patient a gender, but $S_2$ does not.  Finally, entities with the same meaning in both schemas can have different names ({\sf Person} vs {\sf Patient}, for example).

We construct the overlap schema $S$ and mappings $F_1,F_2$ (Figure~\ref{fig1}) by thinking about the meaning of $S_1$ and $S_2$; alternatively, schema-matching techniques~\cite{Doan:2012:PDI:2401764} can be used to construct overlap schemas.  In this example, it is clear that $S_1$ and $S_2$ share a common span $P \ { }_f\leftarrow \ O \ \to_g \ T$ relating patients, observations, and observation types; in $S_1$, this span appears verbatim but in $S_2$, the path $g$ corresponds to $g_2 \circ g_1$.  This common span defines the action of $F_1$ and $F_2$ on entities and foreign keys in $S$, so now we must think about the attributes in $S$.  For purposes of exposition we assume that the names of observation types (``BP'', ``Weight'', etc.) are the same between the instances we are integrating.  Hence, we include an attribute for observation type in the overlap schema $S$.  On the other hand, we do not assume that patients have the same names across the instances we are integrating; for example, we have the same patient named ``Pete'' in one database and ``Peter'' in the other database.  Hence, we do not include an attribute for patient name in $S$.  If we did include an attribute for patient name, then the pushout schema would have a single attribute for patient name, and the integrated instance would include the equation ``Pete'' $=$ ``Peter'' $: {\sf String}$.  We would violate the conservative extension property (see section~\ref{consalg}), which is not a desirable situation~\cite{Ghilardi06didi}.  So, our design pattern explicitly recommends that when two entities in $S_1$ and $S_2$ are identified in an overlap schema, we should only include those attributes which appear in both $S_1$ and $S_2$ for which the actual values of these attributes will correspond in the overlap instance.  As another example of this phenomenon, to a first approximation, attributes for globally unique identifiers such as social security numbers can be added to overlap schemas, but attributes for non-standard vocabularies such as titles (e.g., CEO vs Chief Executive Officer) should not be added to overlap schemas.  

\begin{figure}
\vspace{.1in}
\begin{center}
\begin{tabular}{ccc}
\parbox{1.5in}{
\fbox{
\hspace{.2in} 
\xymatrix@=25pt{
S:= &  \LTO{O} \ar[dl]_{f} \ar[dr]^g & \\
\LTO{P}   &   & \LTO{T} \ar@{-}[d]  \\
& & \LTOO{Att} 
} \hspace{.2in} }
}
& 
 \hspace{-.7in}$\Too{F_1}$ \hspace{-.8in}
&
\hspace{-.9in}
\parbox{1.5in}{
\fbox{
\xymatrix@=25pt{
S_1:= &  \LTO{Observation} \ar[dl]_{f} \ar[dr]^g & \\
\LTO{Person} \ar[r]_{h} \ar@{-}[d] &   \LTO{Gender} \ar@{-}[d]  & \LTO{ObsType} \ar@{-}[d] \\
\LTOO{Att} &   \LTOO{Att} &  \LTOO{Att} 
}}}
\vspace{.1in}
 \\

  $\downarrow F_2$ & & $\downarrow G_1$ \vspace{.1in}
 \\
  
\parbox{1.5in}{
\fbox{
\xymatrix@=25pt{
 S_2:=&  \LTO{Observation} \ar[dl]_{f} \ar[r]^{g_1} &  \LTO{Method}  \ar[d]^{g_2}    \\
\LTO{Patient} \ar@{-}[d] \ar@{-}[d] &  &   \LTO{Type} \ar@{-}[d] \\
\LTOO{Att} & &  \LTOO{Att} 
}}}
& \hspace{.9in} $\Too{G_2}$ & \hspace{-.5in}
\hspace{.55in}
\parbox{1.5in}{
\fbox{ 
\xymatrix@=25pt{
 T := &  \LTO{O} \ar[dl]_{f} \ar[r]^{g_1} & \LTO{M} \ar[d]_{g_2}  \\
 \LTO{P} \ar[r]_{h} \ar@/^/@{-}[d] \ar@/_/@{-}[d]  &   \LTO{G} \ar@{-}[d]  & \LTO{T} \ar@{-}[d]  \\
\LTOO{Att1} \ \LTOO{Att2}  &   \LTOO{Att} &  \LTOO{Att} 
} }}
\end{tabular}
\end{center}
\vspace{.2in}
\caption{Medical Records Schema Integration}
\label{fig1}


\end{figure}

With the overlap schema in hand, we now turn toward our input data.  We are given two input instances, $I_1$ on $S_1$ and $I_2$ on $S_2$.  Entity-resolution (ER) techniques~\cite{Doan:2012:PDI:2401764} can be applied to construct an overlap instance $I$ automatically.  Certain ER techniques can even be implemented as queries in the CQL tool, as we will describe in the next section.  But for the purposes of this example we will construct the overlap instance by hand.  We first assume there are no common observations across the instances; for example, perhaps a cardiologist and nephrologist are merging their records.  We also assume that the observation type vocabulary (e.g., ``BP'' and ``Weight'') are standard across the input instances, so we put these observation types into our overlap instance.  Finally, we see that there is one patient common to both input instances, and he is named Peter in $I_1$ and Pete in $I_2$, so we add one entry for Pete/Peter in our overlap instance.  We have thus completed the input to our design pattern (Figure~\ref{fig2}).

In the output of our pattern (Figure~\ref{fig2}) we see that the observations from $I_1$ and $I_2$ were disjointly unioned together, as desired; that the observation types were (not disjointly) unioned together, as desired; and that Pete and Peter correspond to the same person in the integrated instance.  In addition, we see that Jane could not be assigned a gender.  

\subsection{Pushout-based Data Integration in Practice}

In practice in the CQL tool, we observe several phenomena that are not accounted for in the above theoretical description of the pushout pattern:

\begin{itemize}

\item In those data integration scenarios where a desired integrated schema is given (e.g., a particular star schema motivated by analytics), the process of migrating data from the pushout schema to the target schema seems to invariably be given by a $\Pi \circ \Delta$ migration; i.e., be given by evaluating (rather than co-evaluating) uber-flowers.  

\item The integrated (pushed-out) instance may more ``baggy'' than desired; i.e., there may be multiple rows that are equivalent up to the values of their attributes and foreign keys.  The de-duplication operation discussed in Spivak \& Wisnesky~\cite{relfound} can be used to ``make distinct'' the rows of the integrated instance.

\item When generalizing from pushouts to arbitrary colimits (essentially, $n$-ary pushouts~\cite{BW}) it becomes apparent that there is significant  redundancy in the above span-of-mappings approach to specifying colimits at both the schema and instance level.  We can avoid this redundancy by specifying colimits by means of co-products (disjoint unions) and quotients.  For example, CQL psuedo-code for a quotient-based specification of the schema pushout in Figure~\ref{fig1} is:

\begin{verbatim}
schema T = S1 + S2 / 
  entity equations
    S1_Observation = S2_Observation
    S1_Person = S2_Patient
    S1_ObsType = S2_Type
  path equations 
    S1_f = S2_f
    S1_g = S2_g1.S2_g2
    S1_ObsType_att = S2_Type_att
\end{verbatim}

The above CQL code fragment is six times shorter than the corresponding CQL code for specifying pushouts using spans of schema mappings, and we now always prefer the quotient based approach in practice.  Quotients can also be used to reduce verbosity for instance pushouts.  

\item As mentioned earlier, finding a ``good'' (rather than canonical) presentation of a pushout schema is a subtly difficult problem that occurs in similar form in many places in algebraic specification~\cite{rabe}.  As an alternative to pushouts, so-called pseudo-pushouts may be used.  Conceptually, these are similar to pushouts, but entities are {\it made isomorphic} rather than {\it equated}.  Although pseudo-pushout schemas are ``maximally large'' (the number of entities in the pseudo pushout of $S_1$ and $S_2$ will be the sum of the number of entities in $S_1$ and $S_2$), the entities, attributes, and foreign keys of pseudo-pushout schemas are very natural to name.  Comparing pushouts to pseudo-pushouts for data integration purposes is an important area for future work.

\end{itemize}


\begin{figure}[h]
\begin{center}
\begin{footnotesize}
\begin{tabular}{ccc}
\fbox{
\begin{tabular}[t]{ccc}
$I := \ $ \begin{tabular}[t]{c}
\multicolumn{1}{ c }{{\sf O}}\\ 
{\sf ID}\\\hline 
\end{tabular}
&
\begin{tabular}[t]{c}
\multicolumn{1}{ c }{{\sf P}}\\ 
{\sf ID}\\\hline 
 $p$ \\
\end{tabular}
&
\begin{tabular}[t]{cc}
\multicolumn{1}{ c }{{\sf T}}\\ 
{\sf  ID} & {\sf Att} \\\hline 
$bp$ & BP \\\hline
$wt \ $ & Weight \\
$ \ $ \\
\end{tabular} 
\end{tabular}
}
&$ \Too{(F_1, h_1)}$ & 
\fbox{
\begin{tabular}[t]{cccc}
$I_1$ := \begin{tabular}[t]{cc}
\multicolumn{2}{c}{{\sf Gender}}\\ 
{\sf  ID}&{\sf Att}\\\hline 
$f$ & F \\\hline 
$m$ & M 
\end{tabular}
&
\begin{tabular}[t]{cc}
\multicolumn{2}{c}{{\sf ObsType}}\\ 
{\sf  ID}& \hspace{.1in} {\sf  Att}\\\hline 
$bp$ & BP \\\hline 
$wt \ $ & Weight \\\hline
$hr$ & HR 
\end{tabular}
\\
\begin{tabular}[t]{ccc}
\multicolumn{3}{c}{{\sf Observation}}\\ 
{\sf  ID}& \hspace{.1in} {\sf f} \hspace{.05in} & {\sf g} \\\hline 
$o_5$ & $pe$ & $bp$ \\\hline 
$o_6$ & $pa$ & $hr$ \\\hline
$o_7$ & $pe$ & $wt$ 
\end{tabular}
&
\hspace{.1in}
\begin{tabular}[t]{ccc}
\multicolumn{3}{c}{{\sf Person}}\\ 
{\sf  ID}& \hspace{.1in}{\sf Att}\hspace{.05in} & {\sf h} \\\hline 
$pa$ & Paul & $m$ \\\hline 
$pe$ & Peter & $m$ 
\end{tabular}
\end{tabular}
}
\vspace{.1in}
\\
$\downarrow (F_2, h_2)$ & & $\downarrow (G_2, j_2)$ \vspace{.1in}\\
\fbox{\parbox{2.4in}{
\begin{tabular}[t]{cccc}
$I_2$ := \begin{tabular}[t]{cc}
\multicolumn{2}{c}{{\sf Method}}\\ 
{\sf  ID}&{\sf g2} \\\hline 
$m_1$ & $bp$  \\\hline 
$m_2$ & $bp$ \\\hline 
$m_3$ & $wt$  \\\hline 
$m_4$ & $wt$ 
\end{tabular}
&

\begin{tabular}[t]{cc}
\multicolumn{2}{c}{{\sf Type}}\\ 
{\sf  ID}  &{\sf  Att}\\\hline 
$bp$ & BP \\\hline 
$wt$ & Weight
\end{tabular}
 \\
\begin{tabular}[t]{ccc}
\multicolumn{3}{c}{{\sf Observation}}\\ 
{\sf  ID}  & {\sf f} & {\sf g} \\\hline 
$o_1$ & $p$ & $m_1$ \\\hline 
$o_2$ & $p$ & $m_2$ \\\hline 
$o_3$ & $j$ & $m_3$ \\\hline
$o_4$ & $j$ & $m_1$
\end{tabular}
& 
\begin{tabular}[t]{cc}
\multicolumn{2}{c}{{\sf Patient}}\\ 
{\sf  ID}&{\sf  Att}\\\hline 
$j$ & Jane \\\hline 
$p$ & Pete 
\end{tabular}
\end{tabular}
}}
&$ \Too{(G_1, j_1)}$ &
\fbox{\parbox{2.6in}{
\begin{tabular}[t]{cc} 
\begin{tabular}[t]{cc}
\multicolumn{2}{c}{{\sf M}}\\ 
\hspace{.1in} {\sf  ID} \hspace{.1in} & \hspace{.1in}{\sf  g2} \hspace{.1in} \\\hline 
{\sf g1}($o_5$) & $bp$  \\\hline 
{\sf g1}($o_6$) & $wt$  \\\hline 
{\sf g1}($o_7$) & $hr$  \\\hline 
$m_1$ & $bp$  \\\hline 
$m_2$ & $bp$  \\\hline 
$m_3$ & $wt$  \\\hline 
$m_4$ & $wt$ 
\end{tabular}
&
\begin{tabular}[t]{ccc}
\multicolumn{3}{c}{{\sf O}}\\ 
{\sf  ID}&\hspace{.1in}{\sf f}\hspace{.1in} & \hspace{.1in}{\sf g1}\hspace{.1in} \\\hline 
$o_1$ & $pe$ & $m_1$ \\\hline 
$o_2$ & $pe$ & $m_2$ \\\hline 
$o_3$ & $j$ & $m_3$ \\\hline
$o_4$ & $j$ & $m_1$ \\\hline 
$o_5$ & $pe$ & {\sf g1}($o_5$) \\\hline 
$o_6$ & $pa$ & {\sf g1}($o_6$) \\\hline 
$o_7$ & $pe$ & {\sf g1}($o_7$)
\end{tabular}
 \\
\begin{tabular}[t]{cc}
\multicolumn{2}{c}{{\sf G}}\\ 
{\sf  ID}&{\sf Att}\\\hline 
$f$ & F \\\hline 
$m$& M \\\hline 
{\sf h}$(j) \ $ & {\sf Att(h(}$j))$ \\
\end{tabular}
&
\begin{tabular}[t]{cc}
\multicolumn{2}{c}{{\sf T}}\\ 
{\sf  ID}&{\sf Att}  \\\hline 
$bp$ & BP  \\\hline 
$wt$ & Weight  \\\hline 
$hr$ & HR \\
\end{tabular} \\
\end{tabular} \\ 
\begin{tabular}[t]{cccc}
\multicolumn{4}{c}{{\sf P}}\\ 
{\sf  ID}&{\sf Att1} & {\sf Att2}  & {\sf h}\\\hline 
$j$ & {\sf Att1}($j$) & Jane & {\sf h}($j$) \\\hline 
$pa$ & Paul & {\sf Att2}($pa$) & $m$ \\\hline 
$pe$ & Peter &  Pete & $m$  
\end{tabular} \\  $$\ \ \ \ \ \ \ \ \ \ \ \ \  \ \ \ \ \ \ \ \ \ \ \ \ \ \ \ \ \ \ \ \ \ \ \ \ \ \ =:J$$ }
}
 
\end{tabular}
\end{footnotesize}
\end{center}
\caption{Medical Records Data Integration}
\label{fig2}
\end{figure}
\subsection{Entity-resolution Using Uber-flowers}
\label{er}


Let schemas $S, S_1, S_2$, mappings $F_1 : S \to S_1, F_2 : S \to S_2$, and $S_1$-instance $I_1$ and $S_2$-instance $I_2$ be given. In practice, we anticipate that sophisticated {\it entity-resolution}~\cite{Doan:2012:PDI:2401764} techniques will be used to construct the overlap $S$-instance $I$ and transforms $h_1 : \Sigma_{F_1}(I) \to I_1$ and $h_2 : \Sigma_{F_2}(I) \to I_2$.  However, it is possible to perform a particularly simple kind of entity resolution directly by evaluating uber-flowers. 

Technically, the overlap instance used in the pushout pattern should not be thought of as containing resolved (unified) entities; rather, it should be thought of as containing the {\it record linkages} between entities that will resolve (unify)~\cite{Doan:2012:PDI:2401764}.  The pushout resolves entities by forming equivalence classes of entities under the equivalence relation induced by the links.  As the size of the overlap instance gets larger, the size of the pushout gets smaller, which is the opposite of what would happen if the overlap instance contained the resolved entities themselves, rather than the links between them.  For example, let $A$ and $B$ be instances on some schema that contains an entity ${\sf Person}$, and let $A({\sf Person}) := \{a_1, a_2\}$ and $B({\sf Person}) := \{b_1, b_2\}$.  If the overlap instance $O$ has $O({\sf Person}) := \{(a_1,b_1), (a_1, b_2), (a_2, b_1), (a_2,b_2)\}$, then this does not mean that the pushout will have four people; rather, the pushout will have one person corresponding to $\{a_1,a_2,$ $b_1,b_2\}$, because these four people are linked.  Intuitively, the overlap instance $I$ constructed by the technique in this section is (isomorphic to) a sub-instance of $\Delta_{F_1}(I_1) \times \Delta_{F_2}(I_2)$, where $\times$ denotes a kind of product of instances which we will not define here.  

Let $inc_1 : S_1 \to S_1+S_2$ and $inc_2 : S_2 \to S_1+S_2$ be inclusion schema mappings, and define the $S_1+S_2$ instance $I' := \Sigma_{inc_1}(I_1) + \Sigma_{inc_2}(I_2)$.  This instance will contain $I_1$ and $I_2$ within it, and will contain nothing else.  (Here $X + Y$ means co-product, which is equivalent to the pushout of $X$ and $Y$ over the empty schema or instance).

\begin{figure}[h]
\begin{footnotesize}   
\begin{tabular}{lll}
$Q_1 : S_1 + S_2 \to S :=$
\begin{minipage}{2in}
\begin{verbatim}
 O := for o1 in Observation1 
      keys f->[p1->o1.f], g->[t1->o1.g]
 P := for p1 in Person
 T := for t1 in ObsType, return att->t1.att 
\end{verbatim}
\end{minipage}
\\
 $q_1 \downarrow$ \vspace{.1in}
\\
$Q : S_1 + S_2 \to S:=$
\begin{minipage}{3in}
\begin{verbatim}
  O := for o1 in Observation1, o2 in Observation2 
       where 1 = 2        
       keys f->[p1->o1.f, p2 -> o2.f], 
            g->[t1->o1.g, t2 -> o2.g1.g2]
  P := for p1 in Person, p2 in Patient
       where true = strMatches(p1.PatientAtt, p2.PersonAtt)
  T := for t1 in ObsType, t2 in Type
       where t1.ObsTypeAtt = t2.TypeAtt
       return att->t1.ObsTypeAtt
\end{verbatim}
\end{minipage}
\\
 $q_2 \uparrow$  \vspace{.1in}
\\
$Q_2 : S_1 + S_2 \to S :=$
\begin{minipage}{2in}
\begin{verbatim}
 O := for o2 in Observation2
      keys f->[p1->o2.f], g->[t1->o2.g1.g2]
 P := for p2 in Patient
 T := for t2 in Type, return att->t2.att
\end{verbatim}
\end{minipage}
\end{tabular}
\end{footnotesize}

\vspace*{.1in}
$$
\xymatrix@=50pt@R-2pc{
 S_1 \ar@{^{(}->}[r]^{inc_1} & S_1 + S_2 & \ar@{_{(}->}[l]_{inc_2} S_2 \\
}
$$
$$
  I' \in S_1 + S_2 \iinst:= \Sigma_{inc_1}(I_1) + \Sigma_{inc_2}(I_2) \ \ \ \ \ \ \ \ \ \ I \in S\iinst := eval(Q)(I') 
$$

\vspace*{.1in}

\xymatrix@=50pt@R-3pc{
 I \ar[r]^{eval(q_n)(I')} & eval(Q_n)(I') & & \\
\Sigma_{F_n}(I) \ar[r]^{\Sigma_{F_n}(eval(q_n)(I')) \ \ \ \ \ \ } &  \Sigma_{F_n}(eval(Q_n)(I')) \ar[r]^{\cong }  & \Sigma_{F_n}(\Delta_{F_n}(I_n)) \ar[r]^{\epsilon} & I_n
} 

\vspace{.1in}
\caption{Entity-Resolution on Medical Records using Uber-flowers}
\label{erex}
\end{figure}
We construct overlap $S$-instance $I$ by defining a query $Q : S_1 + S_2 \to S$ and evaluating it on the $S_1 + S_2$-instance $I'$.  For each entity $s \in S$, we choose a set of pairs of attributes from $F_1(s)$ and $F_2(s)$ that we desire to be ``close''.  In the medical records example, for $P$ we choose $({\sf PatientAtt}, {\sf PersonAtt})$ and for $T$ we choose $({\sf ObsTypeAtt}, {\sf TypeAtt})$; we choose nothing for $O$.  We next choose a way to compare these attributes; for example, we choose a string edit distance of less than two to indicate that the entities match.  This comparison function must be added to our type-side, e.g., 
$$
{\sf strMatches} : {\sf String} \times {\sf String} \to {\sf Nat} \ \ \ \ \ {\sf true} : {\sf Nat} \ \ \ \ \ {\sf true} = {\sf 1}
$$ 
The function {\sf strComp} can be defined using equations, although the CQL tool allows such functions to be defined using Java code (see section~\ref{impl}).  With the {\sf String}-comparator in hand, we can now define $Q : S_1 + S_2 \to S$ as in Figure~\ref{erex}. The overlap instance $I$ is defined as $eval(Q)(I')$. To construct $h_n : \Sigma_{F_n}(I) \to I_n$ for $n=1,2$, we define projection queries $Q_n : S_1 + S_2 \to S$ and inclusion query morphisms $q_n : Q_n \to Q$ as in Figure~\ref{erex} as follows.  We start with the induced transforms for $q_n$, then apply $\Sigma_{F_n}$, then compose with the isomorphism $eval(Q_n)(I') \cong \Delta_{F_n}(I_n)$, and then compose the co-unit $\epsilon$ of the $\Sigma_{F_n} \dashv \Delta_{F_n}$ adjunction, to obtain $h_n$ as in Figure~\ref{erex}.  

The result of running Figure~\ref{erex} on the medical records data $I_1, I_2$ from Figure~\ref{fig2} is the overlap instance $I$ from Figure~\ref{fig2}.  To compute the isomorphism $eval(Q_n)(I') \to \Delta_{F_n}(I_n)$, we note that the generators of $eval(Q_n)(I')$ will be singleton substitutions such as
$
[v_n  \mapsto inj_n \ a_n] 
$
where $a_n$ is a term in $\Sigma_{inc_n}(I_n)$ and $inj_n$ means co-product injection.  But $inc_n$ is an inclusion, so $a_n$ is a term in $I_1$.  Because we compute $\Delta_{F_n}$ by translation into an uber-flower similar to $Q_n$, the generators of $\Delta_{F_n}(I')$ will have a similar form:
$
[v'_n  \mapsto a_n] 
$
which defines the necessary isomorphism.  When all schemas are disjoint and variables are chosen appropriately, the isomorphism can be made an equality.


  

%
%

\section{Conclusion}

In this paper we have described an algebraic formalism for integrating data, and work continues.  In the short term, we aim to formalize our experimental ``computational type-sides'', and to develop a better conservativity checker.  In the long term, we are looking to develop other design patterns for  data integration and to study their compositions, and we are developing an equational theorem prover tailored to our needs.  In addition to these concrete goals, we believe there is much to be gained from the careful study of the differences between our formalism, with its category-theoretic semantics, and the formalism of embedded dependencies, with its relational semantics~\cite{Doan:2012:PDI:2401764}.  For example, there is a semantic similarity between our $\Sigma$ operation and the chase; as another example, so far we have found no relational counterpart to the concept of query ``co-evaluation''; and finally, our uber-flower queries may suggest generalizations of comprehension syntax~\cite{monad}.  

\section{Acknowledgments}  

The authors would like to thank David Spivak and Peter Gates.  Patrick Schultz was supported by AFOSR grant FA9550-14-1-0031, ONR grant N000141310260, and NASA grant NNH13ZEA001N. Ryan Wisnesky was supported by NIST SBIR grant 70NANB15H290.

\section{Errata}

\begin{itemize}

\item Corrected the $Q_F$ query of Section~\ref{querytrans}.

\item Added a complete query composition algorithm to Section~\ref{composing}.  To use it to compute co-eval without saturating, Let $Q : S \to T$ be a query and $J$ a given $T$-instance.  Consider $J$ as a query $J'$ on a target schema $1$ with single target table $\star$ with no attributes or foreign keys, so that $J' : T \to 1$.  Then $coeval_Q(J) \cong (Q ; J')(\star)$.  Furthermore, pre-composition with $Q$ has a right adjoint given by using $eval_Q$ on each frozen instance of a given input query.       

\item Fixed typo ($\int S$) in pivot Section~\ref{ppivot}.

\item Switched from $x.a$ notation to $x;a$ notation in many places.

\item Removed JFP formatting.

\item Conexus AI was formerly named Categorical Informatics.  Updated contact info.  Replaced AQL by CQL.  Updated URLs.

\item The paper ``Fast Left Kan Extensions Using the Chase'' shows that iterated pushouts are a sound, but not complete, method to chase theories in regular logic.  \url{https://arxiv.org/abs/2205.02425}

\item The paper ``Presenting Pro-functors'' and the errata to ``Algebraic databases'' describe additional details of how the formalism in ``Algebraic Databases'' relates to this paper.  \url{https://arxiv.org/abs/2404.01406}

\item Pseudo-colimits of schemas, rather than schemas, turned out to be easier to implement in practice, as discovered when integrating spreadsheets.  \url{https://arxiv.org/abs/2209.14457} 

\end{itemize}


\newpage

\setcounter{tocdepth}{3}
\begin{small}
\tableofcontents
\end{small}
\newpage 
\listoffigures

\label{lastpage}

\end{document}